\pdfoutput=1
\documentclass[12pt,a4paper]{article}

\usepackage{ifthen} 
\newboolean{pdflatex}
\setboolean{pdflatex}{true} 

\newboolean{articletitles}
\setboolean{articletitles}{true} 

\newboolean{uprightparticles}
\setboolean{uprightparticles}{false} 

\def\paperauthors{
  Marta Calvi$^{1}$, 
  Tommaso Fulghesu$^{2}$,
  George Hallett$^{3}$,
  Luca Hartman$^{4,5}$,
  Basem Khanji$^{6}$
  Veronica S. Kirsebom$^{1}$\textcolor{red}{$^\dagger$},
  Thomas Latham$^{3}$,
  Marion Lehuraux$^{3}$,
  Ching-Hua Li$^{2}$\textcolor{red}{$^\dagger$},
  Abhijit Mathad$^{4}$\textcolor{red}{$^\dagger$},
  Matthew Monk$^{3,7,8}$,
  Andy Morris$^{2}$\textcolor{red}{$^\dagger$},
  Matthew Scott Rudolph$^{6}$,
  Francesca Swystun$^{7}$,
  Dorothea vom Bruch$^{2}$
} 
\def\paperasciititle{Minimising Event Size, Maximising Physics: Inclusive Particle Isolation for LHCb's Run 3} 
\def\papertitle{Minimising Event Size, Maximising Physics: Inclusive Particle Isolation for LHCb's Run 3} 

\def\paperkeywords{{High Energy Physics}, {LHCb}} 
\def\papercopyright{CERN on behalf of the LHCb collaboration}
\def\paperlicence{CC BY 4.0 licence}
\def\paperlicenceurl{https://creativecommons.org/licenses/by/4.0/}

\usepackage[top=1in, bottom=1.25in, left=1in, right=1in]{geometry}

\usepackage{microtype}
\usepackage{lineno}  
\usepackage{xspace} 
\usepackage{caption} 

\usepackage{graphicx}  
\usepackage{color}
\usepackage{colortbl}
\graphicspath{{./figs/}} 

\usepackage{amsmath} 
\usepackage{amssymb}
\usepackage{amsfonts}
\usepackage{upgreek} 

\newcommand*\patchAmsMathEnvironmentForLineno[1]{%
\expandafter\let\csname old#1\expandafter\endcsname\csname #1\endcsname
\expandafter\let\csname oldend#1\expandafter\endcsname\csname
end#1\endcsname
 \renewenvironment{#1}%
   {\linenomath\csname old#1\endcsname}%
   {\csname oldend#1\endcsname\endlinenomath}%
}
\newcommand*\patchBothAmsMathEnvironmentsForLineno[1]{%
  \patchAmsMathEnvironmentForLineno{#1}%
  \patchAmsMathEnvironmentForLineno{#1*}%
}
\AtBeginDocument{%
\patchBothAmsMathEnvironmentsForLineno{equation}%
\patchBothAmsMathEnvironmentsForLineno{align}%
\patchBothAmsMathEnvironmentsForLineno{flalign}%
\patchBothAmsMathEnvironmentsForLineno{alignat}%
\patchBothAmsMathEnvironmentsForLineno{gather}%
\patchBothAmsMathEnvironmentsForLineno{multline}%
\patchBothAmsMathEnvironmentsForLineno{eqnarray}%
}

\usepackage[pdftex,
            pdfauthor={\paperauthors},
            pdftitle={\paperasciititle},
            pdfkeywords={\paperkeywords}]{hyperref}
\usepackage{hyperxmp}

\usepackage[colorinlistoftodos,textsize=scriptsize]{todonotes}

\usepackage[bottom,flushmargin,hang,multiple]{footmisc}

\usepackage[all]{hypcap} 

\usepackage{xspace} 
\usepackage{upgreek}

\def\lhcb   {\mbox{LHCb}\xspace}
\def\atlas  {\mbox{ATLAS}\xspace}
\def\cms    {\mbox{CMS}\xspace}

\def\lhc    {\mbox{LHC}\xspace}

\def\MagUp {\mbox{\em Mag\kern -0.05em Up}\xspace}

\ifthenelse{\boolean{uprightparticles}}%
{

 \def\Pmu         {\ensuremath{\upmu}\xspace}

 \def\Ppi         {\ensuremath{\uppi}\xspace}

 \def\Ptau        {\ensuremath{\uptau}\xspace}

 \def\PDelta      {\ensuremath{\Delta}\xspace}                 
 \def\PXi         {\ensuremath{\Xi}\xspace}                 
 \def\PLambda     {\ensuremath{\Lambda}\xspace}                 
 \def\PSigma      {\ensuremath{\Sigma}\xspace}                 
 \def\POmega      {\ensuremath{\Omega}\xspace}                 
 \def\PUpsilon    {\ensuremath{\Upsilon}\xspace}
 \let\oldPi\Pi
 \def\PPi         {\ensuremath{\oldPi}\xspace}

 \def\PB      {\ensuremath{\mathrm{B}}\xspace}                 
                  
 \def\PD      {\ensuremath{\mathrm{D}}\xspace}

 \def\PK      {\ensuremath{\mathrm{K}}\xspace}

 \def\Pb      {\ensuremath{\mathrm{b}}\xspace}                 
 \def\Pc      {\ensuremath{\mathrm{c}}\xspace}                 
                  
 \def\Pe      {\ensuremath{\mathrm{e}}\xspace}

 \def\Pi      {\ensuremath{\mathrm{i}}\xspace}

 \def\Ps      {\ensuremath{\mathrm{s}}\xspace}

 \def\thebaroffset{0.0em}
}
{

 \def\Pmu         {\ensuremath{\mu}\xspace}

 \def\Ppi         {\ensuremath{\pi}\xspace}

 \def\Ptau        {\ensuremath{\tau}\xspace}

 \mathchardef\PDelta="7101
 \mathchardef\PXi="7104
 \mathchardef\PLambda="7103
 \mathchardef\PSigma="7106
 \mathchardef\POmega="710A
 \mathchardef\PUpsilon="7107
 \mathchardef\PPi="7105
                  
 \def\PB      {\ensuremath{B}\xspace}                 
                  
 \def\PD      {\ensuremath{D}\xspace}

 \def\PK      {\ensuremath{K}\xspace}

 \def\Pb      {\ensuremath{b}\xspace}                 
 \def\Pc      {\ensuremath{c}\xspace}                 
                  
 \def\Pe      {\ensuremath{e}\xspace}

 \def\Pi      {\ensuremath{i}\xspace}

 \def\Ps      {\ensuremath{s}\xspace}

 \def\thebaroffset{0.18em}
}
\newcommand{\offsetoverline}[2][\thebaroffset]{\kern #1\overline{\kern -#1 #2}}%

\makeatletter
\ifcase \@ptsize \relax
  \newcommand{\miniscule}{\@setfontsize\miniscule{4}{5}}
\or
  \newcommand{\miniscule}{\@setfontsize\miniscule{5}{6}}
\or
  \newcommand{\miniscule}{\@setfontsize\miniscule{5}{6}}
\fi
\makeatother

\DeclareRobustCommand{\optbar}[1]{\shortstack{{\miniscule (\rule[.5ex]{1.25em}{.18mm})}
  \\ [-.7ex] $#1$}}

\def\epm        {{\ensuremath{\Pe^\pm}}\xspace}

\def\mupm       {{\ensuremath{\Pmu^\pm}}\xspace}

\def\taupm      {{\ensuremath{\Ptau^\pm}}\xspace}

\def\squark    {{\ensuremath{\Ps}}\xspace}

\def\cquark    {{\ensuremath{\Pc}}\xspace}

\def\bquark    {{\ensuremath{\Pb}}\xspace}

\def\pion   {{\ensuremath{\Ppi}}\xspace}
\def\piz    {{\ensuremath{\pion^0}}\xspace}
\def\pip    {{\ensuremath{\pion^+}}\xspace}
\def\pim    {{\ensuremath{\pion^-}}\xspace}

\def\kaon    {{\ensuremath{\PK}}\xspace}

\def\KorKbar {\kern \thebaroffset\optbar{\kern -\thebaroffset \PK}{}\xspace}

\def\Kp      {{\ensuremath{\kaon^+}}\xspace}
\def\Km      {{\ensuremath{\kaon^-}}\xspace}

\def\Dbar    {{\ensuremath{\offsetoverline{\PD}}}\xspace}
\def\D       {{\ensuremath{\PD}}\xspace}

\def\DorDbar {\kern \thebaroffset\optbar{\kern -\thebaroffset \PD}\xspace}
\def\Dz      {{\ensuremath{\D^0}}\xspace}
\def\Dzb     {{\ensuremath{\Dbar{}^0}}\xspace}
\def\Dp      {{\ensuremath{\D^+}}\xspace}
\def\Dm      {{\ensuremath{\D^-}}\xspace}

\def\DpDm    {\ensuremath{\Dp {\kern -0.16em \Dm}}\xspace}

\def\Dstarp  {{\ensuremath{\D^{*+}}}\xspace}
\def\Dstarm  {{\ensuremath{\D^{*-}}}\xspace}

\def\Dsp     {{\ensuremath{\D^+_\squark}}\xspace}
\def\Dsm     {{\ensuremath{\D^-_\squark}}\xspace}

\def\B       {{\ensuremath{\PB}}\xspace}

\def\BorBbar {\kern \thebaroffset\optbar{\kern -\thebaroffset \PB}\xspace}
\def\Bz      {{\ensuremath{\B^0}}\xspace}

\def\Bd      {{\ensuremath{\B^0}}\xspace}

\def\BdorBdbar {\kern \thebaroffset\optbar{\kern -\thebaroffset \Bd}\xspace}
\def\Bu      {{\ensuremath{\B^+}}\xspace}

\def\Bp      {{\ensuremath{\Bu}}\xspace}

\def\Bs      {{\ensuremath{\B^0_\squark}}\xspace}

\def\BsorBsbar {\kern \thebaroffset\optbar{\kern -\thebaroffset \Bs}\xspace}

\def\Y#1S{\ensuremath{\PUpsilon{(#1S)}}\xspace}

\def\Lz          {{\ensuremath{\PLambda}}\xspace}

\def\LorLbar     {\kern \thebaroffset\optbar{\kern -\thebaroffset \PLambda}\xspace}

\def\Lc          {{\ensuremath{\Lz^+_\cquark}}\xspace}

\def\Lb           {{\ensuremath{\Lz^0_\bquark}}\xspace}

\def\to                 {\ensuremath{\rightarrow}\xspace}

\def\AT#1     {\ensuremath{A_{\mathrm{T}}^{#1}}\xspace}           

\def\C#1      {\ensuremath{\mathcal{C}_{#1}}\xspace}                       
\def\Cp#1     {\ensuremath{\mathcal{C}_{#1}^{'}}\xspace}                    
\def\Ceff#1   {\ensuremath{\mathcal{C}_{#1}^{\mathrm{(eff)}}}\xspace}        
\def\Cpeff#1  {\ensuremath{\mathcal{C}_{#1}^{'\mathrm{(eff)}}}\xspace}       
\def\Ope#1    {\ensuremath{\mathcal{O}_{#1}}\xspace}                       
\def\Opep#1   {\ensuremath{\mathcal{O}_{#1}^{'}}\xspace}                    

\newcommand{\aunit}[1]{\ensuremath{\text{\,#1}}}       

\newcommand{\tev}{\aunit{Te\kern -0.1em V}\xspace}
\newcommand{\gev}{\aunit{Ge\kern -0.1em V}\xspace}
\newcommand{\mev}{\aunit{Me\kern -0.1em V}\xspace}
\newcommand{\kev}{\aunit{ke\kern -0.1em V}\xspace}
\newcommand{\ev}{\aunit{e\kern -0.1em V}\xspace}
 
\newcommand{\mevc}{\ensuremath{\aunit{Me\kern -0.1em V\!/}c}\xspace}
\newcommand{\gevc}{\ensuremath{\aunit{Ge\kern -0.1em V\!/}c}\xspace}
\newcommand{\mevcc}{\ensuremath{\aunit{Me\kern -0.1em V\!/}c^2}\xspace}
\newcommand{\gevcc}{\ensuremath{\aunit{Ge\kern -0.1em V\!/}c^2}\xspace}

\newcommand{\chisq}{\ensuremath{\chi^2}\xspace}

\def\gsim{{~\raise.15em\hbox{$>$}\kern-.85em
          \lower.35em\hbox{$\sim$}~}\xspace}
\def\lsim{{~\raise.15em\hbox{$<$}\kern-.85em
          \lower.35em\hbox{$\sim$}~}\xspace}

\def\tell1  {TELL1\xspace}
\def\ukl1   {UKL1\xspace}

\newcommand{\ie}{\mbox{\itshape i.e.}\xspace}

\newcommand{\DsStarm}{\ensuremath{D_s^{*-}}\xspace}
\newcommand{\LcStar}{\ensuremath{\Lambda_c^{*+}}\xspace}
\newcommand{\IMI}{\textsc{IMI}\xspace}

\usepackage{cite} 
\usepackage{mciteplus}
\usepackage{xcolor,booktabs,ragged2e}
\usepackage{tabularx}

\usepackage{longtable} 

\usepackage{booktabs} 

\usepackage{graphicx}
\usepackage{amsmath}
\usepackage{enumitem}
\usepackage{subcaption}
\usepackage{xcolor}
\usepackage{multirow}
\usepackage{float}
\usepackage{listings}
\usepackage{xkeyval}
\usepackage{placeins}

\definecolor{codegreen}{rgb}{0,0.6,0}
\definecolor{codegray}{rgb}{0.5,0.5,0.5}
\definecolor{codepurple}{rgb}{0.58,0,0.82}
\definecolor{backcolour}{rgb}{0.95,0.95,0.92}

\lstdefinestyle{mystyle}{
    backgroundcolor=\color{backcolour},   
    commentstyle=\color{codegreen},
    keywordstyle=\color{magenta},
    numberstyle=\tiny\color{codegray},
    stringstyle=\color{codepurple},
    basicstyle=\ttfamily\footnotesize,
    breakatwhitespace=false,         
    breaklines=true,                 
    captionpos=b,                    
    keepspaces=true,                 
    numbers=left,                    
    numbersep=5pt,                  
    showspaces=false,                
    showstringspaces=false,
    showtabs=false,                  
    tabsize=2
}

\begin{document}


\renewcommand{\thefootnote}{\fnsymbol{footnote}}
\setcounter{footnote}{1}


\begin{titlepage}

\vspace*{-1.5cm}

\noindent
\begin{tabular*}{\linewidth}{lc@{\extracolsep{\fill}}r@{\extracolsep{0pt}}}
\ifthenelse{\boolean{pdflatex}}
{\vspace*{-1.5cm}\mbox{\!\!\!\includegraphics[width=.14\textwidth]{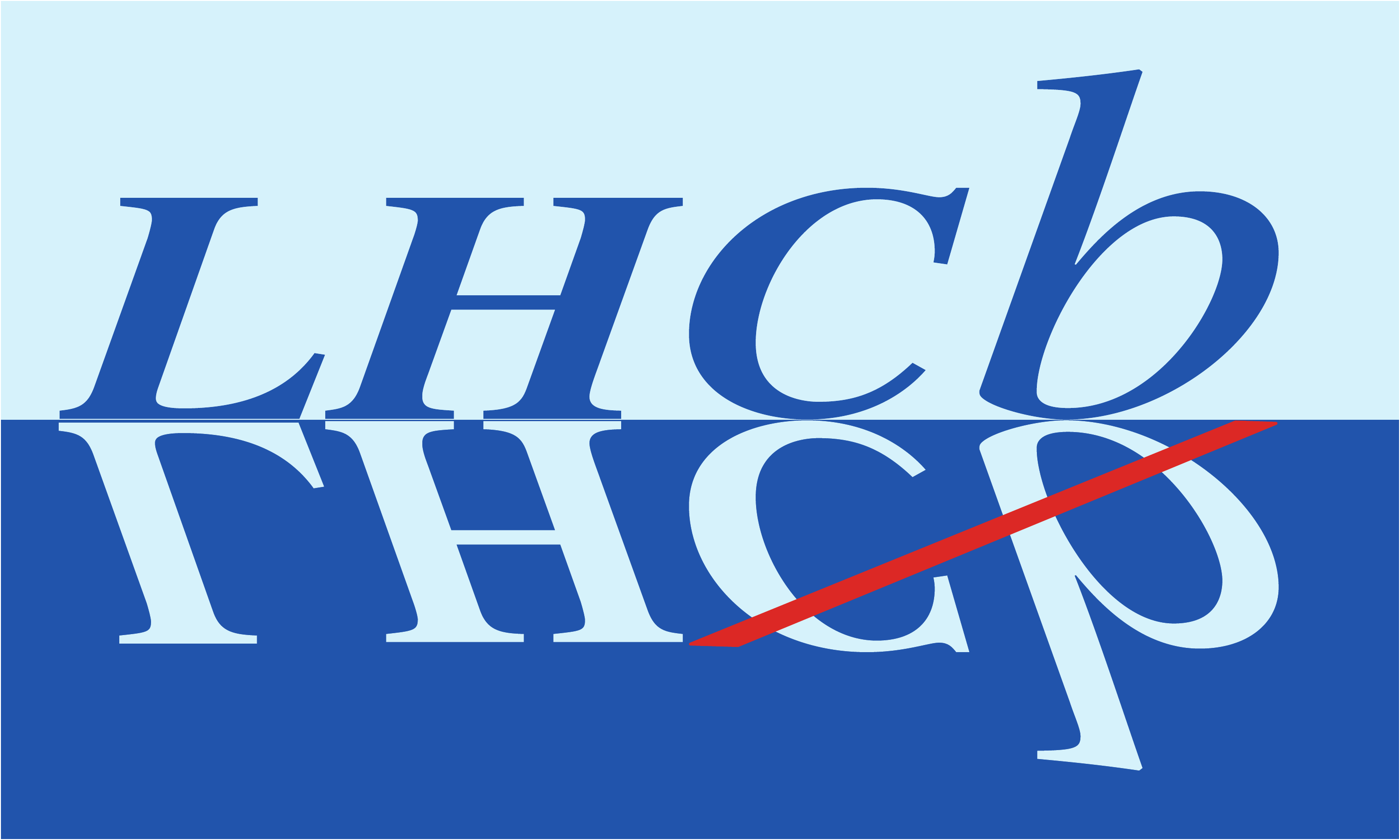}} & &}%
{\vspace*{-1.5cm}\mbox{\!\!\!\includegraphics[width=.12\textwidth]{figs/lhcb-logo.pdf}} & &}
 \\
 & & CERN-LHCb-RD-2025-001 \\  
 \\  
 & & \today \\ 
 & & \\
\hline
\end{tabular*}

\vspace*{1.2cm}

{\normalfont\bfseries\boldmath\huge
\begin{center}
  \papertitle
\end{center}
}

\vspace*{-0.2cm}

\begin{center}
\paperauthors
\bigskip\\
{\normalfont\itshape\footnotesize
$^1$ University of Milan-Bicocca, Milan, Italy\\
$^2$ Aix Marseille Univ, CNRS/IN2P3, CPPM, Marseille, France\\
$^3$ Department of Physics, University of Warwick, Coventry, United Kingdom\\
$^4$ European Organization for Nuclear Research (CERN), Geneva, Switzerland\\
$^5$ Ecole Polytechnique Fédérale de Lausanne (EPFL), Lausanne, Switzerland\\
$^6$ Syracuse University, Syracuse, New York, USA\\
$^7$ Cavendish Laboratory, University of Cambridge, Cambridge, United Kingdom\\
$^8$ School of Physics and Astronomy, Monash University, Melbourne, Australia\\
\vspace{0.5cm}
\textcolor{red}{$^\dagger$} Contact authors (alphabetic): 
\href{mailto:veronica.soelund.kirsebom@cern.ch}{V. S. Kirsebom},
\href{mailto:ching-hua.li@cern.ch}{C. H. Li},
\href{mailto:amathad@cern.ch}{A. Mathad},
\href{mailto:andrew.george.morris@cern.ch}{A. Morris}.
\centerline{Keywords: High-energy-physics, LHCb experiment, Real-time trigger, Data Processing and Offline Analysis}
}

\end{center}


\begin{abstract}
  \noindent 
  The Run~3 of the \lhc brings unprecedented luminosity and a surge in data volume to the \lhcb detector,
  necessitating a critical reduction in the size of each reconstructed event without compromising the physics reach of the heavy-flavour programme.
  While signal decays typically involve just a few charged particles, 
  a single proton-proton collision produces hundreds of tracks, with charged particle information dominating the event size.
  To address this imbalance, a suite of inclusive isolation tools have been developed, including both classical methods and a novel Inclusive Multivariate Isolation (\IMI) algorithm.
  The \IMI unifies the key strengths of classical isolation techniques and is designed to robustly handle diverse decay topologies and kinematics, enabling efficient reconstruction of decay chains with varying final-state multiplicities.
  It consistently outperforms traditional methods, with superior background rejection and high signal efficiency across diverse channels and event multiplicities.
  By retaining only the most relevant particles in each event, the method achieves a 45\% reduction in data-size while preserving full physics performance, selecting signal particles with 99\% efficiency.
  We also validate IMI on Run 3 data, confirming its robustness under real data-taking conditions.
  In the long term, \IMI\ could provide a fast, lightweight front-end to support more compute-intensive selection strategies in the high-multiplicity environment of the High-Luminosity LHC.
\end{abstract}

\vspace*{0.3cm}
{\footnotesize
\centerline{\copyright~\papercopyright. \href{\paperlicenceurl}{\paperlicence}.}}
\vspace*{2mm}
\end{titlepage}

\pagestyle{empty}  




\renewcommand{\thefootnote}{\arabic{footnote}}
\setcounter{footnote}{0}

\tableofcontents
\cleardoublepage


\pagestyle{plain} 
\setcounter{page}{1}
\pagenumbering{arabic}



\section{Introduction}
\label{sec:intro}

A central challenge for the upgraded \lhcb detector~\cite{LHCb-DP-2022-002} during Run~3 (2022--2026) is to minimise the amount of data written to permanent storage while preserving the full physics potential of the experiment. 
The Large Hadron Collider (LHC) now delivers an instantaneous luminosity of up to $\mathcal{L} = 2 \times 10^{33}\,\mathrm{cm}^{-2}\mathrm{s}^{-1}$ to the \lhcb experiment, a five-fold increase compared to Run~2 (2015--2018).
In addition, the upgraded \lhcb detector is fully read out at the 30~MHz non-empty bunch-crossing rate and is processed by a software-only High-Level Trigger (HLT)~\cite{Aaij:2019zbu,Saur:2023dez,Schulte:2023gtt}.

At this input rate, the HLT reconstructs and selects\footnote{The HLT consists of a fast {\sc HLT1} stage with partial event reconstruction, followed by a {\sc HLT2} stage with full event reconstruction.} approximately $\mathcal{O}(250~\text{kHz})$ of physics events. These are subsequently written to tape at a total data rate of around $\mathcal{O}(10~\text{GB/s})$~\cite{LHCb-TDR-018}. These events are split across three data streams: \textsc{Turbo}, \textsc{Turcal}, and \textsc{Full}.

The \textsc{Turbo} stream, inspired by the model introduced in Run~2~\cite{LHCb-DP-2016-001,LHCb-DP-2019-002}, stores only the online-reconstructed information relevant to the triggered signal, yielding a compact, analysis-ready format. 
It is used in two configurations: (i) \emph{minimal} \textsc{Turbo}, which persists only the objects explicitly requested by the selection algorithms (often referred as \emph{selection lines}), typically the signal candidate and essential metadata, and (ii) \textsc{Turbo} with \emph{selective persistency} (\textsc{Turbo}(SP)), which additionally saves a controlled set of associated objects (e.g.\ vertices) according to per-line rules. The latter trades a modest increase in event size for greater offline flexibility.
In contrast, the \textsc{Full} stream retains the entire reconstructed event, enabling detailed offline analysis. 
For instance, in semileptonic $b$-hadron decays (final states containing one or more leptons), controlling backgrounds from partially reconstructed decays is essential. In such cases, the \textsc{Full} stream enables direct reconstruction of these backgrounds, providing strong constraints on their shapes, yields, and associated systematics.
The \textsc{Turcal} stream further supplements the data by including raw detector information required for calibration tasks (e.g.\ particle identification)
Because of their richer event content, \textsc{Full} and \textsc{Turcal} dominate storage bandwidth.
As our work focuses on physics analyses, we specifically target events recorded in the \textsc{Full} and \textsc{Turbo} streams.

To further reduce data volume, the {\sc Sprucing} framework~\cite{Abdelmotteleb:2025kyd} performs a second, centralised selection step that reduces the total output rate to approximately 3.5~GB/s written to disk.
For the \textsc{Full} stream in particular, the Technical Design Report (TDR)~\cite{LHCb-TDR-018} mandates a reduction from 5.9~GB/s at the output of {\sc HLT2} to just 0.8~GB/s after sprucing~\cite{Abdelmotteleb:2025kyd}. 
Achieving this near eight-fold compression across $\mathcal{O}(10^3)$ Sprucing selection lines requires aggressive yet efficient pruning of event content.

The total data rate can be expressed as the product of two factors:
\begin{equation}
\text{Data-rate}\,[\text{MB/s}]
\propto
\underbrace{\text{event rate}}_{\raisebox{0.6ex}{\text{Hlt/Sprucing}}}\,[\text{kHz}]
\times
\underbrace{\text{event size}}_{\raisebox{0.6ex}{\text{Hlt/Sprucing}}}\,[\text{kB}] .
\label{eq:drate}
\end{equation}
Therefore, reduction in the data rate can be achieved by either lowering the number of events retained (event rate) or by decreasing the amount of information stored per event (event size).
Extensive efforts by the \lhcb physics and performance working groups have addressed both strategies, developing high-efficiency reconstruction~\cite{Li:2022tlq} and selection algorithms to retain not only the most ``interesting" events, but also signal candidates well-suited for fast offline analysis.
This work advances the second component in Eq.~\eqref{eq:drate}, aiming to reduce the full event size through isolation algorithms that do more than just isolate signal decays, they selectively retain only the most relevant event information for detailed offline analysis.

\begin{figure}[!htp]
    \centering
    \includegraphics[width=0.6\textwidth]{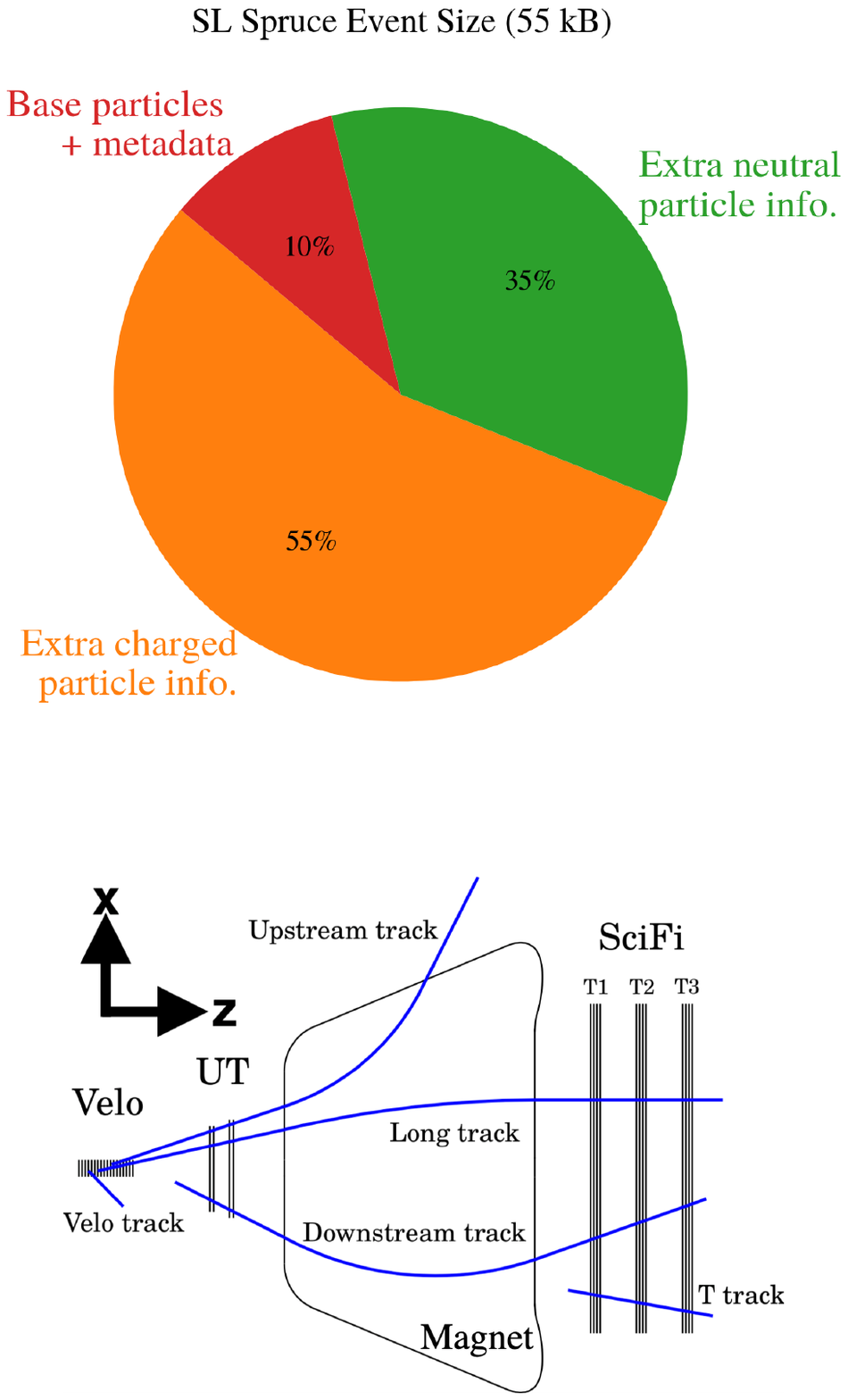}
    \caption{
        (Top) Event size breakdown for semileptonic \textsc{Full} stream events in \lhcb, based on a minimum-bias Run~3 simulation.  
        The \textit{base particles and metadata} include a few particles from a partially reconstructed $b$-hadron decay (e.g., $\mathrm{D}^0$ and a muon), together with primary vertex, trigger, and reconstruction metadata.  
        The \textit{extra charged particle} component covers all other charged tracks (\textit{Long}, \textit{Downstream}, \textit{Upstream}) and RICH particle identification (PID) information. Additional \textit{VELO} and \textit{T} tracks contribute 10\% and 15\% of the event size, respectively, and are not shown as they do not contribute to the Semileptonic event size.  
        The \textit{extra neutral particle} component corresponds to reconstructed neutrals, such as photons and neutral hadrons.  
        (Bottom) Track categories in \lhcb, defined by the tracking sub-detectors where hits are recorded~\cite{LHCb-DP-2022-002}.
    }
    \label{fig:full_stream_size}
\end{figure}

A typical event at \lhcb contains reconstructed charged and neutral particles, along with associated information such as primary vertices, trigger decisions, and a reconstruction summary.
Figure~\ref{fig:full_stream_size} (top) shows the size composition of a representative \textsc{Full} stream event used in semileptonic (SL) analyses.
Only about 10\% of the event size comes from metadata and the few particles of a partially reconstructed $b$-hadron decay. Neutral particles, mainly photons and neutral hadrons measured in the electromagnetic calorimeters, contribute roughly 35\%. The largest share, nearly 55\%, originates from reconstructed charged particles.
As illustrated in figure~\ref{fig:full_stream_size} (bottom), these charged particles are classified into track categories according to the sub-detectors in which they leave hits.
\textit{Long} tracks are reconstructed with hits in the vertex locator (VELO), a high precision silicon detector surrounding the interaction region, and in the downstream tracking stations, providing the best momentum resolution and vertex association.
\textit{Upstream} tracks have hits in the VELO and upstream tracking stations only and typically correspond to low momentum particles that do not traverse the full spectrometer.
\textit{Downstream} and \textit{T} tracks are reconstructed without VELO information and are characteristic of long lived decays such as $K_S^0$ or $\Lambda$, with \textit{T} tracks using only downstream station hits.
\textit{VELO} tracks use hits only in the VELO and are essential for tracking and vertexing close to the interaction point, in particular for reconstructing primary vertices.
Together with particle identification from the Ring Imaging Cherenkov detectors, the tracking information dominates the event size.
For a more detailed information on the track types and their reconstruction, see Ref.~\cite{LHCb-DP-2022-002}.
This consideration leads us to the central question:
\begin{quote}
\itshape
How can the relevant components of an event be identified efficiently, in a fast and inclusive manner, without introducing sensitivity to specific decay topologies or kinematic properties?
\end{quote}

The core difficulty arises from \emph{pileup}: multiple $pp$ interactions per bunch crossing produce several primary (PVs) and secondary vertices (SVs), and the challenge is to associate reconstructed objects with the PV/SV that produced the signal hadrons. Inside the VELO, precise impact-parameter and vertexing information enable a largely geometric association to the correct vertex. Outside the VELO, however -- after the magnet and into the downstream tracking stations, or when using RICH, calorimeter, or muon information -- the association becomes far more difficult: longitudinal resolution degrades, magnetic deflection complicates back-extrapolation, and neutral objects lack track parameters altogether. Any strategy that reduces event size while preserving physics sensitivity must therefore identify and retain only the subset of objects \emph{compatible with the signal PV/SV}, while rejecting contributions from other vertices and pileup activity.

This paper introduces an \emph{Inclusive Multivariate Isolation} (\IMI) algorithm specifically designed to address this association problem. 
Unlike classical isolation methods based on cones or vertexing, \IMI\ evaluates and scores additional charged particles in the event, selecting only those most likely to originate from the same PV and decay chain as the signal candidate, while discarding the rest. 
As an illustrative example, consider the decay
\begin{equation}
\Bz \to D^{*-} \, \textcolor{blue}{\mu^{+}} \, \nu_{\mu}, 
\qquad 
D^{*-} \to \textcolor{blue}{\bar \Dz} \, \textcolor{red}{\pi^{-}}.
\label{eq:example_decay}
\end{equation}

In this example, the $\textcolor{blue}{\bar D^0}$ and $\textcolor{blue}{\mu^{+}}$ constitute the \textit{base} signal particles: they are selected with loose track-quality, PID, and vertex requirements to provide a minimal representation of the $\bquark$-hadron decay vertex. The $\textcolor{red}{\pi^{-}}$ is an additional \emph{non-isolated} particle that belongs to the same decay chain and that \IMI\ is designed to retain with high background rejection, thereby enabling the reconstruction of the $D^{*-}$ state.
By carefully selecting only the most relevant extra particles, \IMI\ can significantly reduce the per-event size.  
Its design is therefore guided by the following key objectives:
\begin{itemize}
    \item Deliver better background rejection than classical isolation techniques, particularly in high-pileup environments where traditional algorithms tend to degrade.
    \item Enable applications beyond standard combinatorial suppression, including:
    \begin{enumerate}
        \item Reconstruction of excited charm or charmless states involving particles from secondary decays, e.g.\ $B^0 \to D^{*-} \mu^+ \nu_\mu$, essential for precision tests of lepton-flavour universality~\cite{LHCb:2023uiv,LHCb:2023zxo};
        \item Selection of data-driven control samples, such as the dominant background mode $\Lambda_b \to \Lambda_c^+ \mu^- \nu_\mu$, to isolate suppressed signal decays like $\Lambda_b \to p \mu^- \nu_\mu$, critical for probing CKM unitarity~\cite{LHCb:2015eia};
        \item Reconstruction of excited beauty-hadron states involving particles consistent with originating from the PV associated with the beauty hadron of interest, e.g., kaons in $B^*_{s2}(5840) \to B^+ K^-$, relevant for studies of lepton-flavour violation (LFV)~\cite{LHCb:2023zxo} and relative branching-fraction measurements~\cite{LHCb:2018azb}.
    \end{enumerate}
    \item 
    minimizing its impact on memory usage and on the overall event processing time (throughput),  while remaining robust to variations in decay topology and kinematics.
\end{itemize}

\begin{figure}[!htp]
    \centering
    \includegraphics[width=1\textwidth]{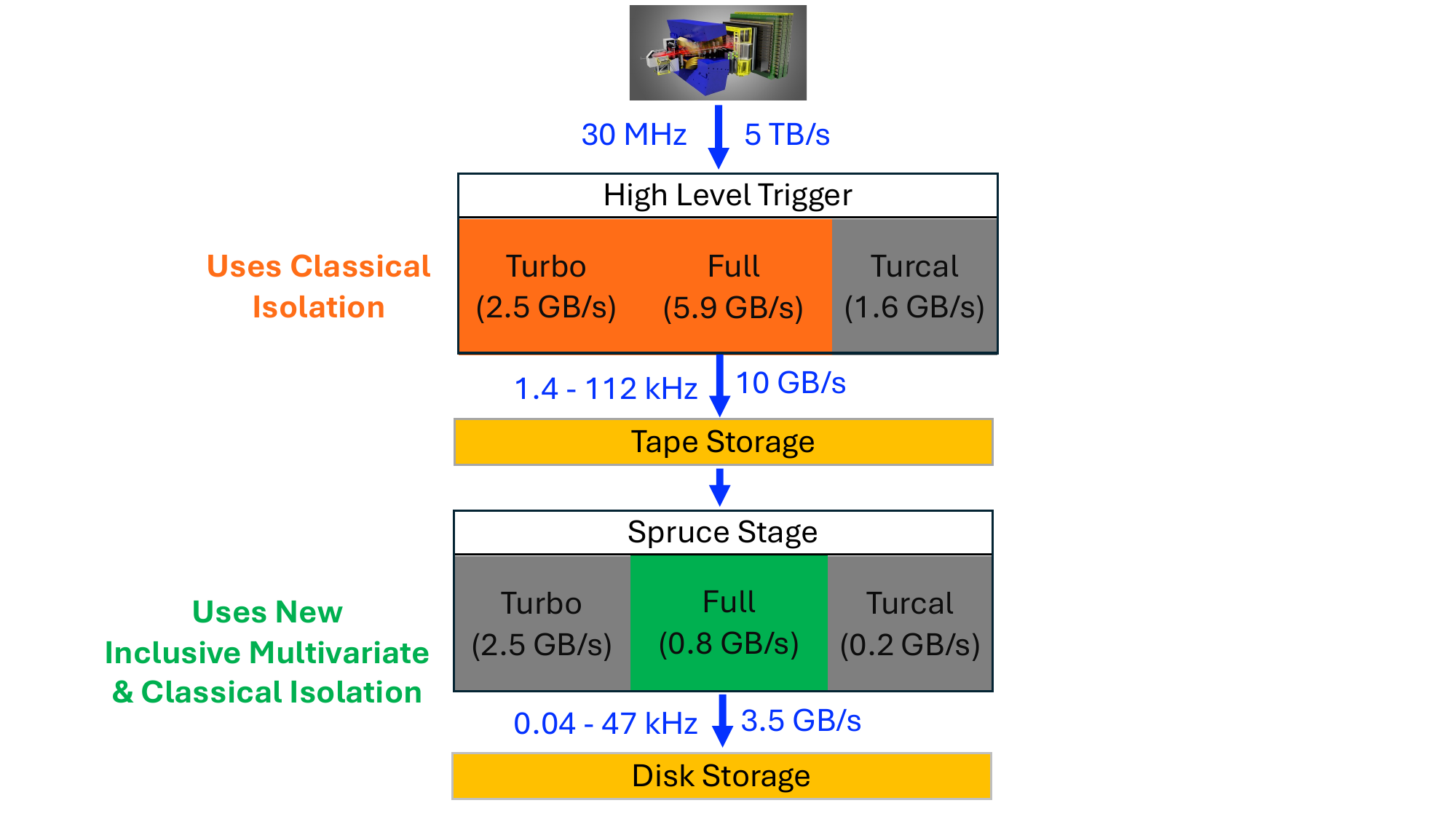}
    \caption{
        The \lhcb data flow during Run~3, illustrating the throughput and event rates for different streams, and indicating where the classical and \IMI isolation algorithms are applied. Note that none of the Spruce selection lines using \IMI rely on candidates preselected with the classical isolation tool. 
        The quoted event rates include all isolation-based selection lines, and the bandwidths correspond to the maximum stream allocations from Ref.~\cite{LHCb-TDR-018}.
    }
    \label{fig:dataflow}
\end{figure}

As part of this work, classical isolation techniques were also developed for Run~3, including cone-based and vertex-based isolation, which remain widely used in many physics selections. Figure~\ref{fig:dataflow} shows the Run~3 data flow, including throughput and event rates for the different streams, and indicates where the classical isolation and \IMI\ algorithms operate: the classical isolation runs in {\sc HLT2}, whereas \IMI\ is applied conservatively at the {\sc Sprucing} stage. 
The two approaches are independent, \ie, selection lines using \IMI\ do not rely on candidates preselected with the classical tool. 
As of the 2025 data-taking period, about $30\%$ of selection lines across the \textsc{Turbo} and \textsc{Full} streams use the classical isolation developed here, while \IMI\ is employed in roughly $20\%$ of {\sc Sprucing} selection lines targeting the \textsc{Full} stream.

The paper is organised as follows.
Section~\ref{sec:previous_isolation} reviews classical isolation techniques, 
Section~\ref{sec:multivariate_isolation} describes the \IMI algorithm, compares its performance with classical methods, and studies its impact on key offline kinematic observables.
Section~\ref{sec:implementation} details the implementation of the classical isolation variants and the \IMI algorithm, and quantifies their effect on data reduction.
Section~\ref{sec:validation} presents validation of the \IMI algorithm using Run~3 data.
Conclusions and outlook are given in Section~\ref{sec:conclusion}.

\section{Classical Isolation Algorithms}
\label{sec:previous_isolation}

The forward geometry of the \lhcb detector, combined with its focus on low transverse momentum physics, results in approximately $\mathcal{O}(200)$ reconstructible charged tracks per inelastic $pp$ interaction at Run~3 instantaneous luminosity.
Isolating the few tracks, typically $\mathcal{O}(2$--$7)$, that originate from a heavy flavour signal decay in this dense environment is therefore essential, both for physics performance and for reducing data volume.  

\begin{figure}
    \centering
    \includegraphics[width=0.6\textwidth]{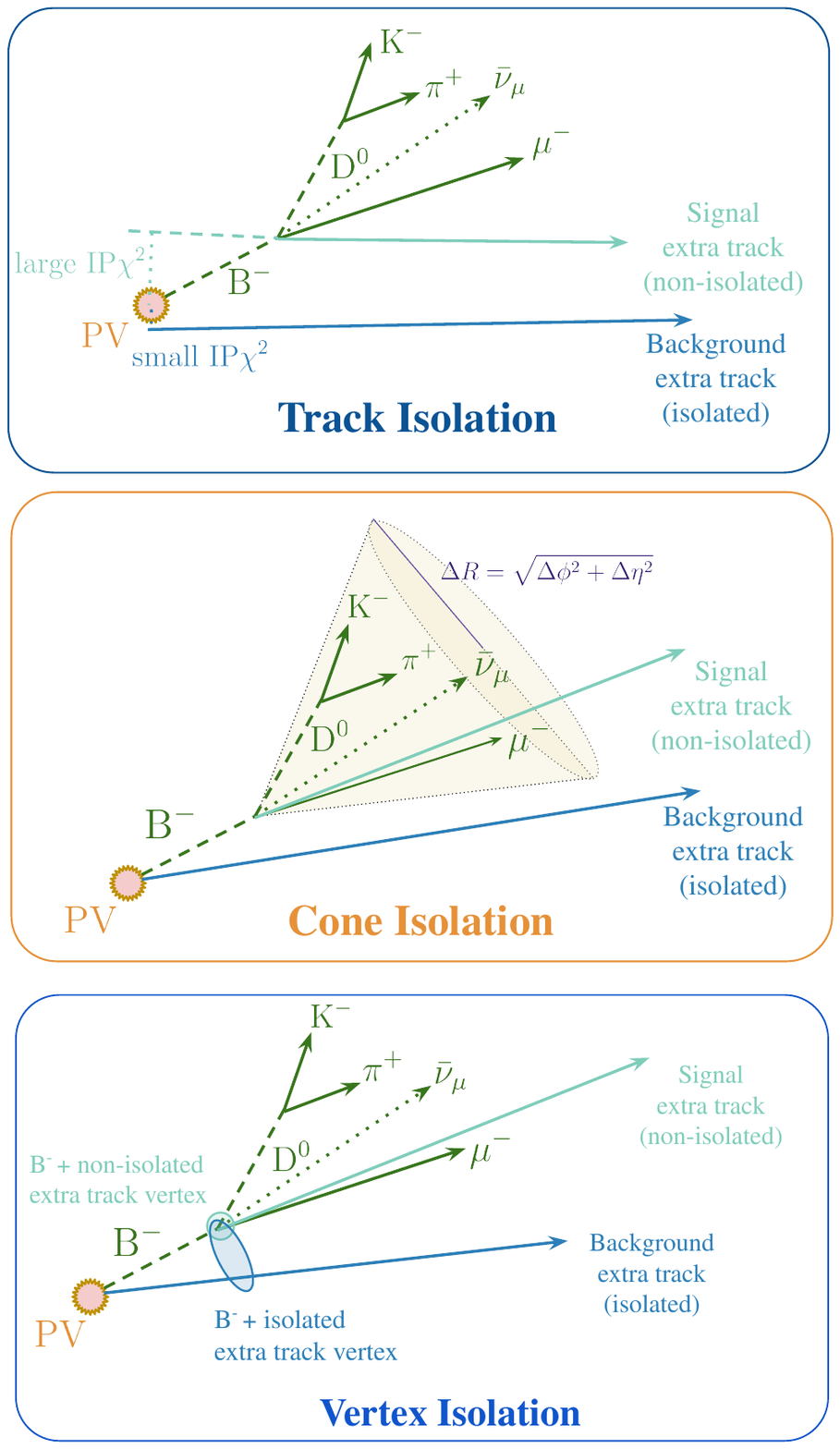}
    \caption{
    Schematic illustration of the three classical isolation strategies:  
    track isolation (top), based on large impact-parameter significance;  
    cone isolation (middle), based on the track's proximity to the signal candidate;  
    and vertex isolation (bottom), based on compatibility with the signal decay vertex.  
    }
    \label{fig:cone_isolation}
\end{figure}

Over the past decade, three complementary charged particle isolation strategies have been developed and routinely employed at \lhcb, 
as well as at the general-purpose LHC experiments and in other high energy physics experiments\footnote{ 
     Neutral-particle isolation can also be built from electromagnetic (ECAL)
     and Hadronic (HCAL) clusters, but its contribution is not as large as charged-particle isolation (see Fig.~\ref{fig:full_stream_size}).
     We therefore restrict the present \IMI study to charged-particle isolation and leave the neutral case for future work, noting that classical isolation algorithms already support it.
}.
While traditionally used to \emph{isolate} signal candidates, defined here as reconstructed candidates for the decay of interest, by suppressing nearby activity, these techniques can also be used in the complementary mode of \emph{selecting and retaining} nearby, decay related (\emph{non-isolated}) particles, and this work focuses on the latter. Accordingly, we refer to \emph{isolated} background particles as those unlikely to be associated with the signal decay, and to \emph{non-isolated} signal particles as those likely to be associated with it (see Fig.~\ref{fig:cone_isolation} for clarity). 
The three classical isolation strategies considered here are: 

\begin{description}

\item[\textbf{Track isolation:}]  
Particles clearly associated with a primary vertex (PV) are rejected by requiring a large impact-parameter significance with respect to  all the reconstructed PVs in the event, $\chi^{2}_{\mathrm{IP\,wrt\,PV}}$, \emph{or} an association with the same PV as the signal candidate (\texttt{samePV} flag). 
The impact parameter (IP) is defined as the minimum distance between the particle's trajectory and the given PV, while $$\chi^{2}_{\mathrm{IP\,wrt\,PV}}$$ quantifies the change in the vertex-fit $\chi^{2}$ when the track is included in the PV fit. 
It is also common to impose a second requirement: that the particle has a small impact-parameter significance with respect to the secondary vertex (SV) of the signal candidate,$\mathrm{IP}\,\chi^{2}_{\mathrm{SV}}$. 
Simple cut-based implementations of this approach were first developed by the D\O{}~\cite{D0:2007vto} and CDF~\cite{CDF:2013ezj} experiments, followed by ATLAS~\cite{ATLAS:2012qdc} and CMS~\cite{CMS:2011spw}. The method was later adopted at \lhcb for rare decay searches~\cite{Martinez:1027522,LHCb:2009zny} and for flavour-tagging algorithms~\cite{LHCB-PAPER-2011-027}.
Multivariate extensions, incorporating additional topological and kinematic features, as well as track-type information (\textit{Long} and \textit{VELO})~\cite{LHCB-DP-2014-002}, were first introduced in $B^0_{(s)} \to \mu^+ \mu^-$ analyses~\cite{LHCb-PAPER-2011-001} and LFV studies~\cite{Gavardi:2013kvq}, and 
they are also used in rare decay searches at \atlas and \cms~\cite{ATLAS:2018cur,CMS:2022mgd}.
A schematic representation of how the track isolation strategy works is shown in Fig.~\ref{fig:cone_isolation}\,(top).

\item[\textbf{Cone isolation:}]  
All reconstructed particles within a cone of radius
\begin{equation*}
R = \sqrt{(\Delta\eta)^2 + (\Delta\phi)^2},
\qquad R \in [0.4,0.5],
\label{eq:cone}
\end{equation*}
around the momentum direction of the signal candidate are considered.  
Here, $\Delta\eta$ and $\Delta\phi$ are the differences in pseudorapidity and azimuthal angle, respectively, between the signal candidate and other particles produced in the $pp$ collision, defined in the laboratory frame ($z$ along the beam direction, $y$ vertically upward, and $x$ completing a right-handed coordinate system).
Typical discriminants include the particle multiplicity inside the cone, the transverse momentum of the leading track $p_{\mathrm{T}}^{\text{lead}}$, and the $p_{\mathrm{T}}$ asymmetry,
\begin{equation*}
A_{p_{\mathrm{T}}}
= \frac{p_{\mathrm{T}}(\text{sig}) - \left|\sum_{i \in \text{cone}} \vec{p}^{\,i}\right|_{\mathrm{T}}}
       {p_{\mathrm{T}}(\text{sig}) + \left|\sum_{i \in \text{cone}} \vec{p}^{\,i}\right|_{\mathrm{T}}},
\label{eq:asy}
\end{equation*}
where the sum runs over the three-momenta $\vec{p}^{\,i}$ of all particles in the cone, and the transverse component is taken only after the sum is formed. 
The variable approaches $+1$ for a perfectly isolated signal candidate.
Cone-based variables are fast and robust, and have seen widespread adoption in electroweak, jet, and heavy-flavour analyses~\cite{LHCb-PAPER-2021-037,CDF:2008hmn,LHCb:2021dlw,LHCb-PAPER-2015-031}, as well as in $\tau$ identification at ATLAS and CMS~\cite{ATLAS-CONF-2017-029,CMS-PAS-PFT-08-001}.
However, their discriminating power tends to degrade in high-occupancy environments.  
Figure~\ref{fig:cone_isolation}\,(middle) provides a visual illustration of cone isolation.

\item[\textbf{Vertex isolation:}]  
Reconstructed tracks in the event that are not part of the signal candidate, but pass loose track-quality requirements, are tested for compatibility with the signal decay by fitting them together with the signal candidate to a common secondary vertex. 
If the combined fit yields a $\chi^{2}/\mathrm{ndf}$ below a chosen threshold and the vertex is significantly displaced from the PV, the additional track is considered consistent with the decay chain.
In fully reconstructed decays, a simple cut-based vertex isolation is often sufficient: any extra track compatible with the decay vertex typically indicates a partially reconstructed background, whereas genuine signal candidates are expected to have no such additional tracks. 
For partially reconstructed decays, however, missing final-state particles make this logic less direct, and multivariate strategies have therefore become standard~\cite{LHCB-PAPER-2015-025,Braun:2020hku}.
Finally, we note that the change in vertex-fit quality upon adding the extra track, $\Delta \chi^{2}_{\mathrm{SV\,fit}}$, is highly correlated with the track impact-parameter significance with respect to the signal vertex, $\chi^{2}_{\mathrm{IP\,wrt\,SV}}$, as both quantify the compatibility of the track with the signal decay vertex; the two variables are thus often used interchangeably across analyses. 
A schematic illustration of the vertex-isolation strategy is shown in Fig.~\ref{fig:cone_isolation} (bottom).
\end{description}

Most \lhcb analyses use classical techniques as standalone isolation tools, though in some cases two are combined: fast track-based isolation is typically deployed in the trigger, while the more computationally demanding vertex-based methods are applied offline, where their complementary strengths can be exploited. 
Each of the classical methods provides good signal efficiency, but none achieves the required background rejection in the denser Run~3 environment, where the average number of reconstructible tracks per event has tripled compared to Run~2~\cite{LHCb-DP-2022-002}.

\section{Inclusive Multivariate Isolation (\IMI)}
\label{sec:multivariate_isolation}

Due to the limitations of classical isolation algorithms under high-pileup conditions, a multivariate isolation tool was developed integrating key features from all the three traditional approaches into a single classifier.
As illustrated in Eq.~\eqref{eq:example_decay}, this classifier assigns a score to each combination constructed from a few \textit{base} particles and an \textit{extra} particle, to determine whether the extra particle should be classified as \textit{isolated} or \textit{non-isolated}.
Non-isolated particles (those close to base particles) are more likely to originate from the signal decay  and are retained for further analysis, particularly for reconstructing excited charm or charmless states, or for defining background-enriched control samples.
Conversely, the \IMI score can also be used as a discriminating variable to suppress backgrounds, such as combinatorial and partially reconstructed decays.

To ensure compatibility with the computational constraints of the Sprucing framework, several lightweight machine learning (ML) algorithms were explored, including Multi-Layer Perceptrons (MLPs), Random Forests, and Gradient Boosted Decision Trees (GBDTs).
The Extreme Gradient Boosting (XGBoost) algorithm~\cite{Chen:2016btl} was selected for its optimal balance between classification performance, computational efficiency, interpretability and strong performance on tabular high-energy physics data.

In the following sub-sections, we describe the data samples used for the training and validation of the \IMI model, the selection of input features, and the performance of the \IMI tool compared to classical isolation methods.

\subsection{Data samples}
\label{subsec:data_samples}

To ensure maximal inclusivity, the training sample consists of a broad class of simulated semileptonic beauty-hadron decays, generated under nominal Run~3 LHCb conditions with an instantaneous luminosity of $2\times10^{33}\,\text{cm}^{-2}\text{s}^{-1}$, corresponding to an average pile-up of $\mu=5.3$ (visible interactions) or $\nu=7.6$ total proton-proton interactions per bunch crossing.
The sample spans a variety of initial states, (\Bz), (\Bp), (\Bs), and (\Lb), and final states containing electrons ((\epm)), muons ((\mupm)), or taus ((\taupm)), covering a wide kinematic phase space to avoid biasing the algorithm toward specific kinematic properties of the signal decay. Throughout this paper, references to a decay mode implicitly include its charge-conjugate process.
These decay chains also include both short-lived intermediate states (e.g., $D^{*}_{(s)}$, $D^{**}$, and $\Lambda_c^*$) and long-lived particles (e.g., $\Dz$, $\tau$), enabling the isolation tool to learn how to disentangle contributions from both secondary and tertiary vertices.
This diversity is essential to ensure that the \IMI algorithm generalises robustly to the complex topologies expected in real LHCb data.
A summary of the simulated decay modes used to train \IMI, together with the corresponding \textit{base} and \textit{non-isolated} particles, is given in Table~\ref{tab:samples_summary} in Appendix~\ref{app:simulation_samples}.

In all simulated decays, the base particles are required to be reconstructed as \textit{Long} tracks and are selected with loose vertex, track quality and PID criteria to ensure they originate from a common decay vertex and are consistent with signal decays.
The \IMI\ tool currently targets non-isolated Long and Upstream tracks. 
Downstream tracks, which make up a relatively small fraction of all reconstructed tracks, were excluded due to their poorer momentum and vertex resolution, which limit their contribution to isolation performance. 
VELO-only tracks have no momentum measurement and are therefore not typically used as analysis objects, so they were not included in the baseline IMI. Their main value is as an extra handle for very rare decays by capturing additional charged activity near the b-hadron decay region in the VELO; since VELO tracks are already persisted, they can be added to the IMI training in a future iteration to further improve background rejection if bandwidth becomes a limiting factor.

\subsection{Signal and background classes}
\label{subsec:sig_bkg_definition}

\begin{figure}[t] 
\centering 
\includegraphics[width=0.75\textwidth]{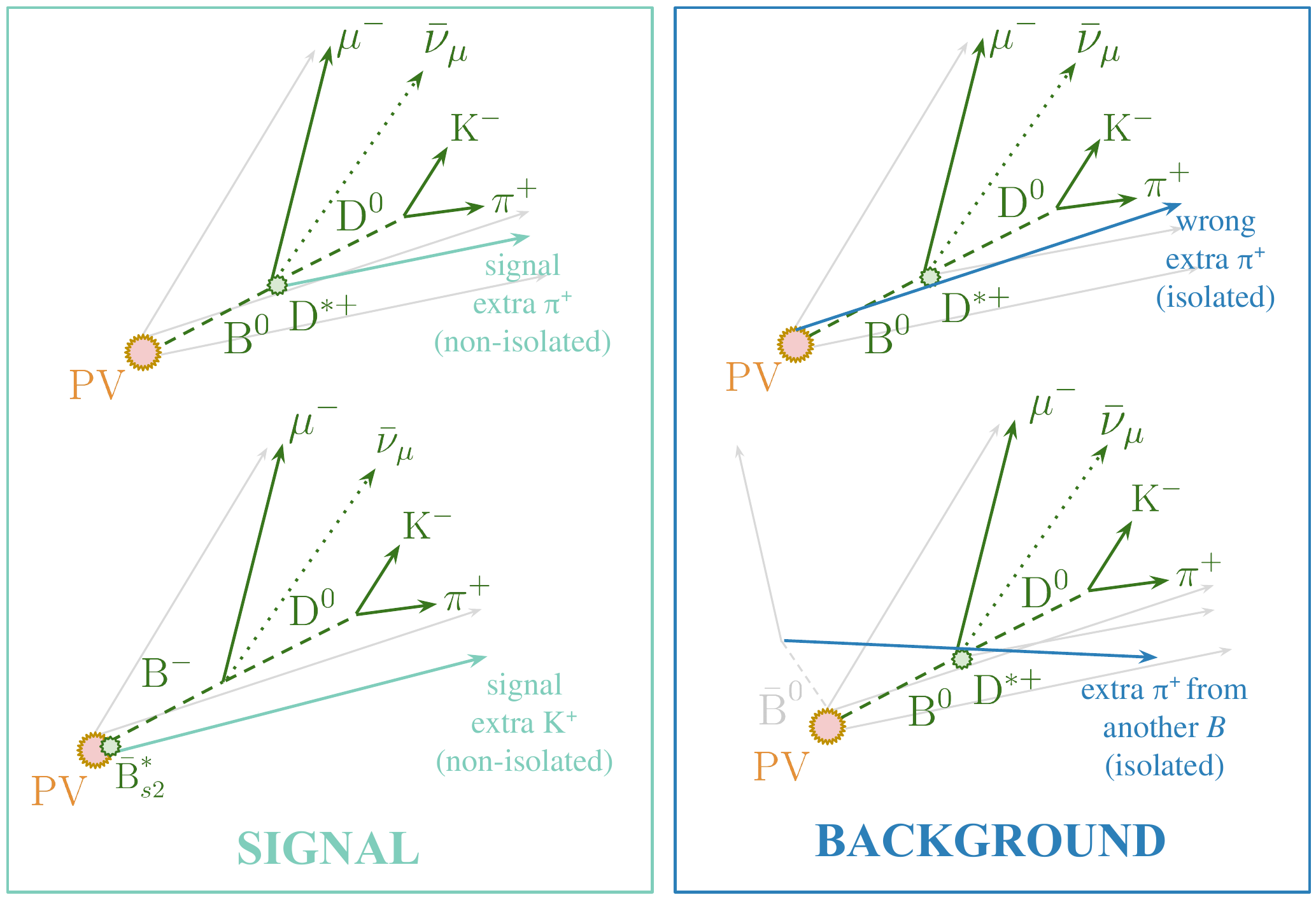}
\caption{
  Illustration of typical signal- and background-like topologies relevant to the \IMI\ tool. The depicted $B_s^{*} \to B K$ decay is shown as an illustrative example and was used only in the development of the cut-based method.
  In the left diagram, the signal candidates include non-isolated particles originating from the decay vertex of a b-hadron or its higher excited state. 
  In the right diagram, the background candidates are defined as those formed by pairing with particles originating from the primary vertex or other b-hadrons in the event.
  }
\label{fig:sig_bkg_definitions} 
\end{figure}

The goal of the \IMI tool is to identify charged particles that genuinely originate from the decay of a heavy-flavour hadron, while rejecting the far more numerous background particles unrelated to the signal decay.
Signal particles are defined as those produced in the decay of a beauty hadron, which typically travels about $\sim 1\,\mathrm{cm}$ from the PV before decaying within the LHCb acceptance. These signal particles fall into two categories:  
\begin{enumerate}[label=(\alph*)]
  \item those produced at the displaced secondary and tertiary vertex of the $b$-hadron decay, and  
  \item those originating from prompt strong decays of excited beauty states, such as $B_{s2}(5840)\!\to\!\Bp\Km$, where the kaon is emitted at the PV, while the $\Bp$ continues to travel and decays further downstream.
\end{enumerate}
The background consists of all other charged particles in the event that are kinematically and topologically uncorrelated with the signal decay. These include:
\begin{enumerate}[label=(\alph*), start=3]
  \item particles produced at or near the PV, predominantly from the soft--QCD hadronisation of light quarks and gluons in the underlying event.
  \item particles originating from the decay of the second $b$-hadron in the event, which often forms the dominant combinatorial background due to their similar displaced-vertex signatures but lack of correlation with the signal vertex.
\end{enumerate}
An illustration of the definitions of signal and background particles, defined in the above manner, is shown in Fig.~\ref{fig:sig_bkg_definitions}.

An architecture was briefly explored categorising outputs into four classes, designed to disentangle all the above categories using inclusive $b\bar b$ simulation samples~\cite{Hartman:2023}.
In practice, the multiclass network delivered no measurable gain: the limited statistics available for categories (b) and (d) prevented the model from learning boundaries more precise than those already captured in a simpler binary formulation.  
Therefore, a two-class \IMI\ classifier was adopted, optimised to separate long-lived $b$-hadron decays from background. Prompt decays of excited $b$-hadrons are instead treated with a dedicated cut-based selection, described in Appendix~\ref{app:excited_b_hadrons}.
Future versions, trained on larger and more diverse simulated samples, may revisit the multiclass strategy to exploit finer distinctions among these categories.

\subsection{Input features}
\label{subsec:training_feature}

The \IMI\ tool was trained on a feature set motivated by classical isolation algorithms and then iteratively refined to maximise signal--background separation while reducing redundancy and limiting correlations with key physics observables (e.g.\ $q^2$). Variables providing only marginal gains in high-occupancy (Run~3-like) conditions were removed, while robust features that generalise across exclusive channels were retained. For readability, we use standard abbreviations throughout: PV (primary vertex), SV (secondary vertex), IP (impact parameter), DOCA (distance of closest approach), POCA (point of closest approach), and DIRA (direction angle between momentum and flight direction).

\begin{description}

\item[\textbf{IP significance wrt\ PV ($\mathbf{\chi^2_{\mathrm{IP\,wrt.\,PV}}}$)}:]
IP $\chi^2$ of the additional particle with respect to the PV; it separates PV-produced (prompt) tracks from displaced tracks originating in long-lived $b$-hadron decays.

\item[\textbf{IP significance wrt\ SV ($\mathbf{\log(\chi^2_{\mathrm{IP\,wrt.\,SV}})}$)}:]
As above, but computed with respect to the SV of the base candidate. Signal-like particles are compatible with the SV (small values), whereas unrelated tracks (including from the other $b$ hadron) tend to be less compatible; tertiary decays (e.g.\ charm, $\tau$) populate intermediate values.

\item[\textbf{Cone separation (\textit{Transformed} $\mathbf{\Delta R}$)}:]
$\Delta R = \sqrt{(\Delta\eta)^2 + (\Delta\phi)^2}$ between the additional and base particles in the laboratory frame (see Sec~\ref{sec:previous_isolation} for definition). Signal-like particles are typically close in angle; to enhance discrimination at small angles we use $(\Delta R)^{0.2}$.

\item[\textbf{Flight-direction alignment (\textit{Transformed} DIRA)}:]
With $\hat{p}$ the momentum direction of the refitted candidate (base+track) and $\hat{FD}$ the PV$\to$SV flight direction,
the direction angle (DIRA) (see Fig.3.4 of Ref~\cite{pescatore_2017}) is defined as
\[
\cos\alpha = \hat{p}\cdot\hat{FD}.
\]
True $b$-hadron candidates are well aligned ($\cos\alpha \to 1$); we use $(1-\cos\alpha)^{0.2}$ to emphasise small misalignments.

\item[\textbf{SV shift (\textit{SV displacement})}:]
To quantify how compatible the additional track is with the base candidate's decay vertex, we refit the base candidate together with the additional track and compare the fitted SV positions. We define
\[
\Delta \vec r_{\rm SV} \equiv \vec r_{\rm SV}^{\,\text{base+trk}} - \vec r_{\rm SV}^{\,\text{base}},
\qquad
d_{\rm SV}^{\rm signed} \equiv \operatorname{sign}(\Delta r_{{\rm SV},z})\,\lvert \Delta \vec r_{\rm SV} \rvert ,
\]
where $\vec r_{\rm SV}^{\,\text{base}}$ is the decay vertex of the base candidate and $\vec r_{\rm SV}^{\,\text{base+trk}}$ is the decay vertex after adding the extra track; $\Delta r_{{\rm SV},z}$ is the $z$-component of the displacement.
For correctly associated tracks (signal-like), the refit leaves the SV essentially unchanged, so $d_{\rm SV}^{\rm signed}$ peaks sharply at $0$, with a width dominated by the SV-fit resolution.
For misassociated tracks (typically prompt PV tracks), the common-vertex fit must compromise between displaced decay tracks and a prompt line pointing back to a PV; this frequently pulls the refitted SV upstream in 
$z$-direction (along the beam axis), yielding a broader distribution with an enhanced negative tail ($d_{\rm SV}^{\rm signed}<0$), although positive shifts are possible depending on the geometry and track covariances.

\item[\textbf{DOCA significance ($\mathbf{\log\chi^2_{\mathrm{DOCA}}}$)}:]
The DOCA (distance of closest approach) between two reconstructed objects is the minimum spatial separation of their trajectories. We use the \emph{significance} of this separation, expressed as
\[
\log\chi^2_{\mathrm{DOCA}},
\]
where $\chi^2_{\mathrm{DOCA}}$ can be interpreted as the squared separation at the POCA (point of closest approach), weighted by the associated covariance matrices; equivalently, it corresponds to the increase in vertex-fit $\chi^2$ when constraining the two objects to originate from a common vertex.
When the base object is a track, both objects are approximated locally as straight lines using their fitted positions and directions in the vicinity of the POCA.
When the base object is composite (\ie decays to two or more particles), it is represented by its flight line from the composite decay vertex along its reconstructed momentum.
This variable provides good discrimination: tracks produced at (or very near) the SV yield small values, tracks from displaced intermediate decays (e.g.\ charm or $\tau$) tend to populate intermediate values, and unrelated combinations (typically prompt PV activity) give the largest values.

\item[\textbf{Transverse momentum ($\mathbf{\log(p_T)}$)}:]
$\log(p_T)$ of the additional particle. It helps reject soft-QCD PV activity; despite some kinematic dependence, its discriminating power motivates inclusion.

\item[\textbf{Signed SV--PV flight distance}:]
For the refitted candidate (base+track), we define the signed displacement between its SV and the associated best PV as
\[
\Delta \vec r_{\rm PV} \equiv \vec r_{\rm SV}^{\,\text{base+trk}} - \vec r_{\rm PV},
\qquad
d_{\rm PV}^{\rm signed} \equiv \operatorname{sign}(\Delta r_{{\rm PV},z})\,\lvert \Delta \vec r_{\rm PV} \rvert ,
\]
where $\vec r_{\rm PV}$ is the position of the best PV (chosen as the PV that minimises the candidate’s IP with respect to the refitted candidate momentum), and $\Delta r_{{\rm PV},z}$ is the $z$-component of the SV--PV displacement.
For genuine long-lived decays, the SV lies downstream of the PV, so $d_{\rm PV}^{\rm signed}$ is predominantly positive, with a tail governed by the $b$-hadron boost.
For background combinations formed by attaching a \emph{prompt} PV track to a genuine displaced base candidate, the refit tends to pull the SV toward the prompt track's PV. Two typical behaviours follow:
(i) if the refitted SV shifts upstream while the best-PV association remains downstream, then $z_{\rm SV}^{\text{base+trk}}<z_{\rm PV}^{\text{best}}$ and $d_{\rm PV}^{\rm signed}$ becomes strongly negative, generating the pronounced negative tail;
(ii) if the best PV flips to the prompt track’s PV, the SV lies close to that PV and $d_{\rm PV}^{\rm signed}$ populates the near-zero (or small positive) region.
These effects are amplified in multi-PV environments, where PV--PV separations along $z$-direction can be several millimetres.

\item[\textbf{Momentum opening angle (\textit{Transformed} $\mathbf{\cos\theta}$)}:]
Cosine of the opening angle between the base and additional-particle momenta. Signal particles are more aligned (large $\cos\theta$); we use $1-(1-\cos\theta)^{0.2}$ to enhance separation where the distributions overlap.

\end{description}

The distributions of the input features for signal and background particles are shown in Fig.~\ref{fig:input_variables}, with the corresponding correlation matrix provided in Appendix~\ref{app:correlation}. An anti-correlation is observed between the \textit{Transformed $\Delta R$} and \textit{Transformed $\cos\theta$} variables for both signal and background categories, along with a moderate correlation between $\log(\chi^2_{\mathrm{IP\,wrt\,SV}})$ and $\log(\chi^2_{\mathrm{DOCA}})$. Apart from these, the features exhibit only weak correlations, suggesting that they offer largely complementary information for the classification task.
The relative importance of each feature is discussed in the following section.

\begin{figure}[!htbp]
    \centering
    \includegraphics[width=0.85\textwidth]{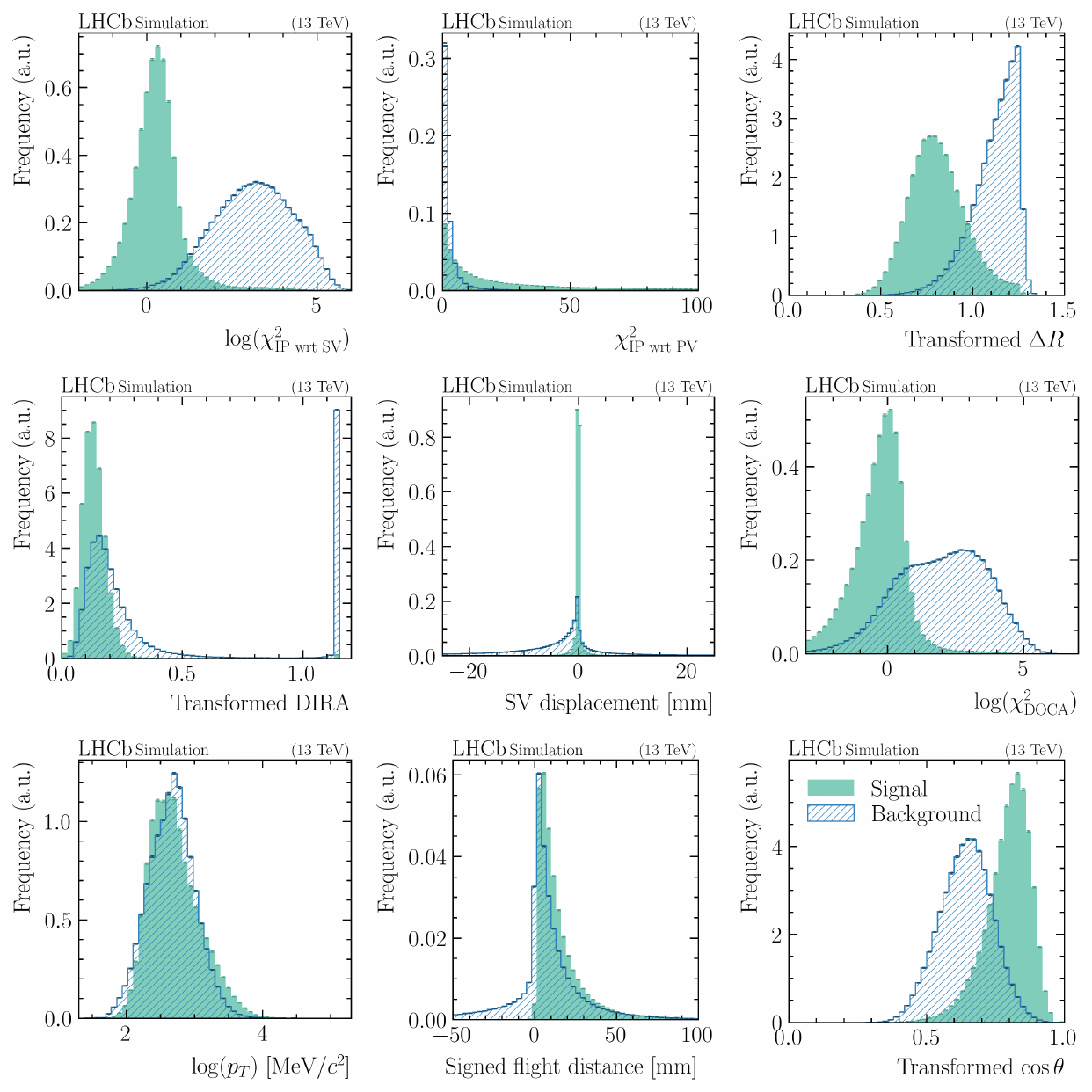}
    \caption{
    Distributions of the input features used to train the \IMI.
    The curves for signal particles (non-isolated particles) are shown as red histograms, while background particles (isolated particles) are represented by green histograms.
    These variables serve as inputs to the multivariate classifier.
    See Sec.~\ref{subsec:training_feature} for a detailed description of each feature.
    }
    \label{fig:input_variables}
\end{figure}

\subsection{Training and performance}
\label{subsec:model_performance}

On average, each simulated decay contains \((3\text{--}4)\times10^{4}\) \emph{non-isolated} signal particles, accompanied by a substantially larger number of \emph{isolated} background particles -- typically 50 to 100 times more numerous than the signal.
To train the \IMI classifier, each sample is split into three statistically independent subsets: 70\% for training, 15\% for validation, and 15\% for evaluation.
During training and validation, the background class is \emph{randomly down-sampled} to match the number of signal particles, enforcing a balanced class ratio that stabilises gradient updates. 
In contrast, the evaluation set retains the full class imbalance (\(\gtrsim 50{:}1\)), thus providing a realistic performance estimate under Run~3-like conditions at \lhcb.

The classifier performance of \IMI is quantified by the area under the receiver-operating-characteristic (ROC) curve, abbreviated AUC.
Hyperparameters, including the number of trees, maximum tree depth, and learning rate, are tuned using the Bayesian optimisation framework \textsc{Optuna}~\cite{Akiba:2019lwq}. 
The objective is to minimise the Kolmogorov-Smirnov (KS) statistic between the classifier outputs on the training and validation sets, thereby suppressing overfitting. The KS statistic obtained from the comparison between the training and validation samples at the optimal hyperparameter point is $k = 8.5\times 10^{-4}$. This very small value indicates that the output-score distributions for the training and validation samples are nearly indistinguishable, implying that the classifier generalises well and that no statistically meaningful overtraining is observed.

\begin{figure}[t]
    \centering
    \includegraphics[width=0.65\columnwidth]{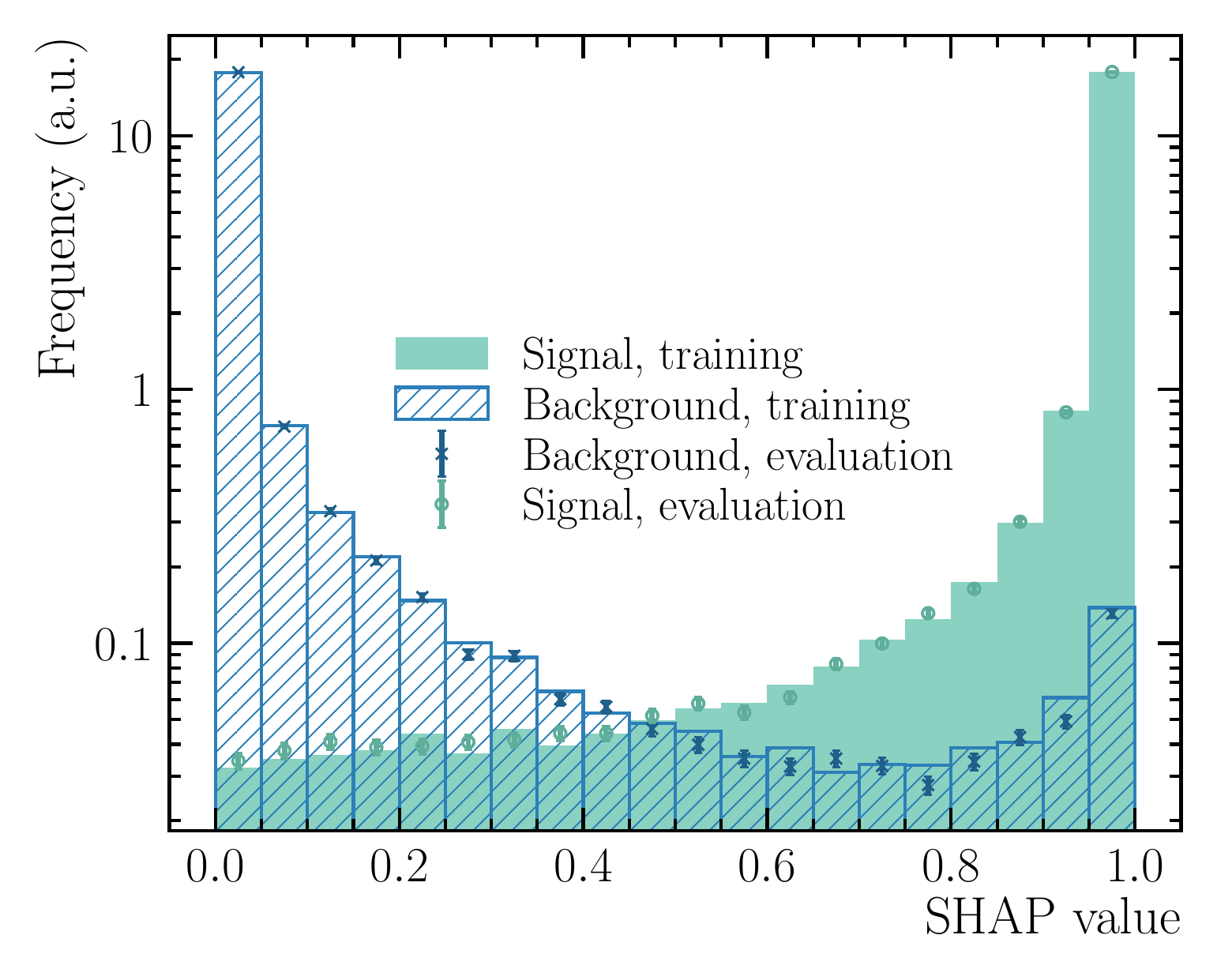}
    \caption{
    Distributions of the \IMI classifier output for non-isolated signal (red) and isolated background (green) particles in the training sample (filled histograms) and evaluation sample (markers). The strong agreement confirms the absence of overtraining and the model's ability to generalise to unseen data.
    }
    \label{fig:xgboost_output}
\end{figure}

The \IMI output for non-isolated (signal) and isolated (background) particles is shown in Fig.~\ref{fig:xgboost_output}. In this figure, the filled histograms represent the training sample, while the markers indicate the evaluation sample. The close agreement between the two confirms that the model generalises well and exhibits no signs of overtraining. 
Quantitatively, the classifier achieves an AUC of
\[
\text{AUC} = 0.9964(3)
\]
on the evaluation set, where the uncertainty reflects the variation across a thousand bootstrap replicas.

\begin{figure}[t]
    \centering 
    \includegraphics[width=0.85\columnwidth]{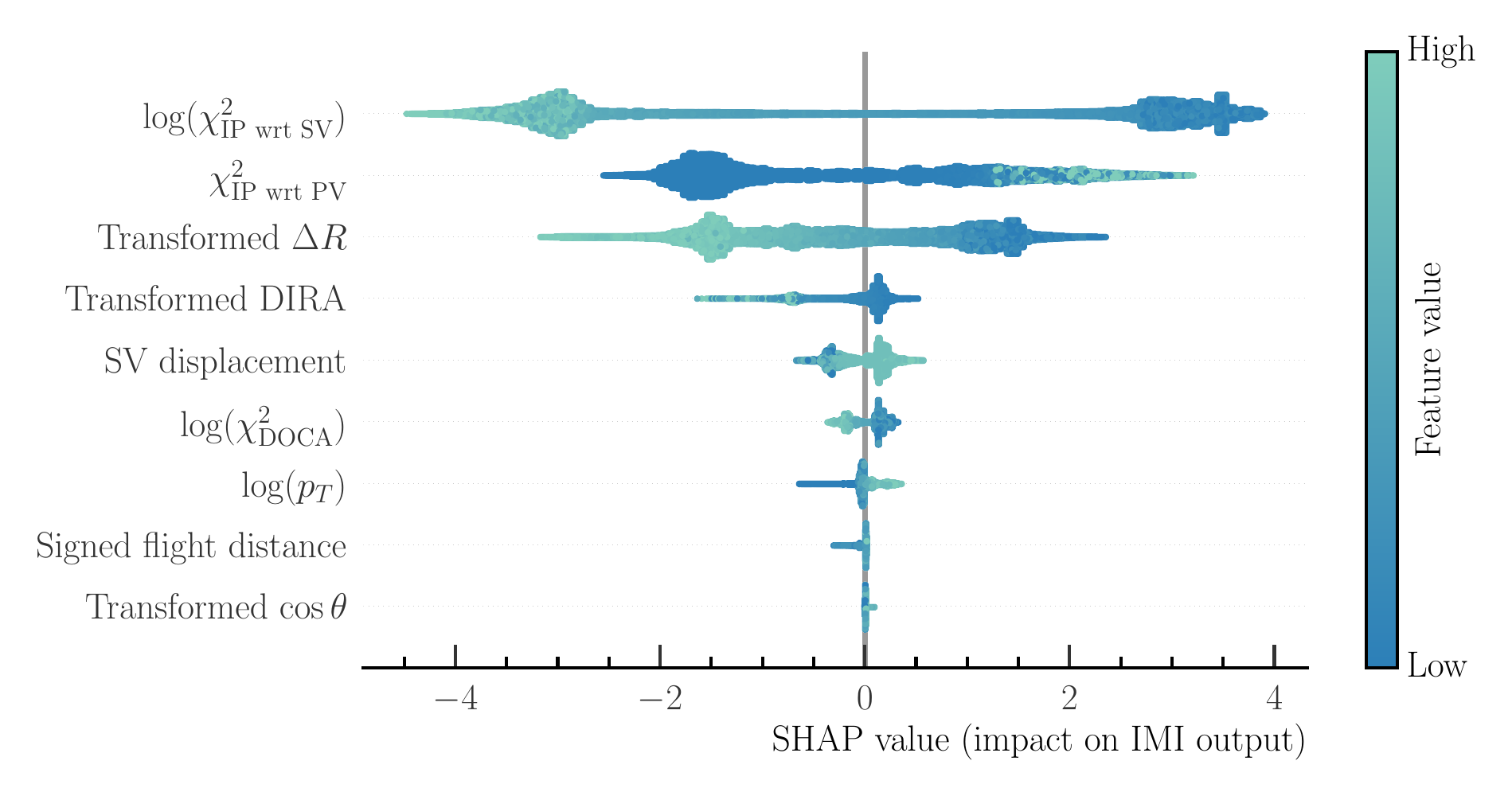}
    \caption{
    SHAP summary plot for the \IMI classifier~\cite{Lundberg:2017}. The y-axis lists input features in order of decreasing importance. The x-axis indicates each feature's SHAP value, that is, its contribution to the classifier's output, while the color encodes the raw feature value (green = low, red = high). Positive SHAP values push the classifier toward signal-like predictions, while negative values indicate background-like behaviour.
    }
    \label{fig:xgboost_feature_ranking}
\end{figure}

To elucidate the classifier's internal logic, we compute SHAP (SHapley Additive exPlanations) values~\cite{Lundberg:2017}, which quantify the contribution of each feature to the model output on a per-particle basis.  
The summary plot in Fig.~\ref{fig:xgboost_feature_ranking} orders the inputs by their mean impact and visualises their effects across the evaluation sample.  
Three variables clearly dominate the decision boundary: \(\log(\chi^{2}_{\mathrm{IP\,wrt\,SV}})\), \(\chi^{2}_{\mathrm{IP\,wrt\,PV}}\), and \(\Delta R\).
Particles with \emph{small} values of \(\log(\chi^{2}_{\mathrm{IP\,wrt\,SV}})\) or \(\Delta R\) (green points) push the SHAP value toward positive numbers, yielding a signal-like prediction, whereas \emph{large} values (red points) shift the output toward background-like.  
Conversely, a large \(\chi^{2}_{\mathrm{IP\,wrt\,PV}}\) indicates significant displacement from the primary vertex and therefore increases the signal score, while small values suppress it, complementary behaviour to the other two variables.  
Secondary inputs, including the \textit{Transformed DIRA}, \textit{SV displacement}, and \(\log(\chi^{2}_{\mathrm{DOCA}})\), introduce fine-grained topological information that sharpens the separation between classes. 
Kinematic observables such as \(\log p_{T}\) play a supportive, albeit less critical, role.  
The narrow SHAP ranges observed for \textit{Signed flight distance} and the \textit{Transformed \(\cos\theta\)} demonstrate that the model does not over-rely on weakly informative features.  
Overall, the SHAP analysis confirms that the classifier's decisions are governed by physically meaningful observables and remain fully aligned with the underlying isolation logic.

The benchmarking of the \IMI tool against classical isolation methods is provided in Sec.~\ref{subsubsec:comparison_classical}, while its performance as a function of event multiplicity is discussed in Sec.~\ref{subsubsec:Performance_PV}. 
The dependence of signal efficiency on key kinematic variables is examined in Sec.~\ref{subsubsec:efficiency_q2}.

\subsubsection{Comparison with Classical Isolation}
\label{subsubsec:comparison_classical}

The performance of the \IMI tool is benchmarked against classical isolation techniques described in Sec.~\ref{sec:previous_isolation},  namely, the \emph{track isolation}, \emph{cone isolation} and \emph{vertex isolation} methods. 
For the comparisons presented here and throughout the paper, we use standard working requirements for each method (see Sec.~\ref{sec:previous_isolation} for variable definitions):
\begin{align}
  \text{Track isolation:} \quad & \chi^{2}_{\mathrm{IP\,wrt\,PV}} > x \quad \text{or} \quad \texttt{samePV} = \text{True}, \\
  \text{Cone isolation:} \quad & \Delta R < y, \\
  \text{Vertex isolation:} \quad & \chi^{2}_{\mathrm{IP\,wrt\,SV}} < z, \\
  \text{IMI:} \quad & \text{IMI score} > a.
\end{align}
These requirements for the classical methods are chosen because they are widely deployed online and provide a robust balance between signal efficiency, background rejection, and throughput cost. While channel-specific offline analyses may combine multiple requirements for additional gains, our goal here is a like-for-like comparison using a baseline representative of general-purpose isolation used at the trigger/Spruce level.

\begin{figure}[t] 
\centering 
\includegraphics[width=0.45\columnwidth]{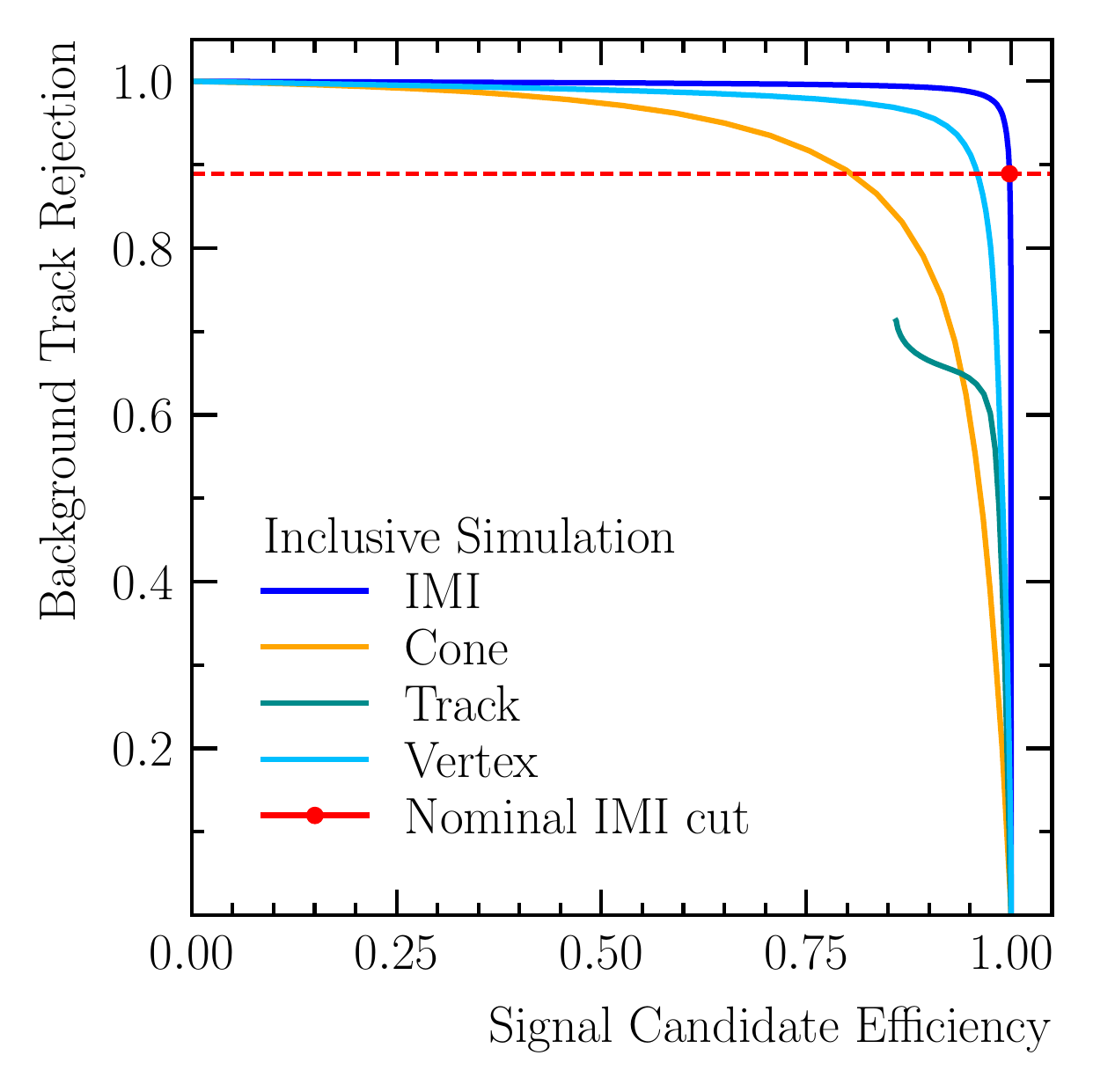}\\
\includegraphics[width=0.325\columnwidth]{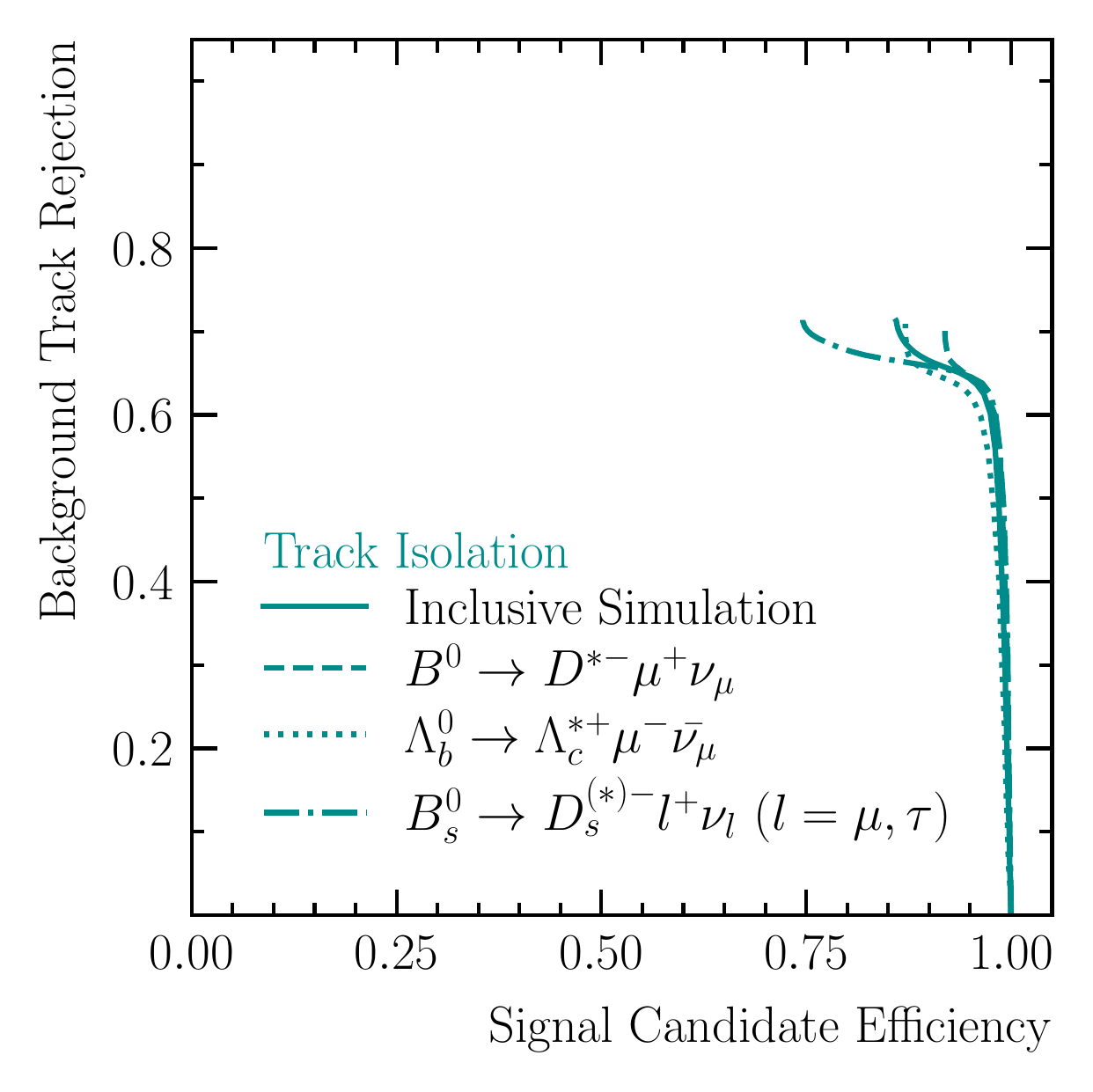}
\includegraphics[width=0.325\columnwidth]{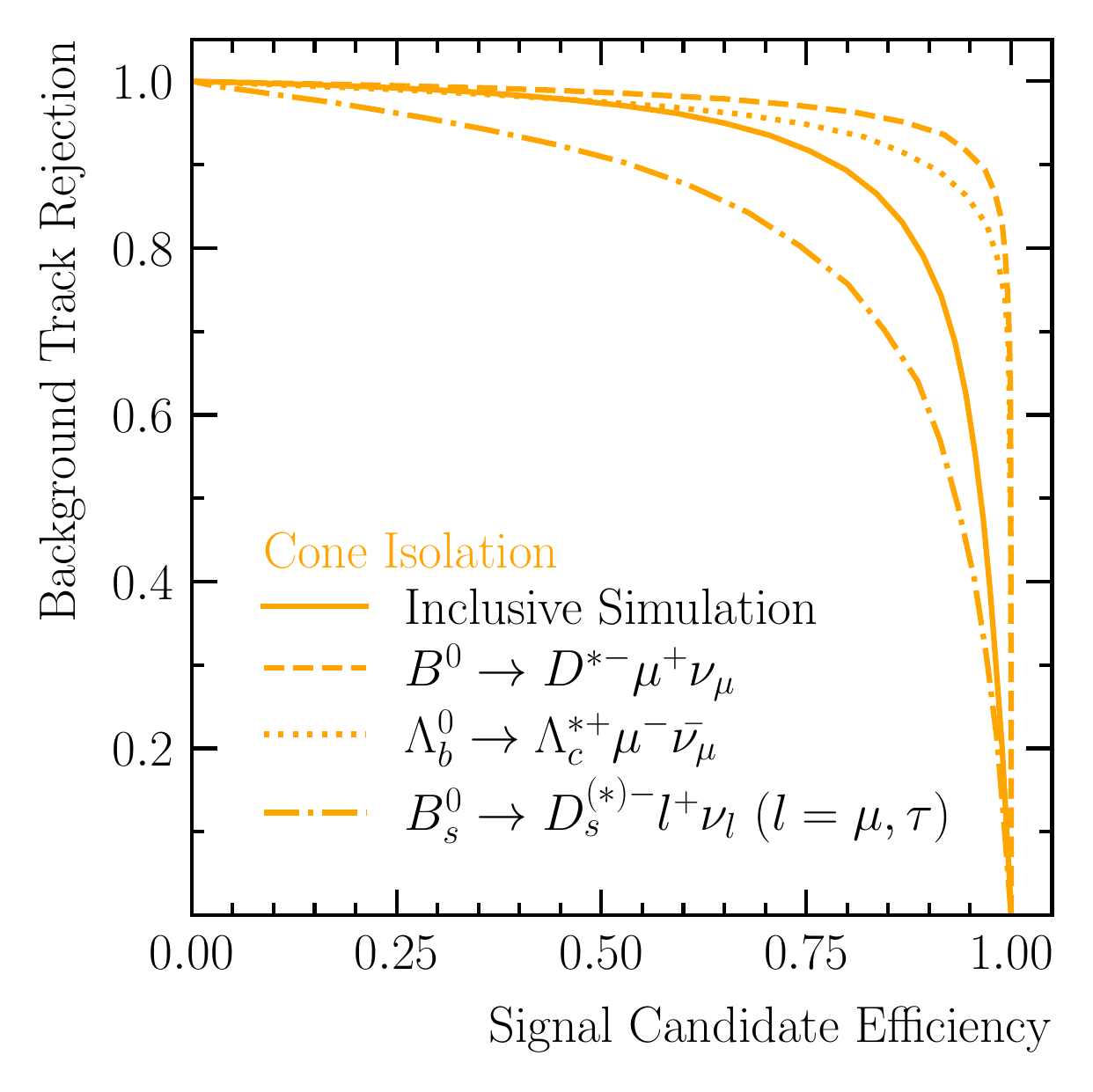}\\
\includegraphics[width=0.325\columnwidth]{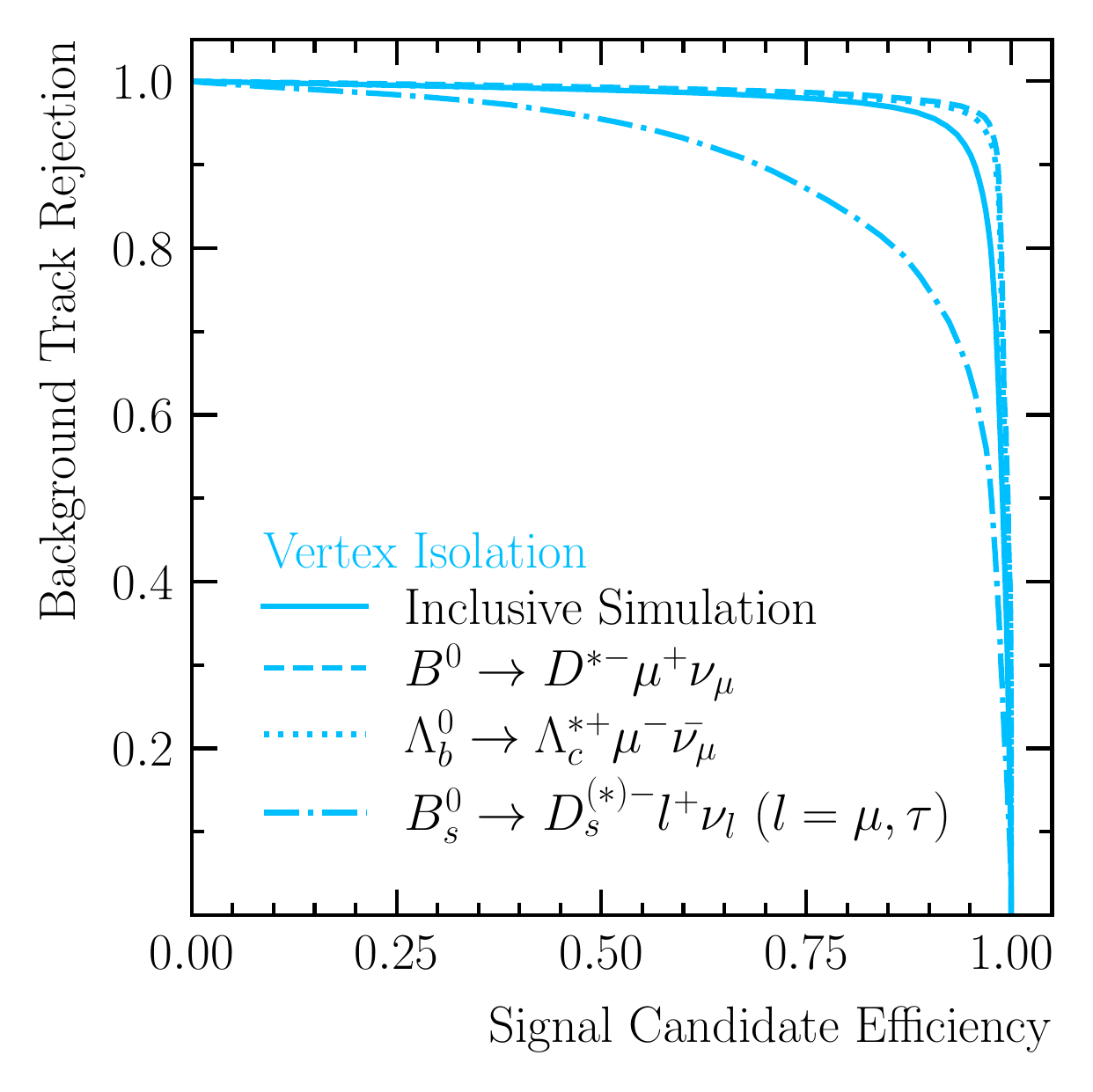}
\includegraphics[width=0.325\columnwidth]{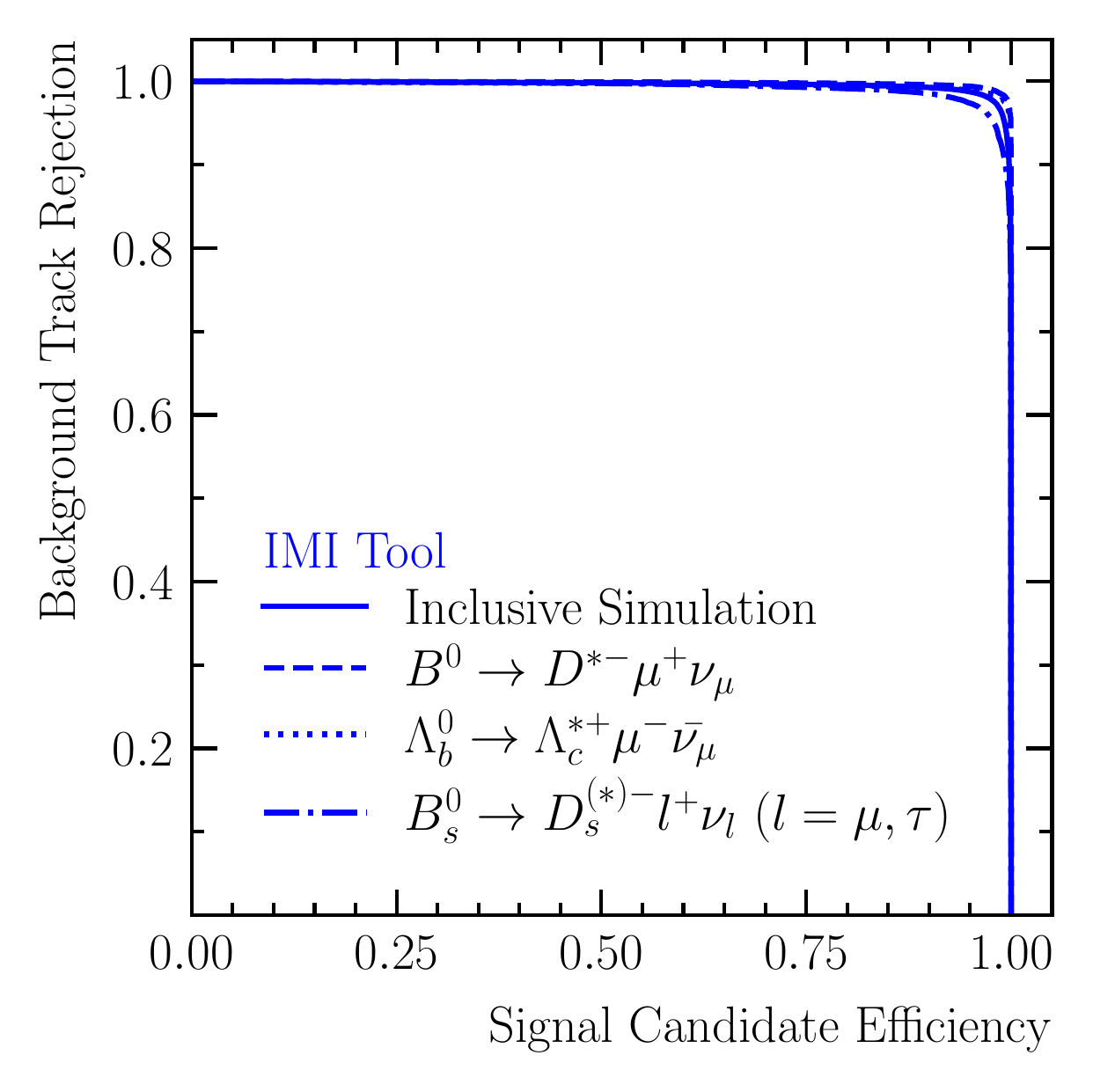}
\caption{
  Signal \emph{candidate} efficiency versus background \emph{track} rejection for the Track (teal), Cone (orange), Vertex (light blue), and \IMI\ (dark blue) isolation methods.
  The top panel shows the performance on the inclusive MC sample; the red marker indicates the chosen \IMI\ working point.
  The four lower panels (clockwise from top-left) show Track, Cone, \IMI, and Vertex isolation, each evaluated on the inclusive sample (solid) and on three exclusive benchmark channels with increasing numbers of non-isolated signal particles (see Table~\ref{tab:samples_summary}):
  $\Bz \to D^{*-}\mu^{+}\nu_{\mu}$ (1 non-isolated particle),
  $\Lambda_b\!\to\!\Lc^*\mu^{-}\bar{\nu}_{\mu}$ (2 non-isolated particles),
  and
  $\Bs\!\to\!D_s^{(*)-}\,\ell^{+}\nu_{\ell}$ (2 to 5 non-isolated particles), where $\ell \in \{\mupm,\taupm\}$.
}
\label{fig:sigEff_vs_tracRejection} 
\end{figure}

Figure~\ref{fig:sigEff_vs_tracRejection} shows the ROC curves obtained from the inclusive simulated samples listed in Table~\ref{tab:samples_summary}. 
In these plots we report the signal \emph{candidate} efficiency, defined as the probability that \emph{all} particles forming the reconstructed signal candidate are retained after applying the isolation requirement. This differs from the signal \emph{track} efficiency, which counts the retention of individual signal particles. Candidate-level efficiency is therefore lower, but is the more relevant metric for analyses that reconstruct full decay chains.
By contrast, the background rejection is defined with respect to background \emph{tracks}, as these particles do not form part of any reconstructed candidate. 
The results demonstrate that \IMI\ provides the best overall performance, reaching about $99\,\%$ background rejection at $95\,\%$ signal \emph{candidate} efficiency.
Beyond this point, the curve drops steeply, indicating a sharp and well-defined operational threshold. 
The \emph{cone isolation} method, which accepts extra particles within a maximum angular separation $\Delta R < x$ from the base particle, can approach similar maximum background rejection, but only in a much narrower efficiency range ($\varepsilon_{\text{sig}} \approx 0$--$50\,\%$). 
This reflects a key limitation of relying solely on geometric separation: particles from the dense underlying event can mimic the topology of genuine signal tracks, reducing discrimination power at higher efficiencies. 
The \emph{track isolation} method, based on requiring either the \texttt{samePV} association with the base particle or $\chi^{2}_{\mathrm{IP\,wrt\,PV}} > y$, is constrained by the binary nature of the \texttt{samePV} selection. 
Once most background particles are assigned to the correct primary vertex, its rejection power saturates at about $70\,\%$, and performance degrades quickly for $\varepsilon_{\text{sig}} \gtrsim 95\,\%$. 
The \emph{vertex isolation} method, based on requiring compatibility with the reconstructed secondary vertex (e.g.\ $\chi^{2}_{\mathrm{IP\,wrt\,SV}} < z$), achieves substantially stronger rejection at high signal efficiencies than the PV-based track isolation, but remains limited by the resolution and stability of the reconstructed secondary vertex and therefore turns over earlier than \IMI. 
Overall, the results highlight that \IMI\ leverages a richer set of features beyond pure geometry or simple PV/SV association, enabling it to maintain both high background rejection and high signal efficiency over a wide operating range.
The signal efficiency and background rejection power for all four methods as function of the threshold values is shown in the Appendix~\ref{app:isolation_score}.

The four lower panels (clockwise from top-left) of Fig.~\ref{fig:sigEff_vs_tracRejection} explore the performance of the track, cone, vertex and \IMI isolation methods when applied to three exclusive decay channels with varying kinematics and numbers of non-isolated signal particles (see Table~\ref{tab:samples_summary}):
\begin{itemize}
  \item $\Bz \to D^{*-}\mu^{+}\nu_{\mu}$, featuring one non-isolated signal particle with soft kinematics;
  \item $\Lambda_b\!\to\!\Lc^*\mu^{-}\bar{\nu}_{\mu}$, featuring two relatively hard non-isolated signal particles;
  \item $\Bs\!\to\!D_s^{(*)-}\,\ell^{+}\nu_{\ell}$, with two to five non-isolated particles, including children of a long-lived $D_s^-$.
\end{itemize}
Across all three benchmark channels, \IMI maintains a robust and consistent performance, demonstrating that it is largely agnostic to the number of non-isolated signal particles and their kinematics in these modes. 
In contrast, the \emph{track isolation} method shows similarly modest performance for each decay, underscoring that its discriminating power is governed almost entirely by the $\chi^{2}_{\mathrm{IP\,wrt\,PV}}$ threshold and saturates once most background tracks are assigned to their correct PV; minor differences in signal efficiency arise from correlations between particle kinematics and impact-parameter significance.
\emph{Cone isolation} exhibits the strongest dependence on the decay topology: for $\Bz \to D^{*-}\mu^{+}\nu_{\mu}$, where the relevant signal activity is relatively collimated, the method performs comparatively well, whereas for topologies with multiple and/or softer non-isolated particles (notably the $D_s^{(*)}$ mode) the performance degrades markedly. This reflects the need for a larger cone to retain all signal-side particles, which unavoidably admits more background.
The \emph{vertex isolation} method, based on requiring compatibility with the reconstructed secondary vertex, provides substantially stronger rejection than PV-based track isolation and shows only mild channel dependence for the $\Bz$ and $\Lambda_b$ benchmarks. A more pronounced degradation is observed for $\Bs\!\to\!D_s^{(*)-}\ell^{+}\nu_{\ell}$, consistent with the presence of additional displaced decay structure from the long-lived $D_s^-$: maintaining high candidate efficiency in this case requires a looser SV-compatibility requirement, which reduces background rejection.

The \IMI\ working point is intentionally set to a conservative response threshold of $0.05$. 
At this setting, the number of selected signal particles per event is reduced from the $\mathcal{O}(200)$ charged particles typically reconstructed to about $\mathcal{O}(10)$, while retaining a signal efficiency of roughly $99\,\%$. 
Although exclusive trainings generally outperform the inclusive ones, as they are optimised for a specific signal topology and kinematics, the low threshold ensures that a broad set of signal-like particles is retained. 
This allows analysts to perform more specialised trainings at the analysis stage while still benefiting from strong background rejection.

\subsubsection{Performance as a function of event multiplicity}
\label{subsubsec:Performance_PV}

We evaluate the robustness of different isolation methods in increasingly busy events by studying their performance as a function of event multiplicity, defined as the number of reconstructed charged particles in the event. 
This variable serves as a proxy for overall event activity, which is particularly relevant in hadronic collisions where occupancy can vary substantially. 
Each method is benchmarked at a fixed signal efficiency of approximately $99\,\%$, and its background rejection power is assessed across the multiplicity spectrum.

\begin{figure}[t]
  \centering
  \includegraphics[width=0.5\columnwidth]{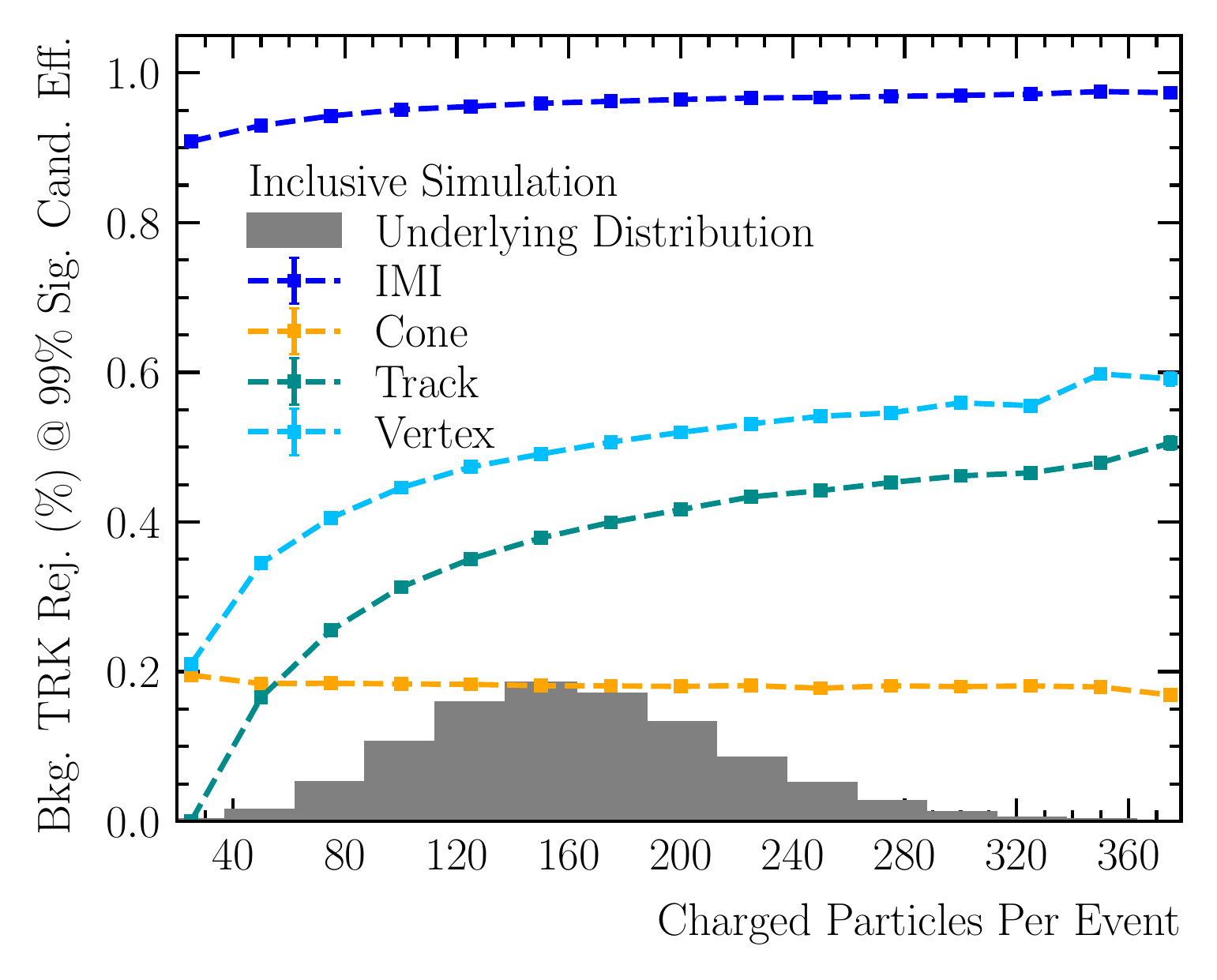}\\
  \includegraphics[width=0.325\columnwidth]{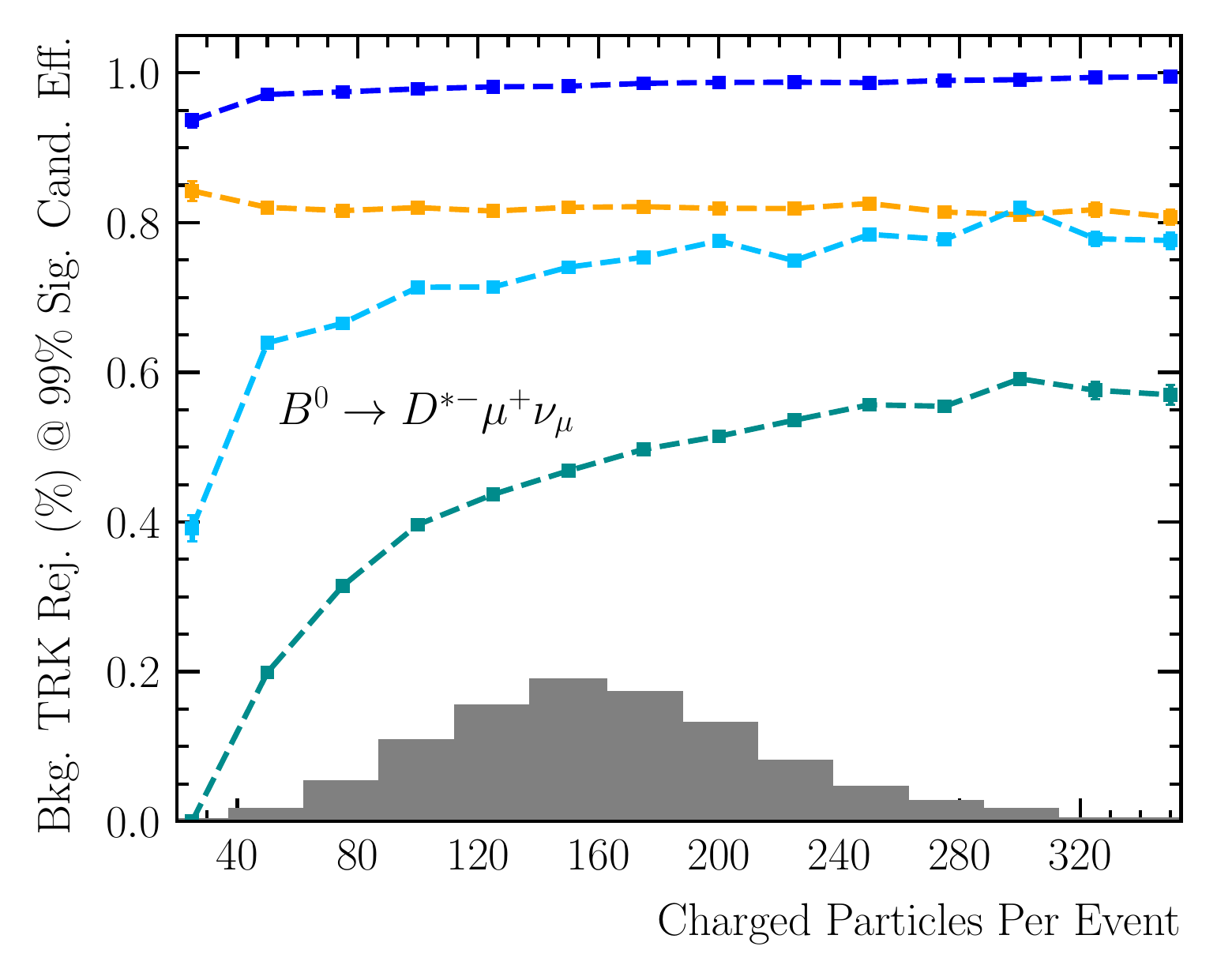}
  \includegraphics[width=0.325\columnwidth]{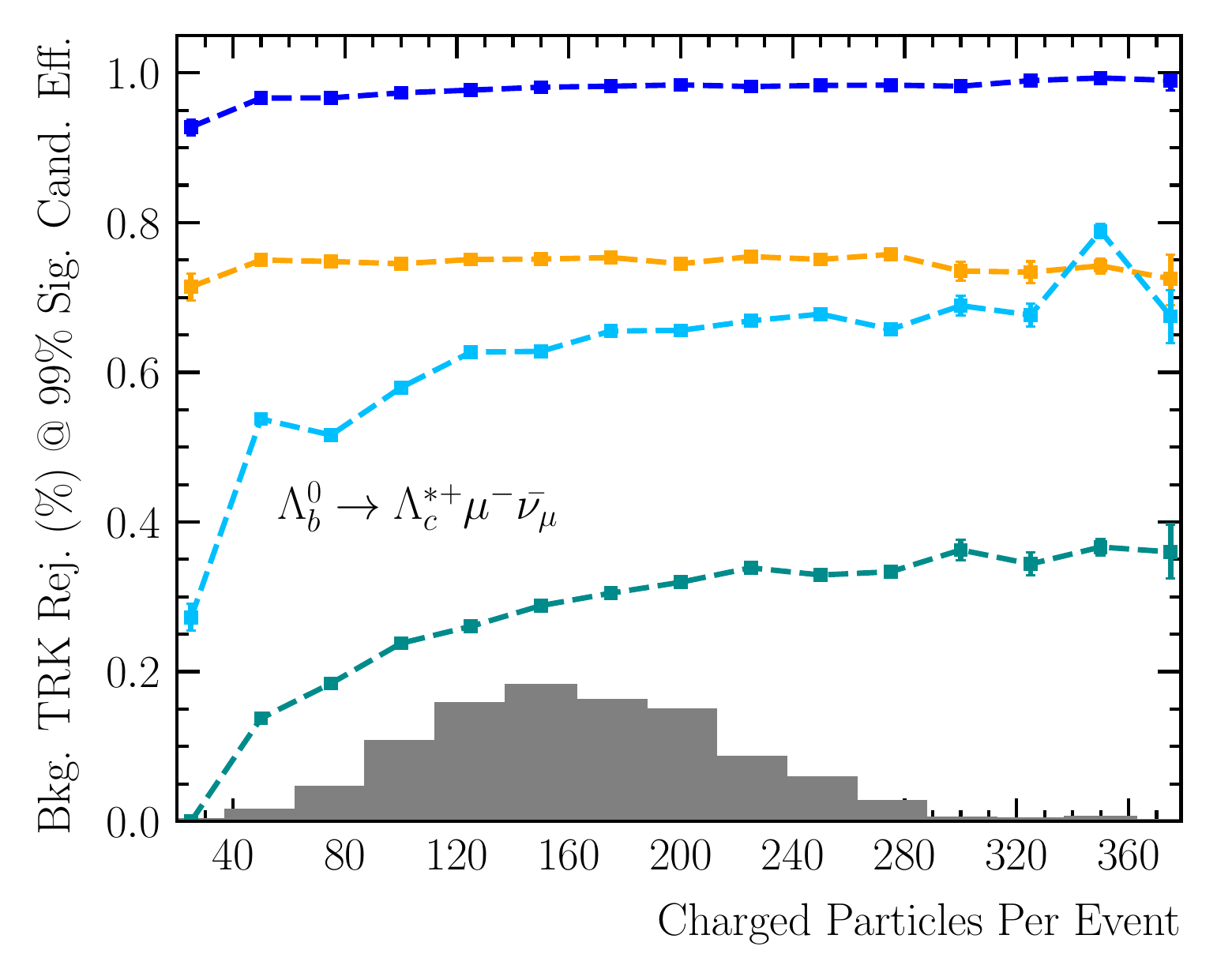}
  \includegraphics[width=0.325\columnwidth]{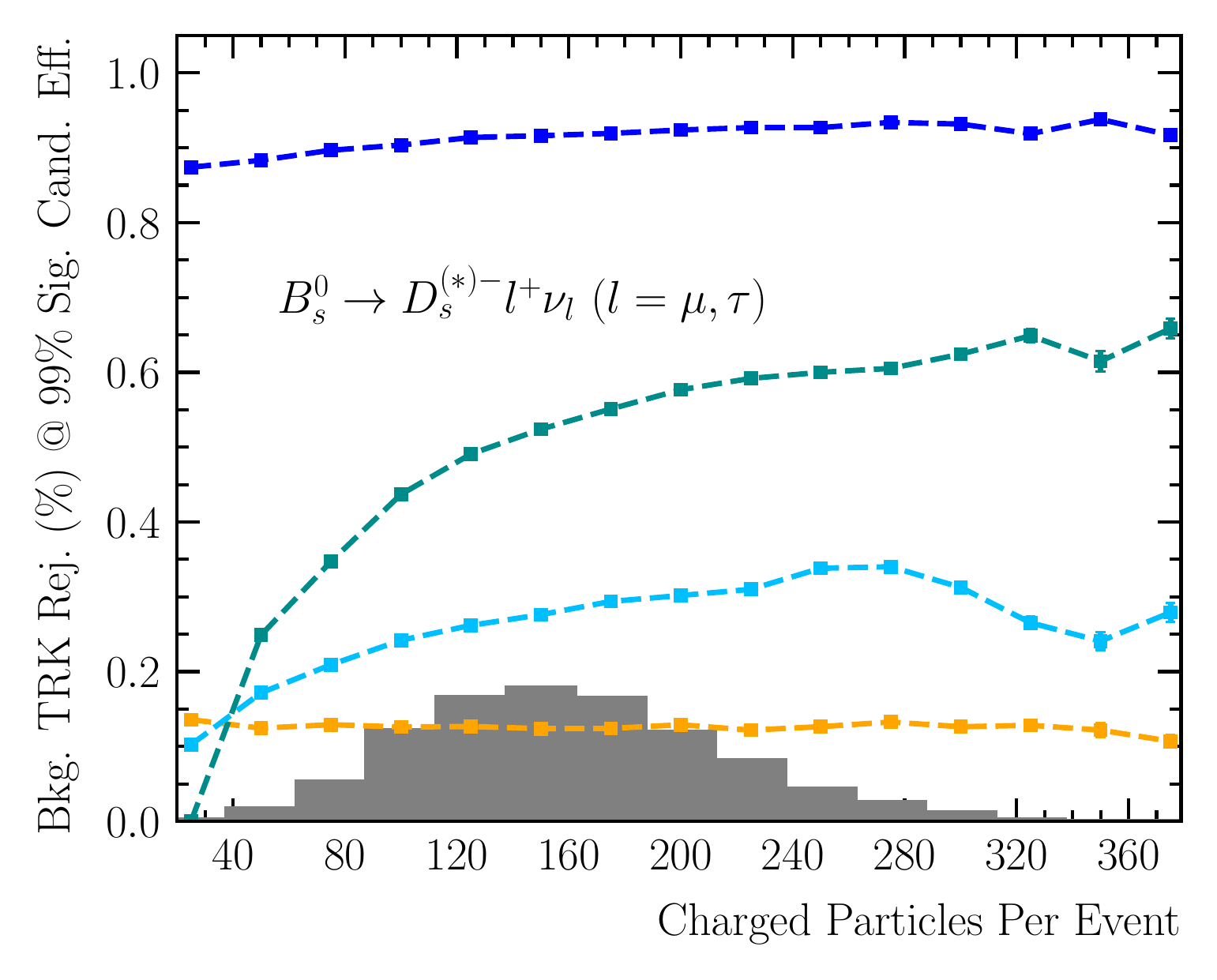}
  \caption{
  Background rejection at a fixed signal efficiency of approximately $99\,\%$ for the Track (teal), Cone (orange), Vertex (light blue), and \IMI\ (dark blue) isolation methods as a function of the number of reconstructed charged particles in the event (event multiplicity).
  The top panel shows performance on the inclusive simulated sample (see Table~\ref{tab:samples_summary}).
  The bottom panels show the performance for three exclusive decay channels with varying kinematics and numbers of non-isolated signal particles:
  $\Bz \to D^{*-}\mu^{+}\nu_{\mu}$ (1 non-isolated particle, left),
  $\Lambda_b\!\to\!\Lc^*\mu^{-}\bar{\nu}_{\mu}$ (2 non-isolated particles, middle), and
  $\Bs\!\to\!D_s^{(*)-}\,\ell^{+}\nu_{\ell}$ (2 to 5 non-isolated particles, right), where $\ell \in \{\mupm,\taupm\}$.
  The \IMI\ tool consistently outperforms the classical methods across all event multiplicities and decay channels.
  }
  \label{fig:multiplicity_performance}
\end{figure}

Figure~\ref{fig:multiplicity_performance} compares four isolation strategies, \emph{track isolation}, \emph{cone isolation}, \emph{vertex isolation}, and \IMI, as a function of event multiplicity, using the inclusive simulation sample described in Table~\ref{tab:samples_summary}.
Across the full range of event multiplicities, the \IMI\ algorithm demonstrates a clear advantage, consistently rejecting around $95\,\%$ of background while retaining $99\,\%$ signal efficiency.
The \emph{track isolation} method shows steadily improving performance with increasing multiplicity, reaching up to $50\,\%$ background rejection in high-occupancy events. This improvement arises because the isolation requirement is particularly effective at rejecting tracks originating from primary vertices other than the one that produced the signal candidate.
The \emph{vertex isolation} method also improves with event multiplicity and provides systematically stronger rejection than PV-based track isolation, reaching about $60\,\%$ in the highest-multiplicity events. 
This reflects the additional constraint of compatibility with the reconstructed secondary vertex: in busy events a larger fraction of unrelated tracks is inconsistent with the candidate SV and is therefore rejected, while the SV resolution remains sufficiently stable to maintain the fixed $99\,\%$ signal efficiency working point.
In contrast, the \emph{cone isolation} method yields relatively flat performance, plateauing at around $20\,\%$ background rejection. 
Its effectiveness is significantly lower than the other methods, especially in high-multiplicity environments where the isolation cone is more likely to contain unrelated particles.

The lower panels of Fig.~\ref{fig:multiplicity_performance} present the same comparison for three exclusive benchmark channels, introduced earlier, that differ in kinematics and in the number of non-isolated particles: 
$\Bz\!\to\!D^{*-}\mu^{+}\nu_{\mu}$ (one soft), 
$\Lambda_b\!\to\!\Lc^*\mu^{-}\bar{\nu}_{\mu}$ (two relatively hard), 
and $\Bs\!\to\!D_s^{(*)-}\,\ell^{+}\nu_{\ell}$ (two to five, including daughters of a long-lived $D_s^-$). 
Across all three channels, \IMI\ achieves the best performance, delivering $90$--$95\,\%$ background rejection while maintaining $99\,\%$ signal efficiency. 
The \emph{track isolation} method performs similarly for the $\Bz$ and $\Bs$ decays, but slightly worse for the $\Lambda_b$ decay. 
This difference arises because the relatively soft non-isolated particles in $\Bz$ and $\Bs$ decays tend to have larger impact parameters with respect to the PV than the harder particles from $\Lambda_b$ decays, settling on a minimal $\chi^{2}_{\mathrm{IP\,wrt\,PV}}$ requirement that is more effective at rejecting background in the former cases. 
The \emph{vertex isolation} method provides strong rejection for the $\Bz$ and $\Lambda_b$ benchmarks and is comparatively stable across multiplicity, but degrades significantly for the $\Bs$ mode. 
This behaviour is consistent with the presence of additional displaced decay structure from the long-lived $D_s^-$: to retain $99\,\%$ candidate efficiency, the SV-compatibility requirement must be loosened to accommodate signal tracks originating from the downstream $D_s^-$ decay, which in turn admits more background and reduces the achievable rejection power.
In contrast, the \emph{cone isolation} method gives comparable results for the $\Bz$ and $\Lambda_b$ decays, but performs significantly worse for the $\Bs$ decay. 
For $\Bs$ decays, reconstructing signal particles from long-lived $D_s^-$ mesons requires a larger isolation cone, which inevitably captures more background and reduces rejection power.

\subsubsection{Signal efficiency as a function of kinematic variables}
\label{subsubsec:efficiency_q2}

We also assess whether the signal efficiency of each isolation method varies as a function of key kinematic variables, particularly the squared four-momentum transfer, $q^2 = (p_B - p_{\text{had}})^2$, which is equivalently the invariant mass squared of the lepton--neutrino system (e.g.\ $q^2 = m^2_{\ell\nu_\ell}$ for $\Bz \to D^{*-} \ell^{+} \nu_{\ell}$ decays).
Ideally, a well-designed selection should yield a flat efficiency across the entire $q^2$ spectrum. 
This is especially important at high $q^2$, where theoretical predictions from lattice QCD are most precise, enabling precision measurements of CKM matrix elements.

\begin{figure}[!htbp]
\centering 
\includegraphics[width=0.8\columnwidth]{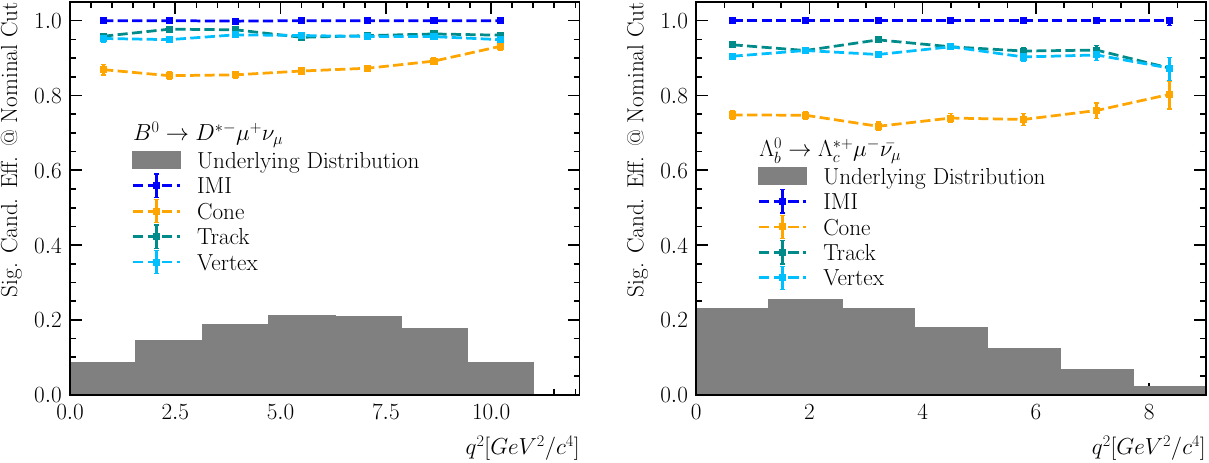}
\caption{
  Signal efficiency as a function of $q^2$ for different isolation methods: \IMI with output $> 0.05$ (blue),
  \emph{track isolation} with \texttt{samePV} and $\chi^{2}_{\mathrm{IP\,wrt\,PV}} > 16$ (teal),
  \emph{cone isolation} with a fixed cone size of $\Delta R = 0.5$ (orange),
  and \emph{vertex isolation} with $\chi^{2}_{\mathrm{IP\,wrt\,SV}} < 1$ (light blue).
  Shown are two decay channels: $\Bp\!\to\!D^{*-}\mu^{+}\nu_{\mu}$ (left), with one non-isolated signal particle, and $\Lambda_b\!\to\!\Lc^*\mu^{-}\bar{\nu}_{\mu}$ (right), with two. Here $\Lambda_c^*$ denotes the $\Lambda_c(2625)$ state.
}
\label{fig:effvariation_q2}
\end{figure}

To study this, we evaluate the signal efficiency as a function of $q^2$ for each isolation method. 
For the \IMI tool, we apply the nominal working point defined earlier, corresponding to a \IMI output threshold of \IMI $> 0.05$ on each signal particle. 
For \emph{track isolation}, we use the \texttt{samePV} flag or $\chi^{2}_{\mathrm{IP\,wrt\,PV}} > 16$.
For \emph{cone isolation}, we use a fixed cone size of $\Delta R = 0.5$ which is typically used to capture the $b$-jet structure.
For \emph{vertex isolation}, we use a common secondary-vertex compatibility requirement of $\chi^{2}_{\mathrm{IP\,wrt\,SV}} < 1$.
Figure~\ref{fig:effvariation_q2} shows the resulting signal efficiencies for two representative decay channels: 
$\Bp\!\to\!D^{*-}\mu^{+}\nu_{\mu}$ (left), with one non-isolated signal particle, 
and $\Lambda_b\!\to\!\Lc^*\mu^{-}\bar{\nu}_{\mu}$ (right), with two non-isolated particles, where $\Lambda_c^*$ denotes the $\Lambda_c(2625)$ state. 
Cone, track, and vertex isolation show modest $q^2$-dependent variations -- reflecting their intrinsic correlations with the signal kinematics -- whereas the \IMI\ tool retains a consistently high and nearly flat efficiency across the full $q^2$ range.

\FloatBarrier

\section{Implementation and data-size reduction}
\label{sec:implementation}

This section describes 
the implementation of both the classical isolation algorithms and the new IMI algorithm
, and their integration into the \lhcb software framework in sub-section~\ref{sec:integration}.
The data-size reduction achieved by the \IMI and its impact on \textsc{Sprucing} throughput are discussed in sub-section~\ref{sec:size_reduction}.

\subsection{Integration into the \lhcb selection framework}
\label{sec:integration}

\begin{figure}[htb!]
  \centering
  \includegraphics[width=0.8\textwidth]{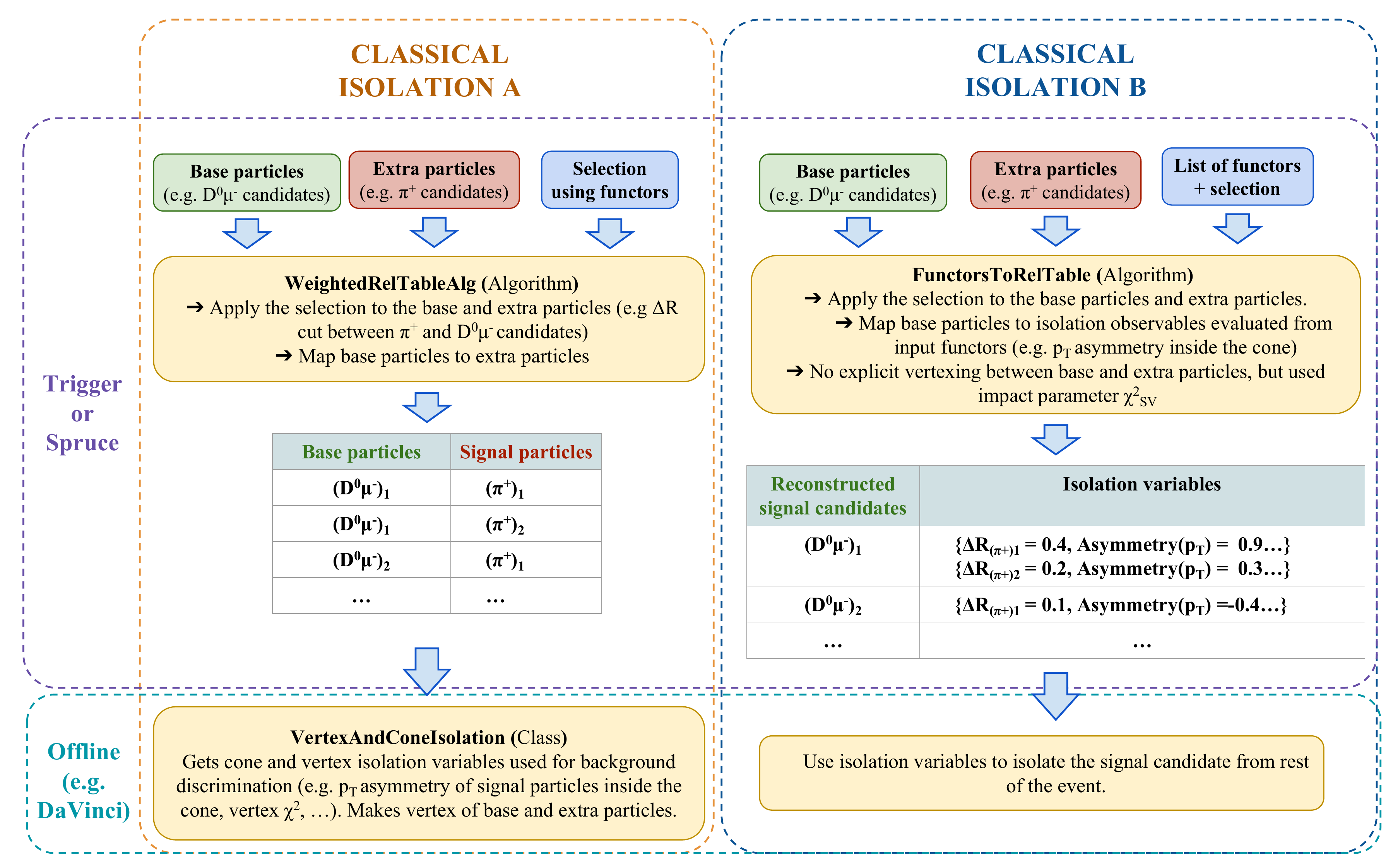}
  \caption{Comparison of the two classical isolation strategies implemented in the \lhcb selection framework.
  Although type ``B'' is designed for the trigger, it can also be run offline, ensuring consistent isolation variable computation across channels using different strategies.
  }
  \label{fig::classical_isolation_sketch}
\end{figure}

The computation of classical isolation variables, introduced in
Section~\ref{sec:previous_isolation}, is implemented in the LHCb software
through two complementary workflows, illustrated in
Figure~\ref{fig::classical_isolation_sketch}.

\paragraph{Classical isolation A:}
In this implementation, combinations are formed between the reconstructed base particles and all other charged particles in the event that could contribute to the isolation assessment.  
The selected additional particles are persisted for offline use, where cone- and vertex-based observables are computed by explicitly fitting a vertex between the base and each extra particle.  
These observables can then be stored~\cite{Mathad:2023zky} and used directly in offline analyses.  
The corresponding algorithms are implemented within the \textsc{Rec} reconstruction framework~\cite{Rec}.

\paragraph{Classical isolation B:}
The second implementation follows the same principle as the first, but computes the isolation variables directly at the trigger level.  
Instead of persisting the full information of the additional particles, only a minimal set of observables is stored, which significantly reduces the average event size.  
To optimise throughput in the case of vertex isolation, no dedicated vertex fit is performed; rather, the observables are derived from the impact-parameter significance of the extra particles with respect to the vertex formed by the base particles.  
This approach is well motivated, since the impact-parameter $\chi^2$ with respect to the decay vertex and the change in the vertex-fit $\chi^2$ when adding the extra track are found to be equivalent. 
The resulting quantities are written to the event record and can be used directly in offline analyses.  
The corresponding algorithm is implemented within the \textsc{Rec} reconstruction framework~\cite{Rec}.

\begin{figure}[!htb]
  \centering
  \includegraphics[width=0.75\textwidth]{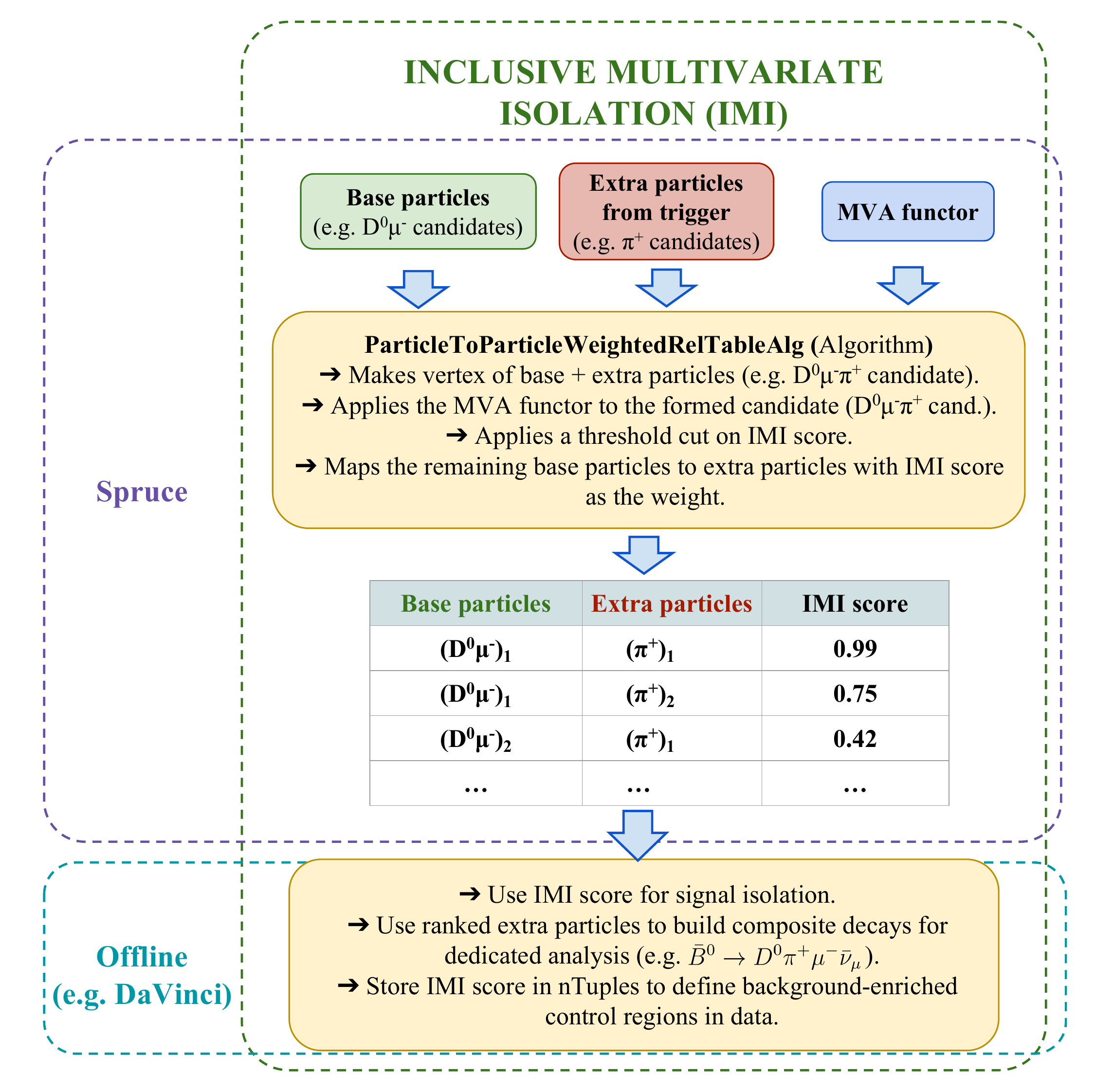}
  \caption{Isolation workflow based on the Inclusive Multivariate Isolation (IMI) approach.}
  \label{fig::mva_isolation_sketch}
\end{figure}

\paragraph{Inclusive multivariate isolation:}
The \IMI strategy follows a conservative data-storage model.  
At the trigger level, all reconstructed particles in the event are written to tape, and the actual isolation decision is deferred to the \textsc{Sprucing} stage, where the fully reconstructed event is available.  
At this point, only those additional particles identified as signal-like are retained on disk, leading to a significant reduction in event size.  
As illustrated in Fig.~\ref{fig::mva_isolation_sketch}, each signal candidate is combined with every other particle in the event, and a vertex fit is attempted.  
To avoid an excessive number of combinations, loose fiducial cuts are applied to the additional particles, such as $\log(\chi^2_{\mathrm{IP\,w.r.t.\,SV}}) < 5$, a minimum 
signed flight distance greater than $-3\,\mathrm{mm}$, and a minimum SV displacement greater than $-5\,\mathrm{mm}$.  
A multivariate classifier then evaluates each base-extra particle pair and assigns an IMI score.  
The resulting relation table, linking base particles to additional particles with their IMI scores, provides a flexible input for offline analyses.  
It can be used to suppress backgrounds, to rank extra particles when reconstructing complex decay chains, or to define background-enriched control regions.
Again all the related algorithms form part of the \textsc{Rec} project~\cite{Rec}.  

\subsection{Data-size reduction and throughput}
\label{sec:size_reduction}

Before integrating the \IMI algorithm into the \lhcb\ selection framework, it was essential to demonstrate that it could deliver a substantial reduction in the volume of data written to disk, \emph{without} compromising on the processing throughput.
To evaluate this trade-off, we processed a large sample of minimum-bias simulation events generated under Run~3 conditions, with all semileptonic \textsc{Sprucing} selection lines enabled. 
These lines typically reconstruct on average two $b$-hadron candidates and write minimal information to disk. 

\begin{figure}[!htp]
    \centering
    \includegraphics[width=0.9\linewidth]{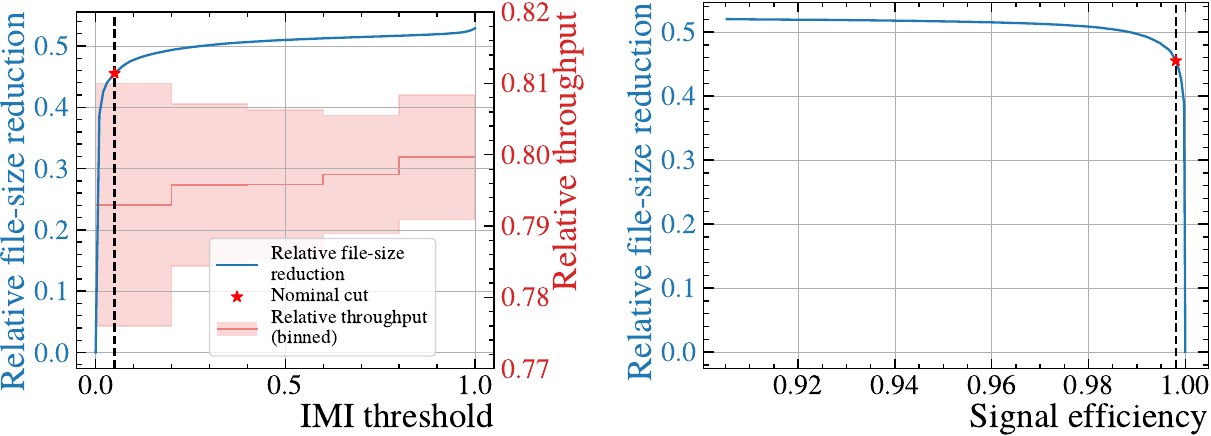}
    \caption{%
      (Left) Relative reduction in output file size (blue) and relative inference throughput (red) as a function of the \IMI\ threshold. The throughput is evaluated in steps of $0.01$ and, for display, the median is shown in bins of width $0.2$; the shaded band indicates the standard deviation. 
      Throughput is measured across all production \lhcb\ sprucing lines relative to a baseline in which isolation is ignored and all extra particles are persisted. 
      As expected, the throughput is only weakly dependent on the \IMI\ threshold, since the candidate combinations (and associated vertexing) are constructed independently of this cut and the threshold primarily controls how much content is written out. 
      (Right) File-size reduction versus signal-candidate efficiency for the inclusive simulation sample. The nominal working point is indicated by the vertical dashed line and the red marker at $\mathrm{IMI}=0.05$.
    }
    \label{fig:eff_throughput_cut}
\end{figure}

For each combination of a base particle and an extra particle in the event, the maximum \IMI score is evaluated, and the combination is retained only if this score exceeds a configurable threshold. 
By scanning the threshold from 0 to 1 in steps of 0.01, we map out the relationship between (i) the relative reduction in output file size, (ii) changes in processing throughput, and (iii) the signal-candidate efficiency. 
The results are presented in Fig.~\ref{fig:eff_throughput_cut}.  
The left panel shows the relative file-size reduction (blue) and throughput variation (red). The filesize measurement is normalised to the baseline where no cut is applied (i.e., all additional particles are retained) and throughput is normalised relative to if all particles are saved without running the \IMI at all.
The right panel shows the direct trade-off between file-size reduction and signal-candidate efficiency, where the signal efficiency is the one obtained using the inclusive simulation sample (Table~\ref{tab:samples_summary}).
In these figures, the reported file-size reductions are evaluated after applying a common set of loose pre-cuts, described in Sec.~\ref{sec:integration}. 
These pre-cuts are applied uniformly in all configurations, including the ``no-isolation'' baseline, and serve only to remove pathological vertex combinations (e.g.\ extreme or ill-defined SV-association values). 
They are intentionally chosen to be essentially lossless for signal candidates ($\gtrsim 99\%$) while rejecting only a small fraction of background tracks ($\lesssim 10--15\%$), so their standalone impact on the overall file-size reduction is minimal compared with the effect of the isolation requirements studied here.

At the nominal working point, defined by $\mathrm{IMI} > 0.05$, the output file size is reduced by 45\%, while preserving more than $99\,\%$ of signal candidates.  
It is worth noting that the maximum achievable file-size reduction asymptotes at roughly 50\%, as the remaining portion consists of indispensable reconstructed content such as primary vertices, base particles, neutral objects, and trigger information (see Fig.~\ref{fig:full_stream_size}).
Crucially, the throughput remains essentially constant, increasing by less than $0.1\,\%$, across the full threshold range. 
This is to be expected as all combinations of base particles and additional charged particles will be made regardless of the chosen cut value.
The slight increase in throughput at higher thresholds simply reflects fewer events being written, which is a comparatively light computational task.

Although \IMI\ is intrinsically lightweight at the inference stage, its integration introduces additional overheads--dominated by the vertex fits needed to evaluate the input features---resulting in an overall throughput reduction of about $20\%$ in \textsc{Sprucing}. This sets the scale of the reduction shown in Fig.~\ref{fig:eff_throughput_cut} (left). Such an overhead is acceptable at \textsc{Sprucing}, where throughput is far less critical than in Hlt2: for example, the \textsc{Full} stream in Hlt2 operates at input rates of $0.5$--$1.5\,\mathrm{MHz}$, while \textsc{Sprucing} handles only about $0.04\,\mathrm{MHz}$, i.e.\ more than a factor of thirty lower. In this regime, the modest throughput penalty is clearly outweighed by the $\sim45\%$ reduction in file size, which substantially improves downstream data handling and storage efficiency. 
We additionally note that the throughput spread decreases as the \IMI\ threshold is tightened. The vertexing needed to evaluate the \IMI\ inputs is performed independently of the threshold and therefore sets an approximately constant baseline cost, while the threshold mainly controls how many extra particles are ultimately persisted. At low thresholds we observe a higher spread driven by a small number of outliers, predominantly due to fluctuations in the load of the machine used for the throughput measurements (i.e.\ periods when the host is more or less busy). An additional contribution arises from event-dependent variability in packing/serialization and I/O time when many more particles are written out.

\section{Validation in data}
\label{sec:validation}

While classical isolation algorithms from Run~2 have been successfully adapted and re-designed for Run~3, the \IMI algorithm represents a completely new development.
Given that much of the current semileptonic physics programme at \lhcb, particularly analyses involving excited charm (and charmless) states and so-called ``double-charm” decays, relies heavily on this algorithm, a robust validation in data is essential.

\subsection{Ranking behaviour in real data}
\label{sec:ranking}

\begin{figure}[!htb]
  \centering
  \includegraphics[width=0.62\linewidth]{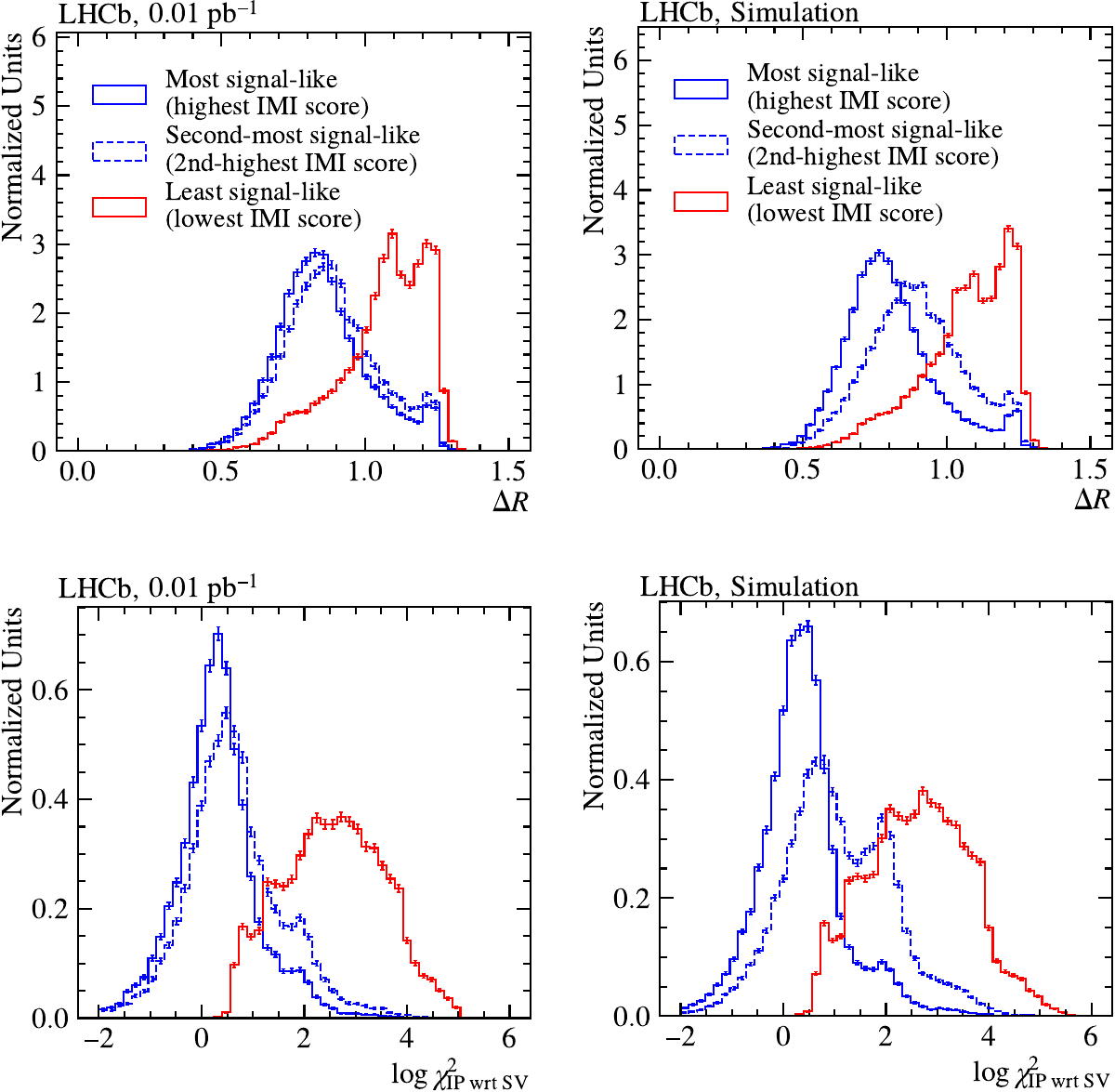}
  \caption{
    Distributions of (top) $\Delta R$ and (bottom) $\log\!\bigl(\chi^{2}_{\mathrm{IP\,w.r.t.\,SV}}\bigr)$ for extra particles in partial Run~3 data (left) and simulation (right) of $B^{0}\!\to D^{*-}\ell^{+}\nu_{\ell}$ candidates. The samples are split by the \IMI rank of the extra particle (highest, second-highest, and lowest \IMI score).
  }
  \label{fig:ranking}
\end{figure}

The \IMI algorithm assigns a score to each extra particle near a selected signal candidate, reflecting how likely it is to originate from the same decay chain as the base particle.
To validate that this ranking is physically meaningful, we compare the highest-, second-highest-, and lowest-ranked extra particles in partial Run~3 data ($0.01\,\mathrm{pb}^{-1}$) used to reconstruct $B^{0}\!\to D^{*-}\ell^{+}\nu_{\ell}$ candidates and in simulation of the same decay.
Figure~\ref{fig:ranking} shows the distributions of two representative input features: the cone angle $\Delta R$ and the IP \chisq\ with respect to the SV, $\log\!\bigl(\chi^{2}_{\mathrm{IP\,wrt\,SV}}\bigr)$.

In both data and simulation, the ordering follows the expected pattern: the most signal-like particles (highest \IMI score) concentrate at smaller $\Delta R$ and low $\chi^{2}_{\mathrm{IP\,wrt\,SV}}$, while the least signal-like particles are shifted to larger $\Delta R$ and larger $\chi^{2}_{\mathrm{IP\,wrt\,SV}}$; the second-most signal-like particles lie in between.
The separation between ranks is well reproduced, with good overall data--simulation agreement.
Differences are visible, with data generally showing broader shapes, leading to slightly more overlap between ranks.
These residual differences are driven by components present in data that are not explicitly included in the signal-only simulation, such as combinatorial and mis-identified background and feed-down from higher orbitally excited charm states ($D^{**}$), together with residual mismodelling of underlying-event activity from soft QCD.
Overall, the observed behaviour supports that \IMI performs the intended physical ranking.
The remaining input features show similar agreement and are presented in Appendix~\ref{app:input_feature_validation}.

\subsection{Reconstructing resonances using \IMI-selected particles}
\label{sec:reconstruction}

A more stringent test of the algorithm's performance is whether the extra particles selected by \IMI can be used to reconstruct well-known resonances, without relying on channel-specific tuning.  

\begin{figure}[!htb]
  \centering
  \includegraphics[width=0.6\columnwidth]{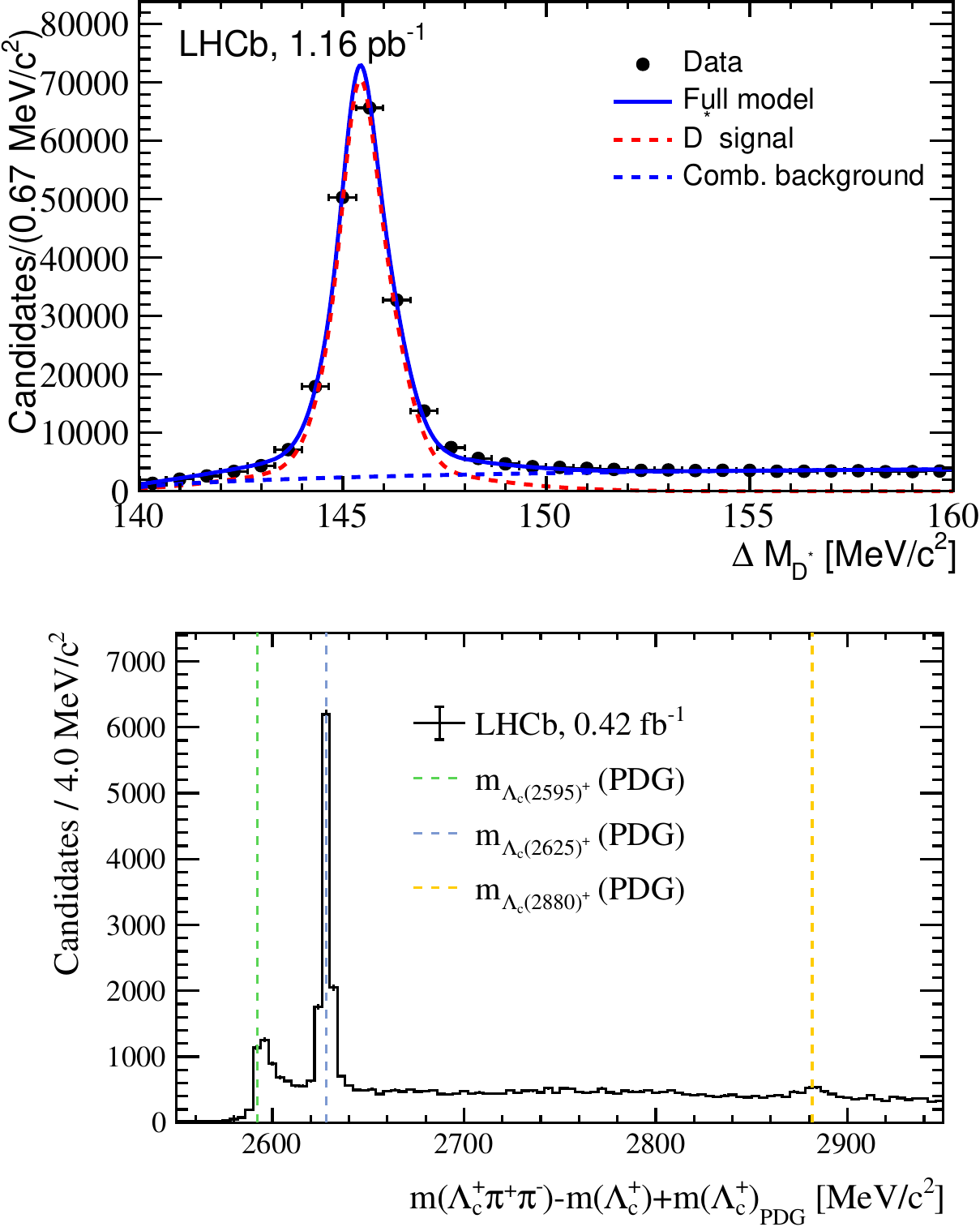}
  \caption{
  (Top) $\Delta M_{D^{*-}}$ distribution in partial Run~3 data, using the highest-ranked $\pi^-$ to reconstruct $D^{*-}$ mesons from $\bar{D}^{0}$ candidates.  
  The signal (red) and background (dashed blue) fit components are overlaid.  
  (Bottom) 
  The $\Delta M_{\Lambda_c^*} = M(\Lambda_{c}^{+}\pi^{+}\pi^{-}) - M(\Lambda_c) - M_{PDG}(\Lambda_c)$ distribution in partial Run~3 data, using the two highest-ranked oppositely charged particles to reconstruct $\Lambda_{c}^{*+}$ baryons from $\Lambda_{c}^{+}$ candidates.
  Vertical lines indicate the known $\Lambda_{c}^{*+}$ masses from PDG.
  }
  \label{fig:Dst_distribution}
\end{figure}

We first consider reconstruction of the $D^{*-} \to \bar{D}^{0} \pi^{-}$ decay.  
Starting from a clean $\bar{D}^{0} \to K^{+}\pi^{-}$ candidate, we combine the single highest-ranked extra particle with negative charge (assumed to be a $\pi^-$) with the base particles to form a $D^{*-}$.  
The resulting $\Delta M_{D^{*-}} \equiv |M_{D^{*-}} - M_{\bar{D}^{0}}|$ spectrum, shown in Fig.~\ref{fig:Dst_distribution} (top), exhibits a clear, narrow peak on top of a small combinatorial background.  
Notably, only loose particle identification requirements are applied to the base particles, further highlighting the discriminating power of \IMI.

A similar validation is performed in $\Lambda_{b}^{0} \to \Lambda_{c}^{*+} \mu^{-} \bar{\nu}_{\mu}$ decays, where excited charm baryons $\Lambda_{c}^{*+} \to \Lambda_{c}^{+} \pi^{+} \pi^{-}$ are reconstructed by combining the two highest-ranked oppositely charged extra particles with a $\Lambda_{c}^{+}$ baryon.  
The resulting $\Delta M_{\Lambda_c^*} = M(\Lambda_{c}^{+}\pi^{+}\pi^{-}) - M(\Lambda_c) - M_{PDG}(\Lambda_c)$ distribution, shown in Fig.~\ref{fig:Dst_distribution} (bottom), reveals peaks corresponding to the $\Lambda_{c}(2595)^{+}$ and $\Lambda_{c}(2625)^{+}$ resonances, as well as a structure around the $\Lambda_{c}(2880)^{+}$.  
Again, these results are obtained with minimal selection, demonstrating that \IMI can reliably recover non-isolated signal decay products in data.

\subsection{Isolation efficiency in data and simulation}
\label{sec:efficiency_comparison}

\begin{figure}[!htb]
  \centering
  \includegraphics[width=0.8\linewidth]{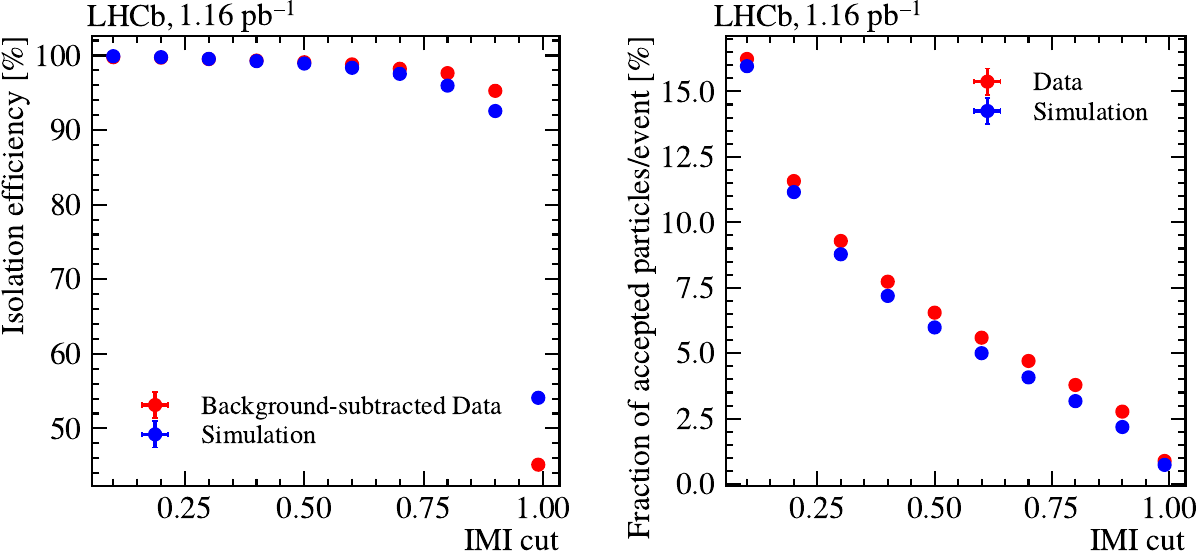}
  \caption{
    Signal efficiency (left) and fraction of charged particles accepted per event (right) as function of \IMI threshold, computed on simulated $B^{0} \to D^{*-} e^{+} \nu_{e}$ events and compared to partial Run~3 data.
  }
  \label{fig:fit_eff}
\end{figure}

While the \IMI algorithm performs well qualitatively, it is also important to evaluate its signal efficiency quantitatively, and compare data to simulation.  
Figure~\ref{fig:fit_eff} shows the efficiency of the \IMI selection (left) and the fraction of charged particles accepted per event (right), as a function of the threshold, computed on simulated $B^{0} \to D^{*-} \ell^{+} \nu_{\ell}$ events and compared to background-subtracted partial Run~3 data that reconstructs the same decay.
At low \IMI thresholds, the data--simulation agreement is good for the isolation efficiency, but not for the fraction of accepted charged particles: data consistently retains more particles at all thresholds. This is expected because the data sample contains sizeable backgrounds that are not present in the signal simulation, including combinatorial contributions (only fake $D^{*}$ has been subtracted), partially reconstructed decays, mis-identified hadronic decays, and feed-down from excited charm states ($D^{**}$), together with residual mismodelling of underlying-event activity from soft and multiple-parton interactions. At tighter \IMI thresholds the isolation efficiency also begins to diverge, pointing to mismodelling of some input-feature distributions in simulation; this could be reduced in future via inclusive simulation-to-data corrections. In practice, the tool is operated with loose working points chosen to retain high signal efficiency, so these discrepancies have minimal impact on downstream physics analyses.

\section{Summary and Outlook}
\label{sec:conclusion}

For Run~3, the \lhcb experiment faces the demanding task of reducing data rates by up to a factor of eight, imposing stringent constraints not only on which events are selected, but also on the size of each recorded event.
While signal decays typically involve just 5--7 charged particles, a typical Run~3 event contains $\mathcal{O}(200)$ reconstructed tracks, with charged particle information alone accounting for over 50\% of the total event size. 
To address this imbalance, a suite of inclusive isolation tools were developed, including classical track-, cone-, and vertex-based methods, alongside a new Inclusive Multivariate Isolation (\IMI) algorithm.

As its name implies, \IMI is inherently both inclusive in scope and multivariate in structure.
It is inclusive in that it has been trained from a wide range of simulated events, encompassing diverse decay topologies and kinematic configurations, including scenarios where signal particles emerge from both short- and long-lived intermediate states within the $b$-hadron decay chain.
Its multivariate nature arises from combining the prominent features of all classical isolation techniques -- cone-, vertex-, and track-based -- into a single classifier.
Built on the fast and lightweight XGBoost framework, \IMI assigns a score to each extra particle based on its compatibility with a signal origin, allowing only the most relevant particles to be retained for downstream analysis.
This selective retention enables the efficient reconstruction of complex decay chains with varying final-state multiplicities and facilitates the definition of background-enriched control regions, both essential for controlling systematic uncertainties in precision measurements.
As of 2025, \IMI plays a central role in the \lhcb semileptonic physics programme involving missing energy and is well suited for broader application to other decay channels.

The \IMI algorithm delivers exceptional performance across a broad range of decay modes and event multiplicities, consistently outperforming classical isolation techniques. It achieves an area under the curve (AUC) of 0.997, demonstrating strong separation power between signal and background particles. 
At its nominal working point, \IMI rejects over 90\% of background while retaining approximately $99\%$ of signal particles, representing a $2$--$5\times$ improvement in background rejection relative to traditional methods in the inclusive sample. 
Crucially, this performance is preserved even in high-multiplicity environments and across diverse decay topologies, without introducing biases in sensitive kinematic observables such as the momentum transfer squared ($q^2$) in semileptonic decays. 
By retaining only $\mathcal{O}(10)$ signal-like particles out of the $\mathcal{O}(200)$ reconstructed tracks per event, \IMI yields a significantly cleaner event representation, leading to a data-size reduction of approximately 45\%.

Both the classical and \IMI\ isolation algorithms are fully integrated into the \lhcb selection framework, each optimised for a different stage of the data-processing pipeline. 
The two also operate independently: selection lines using \IMI\ do not rely on candidates preselected with the classical isolation tool.
Classical isolation is deployed directly at the trigger level through two complementary approaches: one links signal particles to a minimal set of nearby tracks based on cone isolation, enabling isolation observables to be computed offline; the other computes and stores these observables at the trigger stage itself, allowing for immediate event size reduction with minimal impact on processing throughput. 
In contrast, \IMI adopts a more conservative strategy: the full reconstructed event is written to tape after the trigger stage, and \IMI is applied later at the offline \textsc{Sprucing} stage, all while ensuring that the additional memory and throughput cost remains non-significant.
This design offers long-term flexibility, as the \textsc{Sprucing} stage can be re-run on triggered events, enabling future updates to \IMI without the need to modify or reprocess data at the trigger level.

Validation using Run~3 data confirms that the \IMI algorithm performs reliably under real data taking conditions. It produces physically consistent and interpretable rankings of extra particles in the event, enabling the reconstruction of well-known resonances such as the $D^{*+}$ and $\Lambda_c^*$ without requiring dedicated tuning. For the loose working point used in production, the agreement between data and simulation in terms of signal efficiency is excellent across the phase space relevant to most analyses. These results demonstrate good confidence in the use of \IMI as a core isolation tool in the \lhcb selection framework.

In the near term, planned improvements to \IMI\ include extending isolation to neutral particles (Fig.~\ref{fig:full_stream_size}), adding VELO-based features for dense environments, and adopting a multiclass classifier to identify excited heavy-flavour states currently treated with cut-based selections (Appendix~\ref{app:excited_b_hadrons}). Looking further ahead, an attractive possibility is a broader architectural role for \IMI\ as a fast pruning layer in multi-stage reconstruction. In this setup, \IMI\ would ease the load on more complex, compute-intensive stages (e.g., GNN-based approaches~\cite{Sutcliffe:2025arn}), helping to keep data volumes manageable in Run~4 and enabling the $\mathcal{O}(20)$ reductions anticipated for Run~5~\cite{Vagnoni:2025yps,upgradeII}. In conclusion, the demonstrated Run~3 performance and lightweight design make \IMI\ a compelling and forward-compatible building block for scalable reconstruction in future runs.




\addcontentsline{toc}{section}{Acknowledgements}
\section*{Acknowledgements}
We thank the \textit{\lhcb Simulation Project} for the support in producing the simulated samples used in this work.  
We are also grateful to the \textit{Real-Time Analysis (RTA) Project} and \textit{Data Processing and Analysis (DPA) Project} for their valuable contributions throughout the project.  
T.~Fulghesu, C.-H.~Li, A.~Morris, and D.~vom~Bruch are supported by the European Research Council Starting Grant ALPACA (No.~101040710). 
A.~Mathad and L.~Hartman are supported by CERN. 
M.~Calvi and V.S.~Kirsebom are supported by Università di Milano-Bicocca, the Swiss National Science Foundation (SNSF, P500PT\_222273), the Istituto Nazionale di Fisica Nucleare (INFN), and CERN. 
G.~Hallett, T.~Latham, M.~Lehuraux, M.~Monk, and F.~Swystun are supported by the UK Science and Technology Facilities Council (STFC). 
M.~Monk also acknowledges support from the Monash Warwick Alliance. 
M.~Monk and F.~Swystun acknowledge support from UK Research and Innovation under grant \#EP/X014746/2. 
M.~Rudolph is supported by the U.S.\ Department of Energy (DOE) and the National Science Foundation (NSF).

\addcontentsline{toc}{section}{Author Contributions}
\section*{Author Contributions}
This project developed both classical isolation algorithms (variants A and B) and a novel Inclusive Multivariate Isolation (IMI) algorithm. L.H.\ initiated the IMI project as a CERN summer student under the direct supervision of A.M., who conceived the overall strategy and guided all subsequent stages of the work. A.M., V.S.K., A.Mo.\ and C.-H.L.\ then expanded the IMI project from training on realistic inclusive simulation samples through to deployment and validation on Run~3 data. V.S.K.\ was responsible for the IMI training and performance studies, A.Mo.\ led the deployment of the algorithm within the LHCb software framework, and C.-H.L.\ and V.S.K.\ performed the data validation studies, with A.M.\ coordinating and contributing to all aspects of these activities. M.S.R.\ worked on the persistence of tracks from excited beauty hadrons as part of the IMI project. The IMI design was informed by earlier isolation studies on Run~2 data performed by B.K. T.F.\ implemented and validated classical isolation algorithm variant A and, together with C.-H.L., co-developed algorithms for the use of IMI in offline analyses. The development and deployment of classical isolation algorithm variant B at the high-level trigger was led by T.L., with contributions from G.H., M.Le., M.M.\ and F.S. The majority of the manuscript was written by A.M., with supporting figures prepared by V.S.K., A.Mo.\ and C.-H.L.; all authors reviewed and approved the final manuscript. V.S.K.\ was partly supported by funding obtained by M.C., and A.Mo.\ and C.-H.L.\ were supervised by and supported by funding obtained by D.v.B.

\noindent
\textbf{Initials:}
A.M.\ (Abhijit Mathad), L.H.\ (Luca Hartman), V.S.K.\ (Veronica Sølund Kirsebom), A.Mo.\ (Andy Morris), C.-H.L.\ (Ching-Hua Li), M.S.R.\ (Matthew Scott Rudolph), B.K.\ (Basem Khanji), T.F.\ (Tommaso Fulghesu), T.L.\ (Thomas Latham), G.H.\ (George Hallett), M.Le.\ (Marion Lehuraux), M.M.\ (Matthew Monk), F.S.\ (Francesca Swystun), M.C.\ (Marta Calvi), D.v.B.\ (Dorothea vom Bruch).

\appendix

\section{Inclusive simulation samples}
\label{app:simulation_samples}

Table~\ref{tab:samples_summary} provides an overview of the simulated \(b\)-hadron decay modes used to train the \IMI algorithm. The samples cover a broad range of decay topologies, including semileptonic decays with light leptons (\(e\), \(\mu\)) and tau leptons (\(\tau\)), hadronic states involving excited charm mesons and baryons, as well as decays containing charmonium states.   The table also indicates $\textcolor{blue}{\emph{base}}$ particles (in blue) that form a combination relative to which the input features are computed, and $\textcolor{red}{\emph{non-isolated}}$ particles (in red) are those the \IMI algorithm must select.

\begin{table}[!htp]
  \centering
  \small
  \renewcommand{\arraystretch}{1.15}     
  \caption{Simulated decays used to train the \IMI tool.
           The \textcolor{blue}{\emph{base}} particles (blue) form a combination
           relative to which the input features are computed, while the
           \textcolor{red}{\emph{non-isolated}} particles (red) are those the
           algorithm must select.  
           Labels $D^{*}$ and $\Lc^{*}$ represent inclusive simulations
           of excited charm--meson and charm--baryon states, respectively.
           Label $\ell$ represents either an electron ($e$) or a muon ($\mu$).
           The charge conjugate modes are implied.
           }
    \begin{tabularx}{\textwidth}
        {@{\hspace{0pt}}>{\RaggedRight\arraybackslash}p{2.5cm}%
        >{\RaggedRight\arraybackslash}p{2.5cm}%
        X@{\hspace{0pt}}}
    \toprule
    \textbf{Base particles} & \textbf{Non-isolated particles} & \textbf{Decay chain} \\
    \midrule
    $\textcolor{blue}{\Dzb\,\ell^{+}}$
      & $\textcolor{red}{\pim}$
      & $\Bz\!\to\!\Dstarm\,\textcolor{blue}{\ell^{+}}\nu_{\ell},\;
         \Dstarm\!\to\!\textcolor{blue}{\Dzb}\,\textcolor{red}{\pim}$ \\[0.5ex]

    $\textcolor{blue}{\Dzb\,\ell^{+}}$
      & $\textcolor{red}{\pim}$
      & $\Bz\!\to\!\Dstarm\,\tau^{+}\nu_{\tau},\;
         \Dstarm\!\to\!\textcolor{blue}{\Dzb}\,\textcolor{red}{\pim},\;
         \tau^{+}\!\to\!\textcolor{blue}{\ell^{+}}\nu_{\ell}\bar{\nu}_{\tau}$ \\[0.5ex]

    $\textcolor{blue}{\Dzb\,\tau^{+}}$
      & $\textcolor{red}{\pim}$
      & $\Bz\!\to\!\Dstarm\,\textcolor{blue}{\tau^{+}}\nu_{\tau},\;
         \Dstarm\!\to\!\textcolor{blue}{\Dzb}\,\textcolor{red}{\pim},\;
         \tau^{+}\!\to\!\pi^{+}\pi^{-}\pi^{+}(\pi^{0})\bar{\nu}_{\tau}$ \\[0.5ex]

    $\textcolor{blue}{\Dz\,\ell^{+}}$
      & $\textcolor{red}{\pim,\;\pip}$
      & $\Bp\!\to\!D^{**0}\,\textcolor{blue}{\ell^{+}}\nu_{\ell},\;
         D^{**0}\!\to\!\Dstarp\,\textcolor{red}{\pim}\piz,\;
         \Dstarp\!\to\!\textcolor{blue}{\Dz}\,\textcolor{red}{\pip}$ \\[0.5ex]

    $\textcolor{blue}{\Dz\,\ell^{+}}$
      & $\textcolor{red}{\pim,\;\pip}$
      & $\Bp\!\to\!D^{**0}\,\tau^{+}\nu_{\tau},\;
         D^{**0}\!\to\!\Dstarp\,\textcolor{red}{\pim}\piz,\;
         \Dstarp\!\to\!\textcolor{blue}{\Dz}\,\textcolor{red}{\pip},\;
         \tau^{+}\!\to\!\textcolor{blue}{\ell^{+}}\nu_{\ell}\bar{\nu}_{\tau}$ \\[0.5ex]

    $\textcolor{blue}{\Dzb\,\ell^{+}}$
      & $\textcolor{red}{\pip,\;\pim,\;\pim}$
      & $\Bz\!\to\!D^{**-}\,\textcolor{blue}{\ell^{+}}\nu_{\ell},\;
         D^{**-}\!\to\!\Dstarm\,\textcolor{red}{\pip\pim},\;
         \Dstarm\!\to\!\textcolor{blue}{\Dzb}\,\textcolor{red}{\pim}$ \\[0.5ex]

    $\textcolor{blue}{\Dzb\,\ell^{+}}$
      & $\textcolor{red}{\pip,\;\pim,\;\pim}$
      & $\Bz\!\to\!D^{**-}\,\tau^{+}\nu_{\tau},\;
         D^{**-}\!\to\!\Dstarm\,\textcolor{red}{\pip\pim},\;
         \Dstarm\!\to\!\textcolor{blue}{\Dzb}\,\textcolor{red}{\pim},\;
         \tau^{+}\!\to\!\textcolor{blue}{\ell^{+}}\nu_{\ell}\bar{\nu}_{\tau}$ \\[0.5ex]

    $\textcolor{blue}{\Dz\,\ell^{+}}$
      & $\textcolor{red}{\pip,\;\pim,\;\pim}$
      & $\Bp\!\to\!D^{(*,**)}D_{(s)}^{(*)},\;
         D^{(*,**)}\!\to\!\textcolor{blue}{\Dz}\,
         \textcolor{red}{\pim}(\textcolor{red}{\pip\pim}),\;
         D_{(s)}^{(*)}\!\to\!\textcolor{blue}{\ell^{+}}X$ \\[0.5ex]

    $\textcolor{blue}{\Dz\,\ell^{+}}$
      & $\textcolor{red}{\Km,\;2\pim,\;2\pip}$
      & $\Bp\!\to\!D^{(*,**)}D_{(s)}^{(*)}\textcolor{red}{\Km\pip},\;
         D^{(*,**)}\!\to\!\textcolor{blue}{\Dz}\,
         \textcolor{red}{\pim}(\textcolor{red}{\pip\pim}),\;
         D_{(s)}^{(*)}\!\to\!\textcolor{blue}{\ell^{+}}X$ \\[0.5ex]
    \midrule
    $\textcolor{blue}{\Kp\,\ell^{-}}$
      & $\textcolor{red}{\ell^{+}}$
      & $\Bp\!\to\!\textcolor{blue}{\Kp}J/\psi,\;
         J/\psi\!\to\!\textcolor{red}{\ell^{+}}\textcolor{blue}{\ell^{-}}$ \\[0.5ex]

    $\textcolor{blue}{\Kp\,\ell^{-}}$
      & $\textcolor{red}{\Kp,\;\pim}$
      & $\Bs\!\to\!\Dsm\,\textcolor{blue}{\ell^{+}}\nu_{\ell},\;
         \Dsm\!\to\!\textcolor{red}{\Kp\pim}\textcolor{blue}{\Km}$ \\[0.5ex]

    $\textcolor{blue}{\Kp\,\ell^{-}}$
      & $\textcolor{red}{\Kp,\;\pim}$
      & $\Bs\!\to\!\Dsm\,\tau^{+}\nu_{\tau},\;
         \Dsm\!\to\!\textcolor{red}{\Kp\pim}\textcolor{blue}{\Km},\;
         \tau^{+}\!\to\!\textcolor{blue}{\ell^{+}}\nu_{\ell}\bar{\nu}_{\tau}$ \\[0.5ex]

    $\textcolor{blue}{\Kp\,\ell^{-}}$
      & $\textcolor{red}{\Kp,\;2\pim,\;\pip}$
      & $\Bs\!\to\!\DsStarm\,\textcolor{blue}{\ell^{+}}\nu_{\ell},\;
         \DsStarm\!\to\!\Dsp\,\textcolor{red}{\pip\pim},\;
         \Dsp\!\to\!\textcolor{red}{\Kp\pim}\textcolor{blue}{\Km}$ \\[0.5ex]
    \midrule
    $\textcolor{blue}{\Lc\,\ell^{-}}$
      & $\textcolor{red}{\pip\pim}$
      & $\Lb\!\to\!\LcStar\,\textcolor{blue}{\ell^{-}}\bar{\nu}_{\ell},\;
         \LcStar\!\to\!\textcolor{blue}{\Lc}\,\textcolor{red}{\pip\pim}$ \\[0.5ex]
    \midrule
    $\textcolor{blue}{p\,\ell^{-}}$
      & $\textcolor{red}{\Km\pip}$
      & $\Lb\!\to\!\Lc\,\textcolor{blue}{\ell^{-}}\bar{\nu}_{\ell},\;
         \Lc\!\to\!\textcolor{blue}{p}\,\textcolor{red}{\Km\pip}$ \\
    \bottomrule
  \end{tabularx}
  \label{tab:samples_summary}
\end{table}

\section{Selection of prompt particles from excited \texorpdfstring{$b$}{b}-hadron decays}
\label{app:excited_b_hadrons}

Orbitally and radially-excited \(b\)-hadrons (e.g., \(B_{sJ}^{\ast}\), \(B^{\ast\ast}\)) frequently decay promptly at the PV, producing high-momentum final-state particles such as the kaon in \(B_{s2}^{\ast}(5840)\!\to\!B^{+}K^{-}\).  
Although these particles constitute a small fraction of the overall event content, they are critical for precision studies involving missing energy such as searches for LFV in \(B_{s}^{\ast}\) decays~\cite{LHCb:2023zxo} and relative branching fraction measurements~\cite{LHCb:2018azb}.

Owing to their \emph{prompt} origin and \emph{high} momentum, these particles are not efficiently identified by the \IMI classifier, which is specifically trained to target displaced tracks from long-lived \(b\)-hadron decays.  
An initial attempt to incorporate them into the \IMI training as a separate signal category was explored using inclusive $\bquark\bar \bquark$ simulation samples; however, the limited size of available simulation samples at the time prevented meaningful gains in classification performance~\cite{Hartman:2023}.  
While a future version of the \IMI tool may revisit this approach using larger and more diverse training data, the current strategy adopts a simple, cut-based selection that exploits the distinctive kinematic features of these prompt decay products.

Three observables are studied to obtain near-orthogonal separation between signal and background:  
\begin{enumerate}
    \item the transverse momentum, \(p_{T}\), of the track;
    \item the perpendicular momentum, \(p_{\perp}\), defined with respect to the reconstructed \(b\)-hadron flight direction;
    \item particle identification (PID) information, expressed as the logarithm of the likelihood ratio for the kaon versus pion hypotheses, \(\Delta \log(\mathcal{L}_{K/\pi})\).
\end{enumerate}
Excited \(b\)-hadrons are produced in hard QCD processes and are typically highly boosted. Consequently, their decay products carry large transverse momentum and are emitted almost collinearly with the parent \(b\)-hadron trajectory. This effect is particularly pronounced for narrow resonances produced near threshold, where the decay products tend to have small transverse momenta (\(p_{\perp}\)). 
In contrast, background particles originating from the primary vertex through soft QCD interactions generally exhibit lower \(p_{T}\) and a broader angular distribution, resulting in larger \(p_{\perp}\). 
Moreover, excited \(b\)-hadrons containing two heavy quarks (e.g.\ \(B_{s2}^{\ast}\)) often decay into a \(B\) meson and a kaon, producing a strong kaon PID signal, whereas background particles are predominantly pions.  
It should be noted that while the requirements \(p_{T}\) of daughter particles from excited \(b\)-hadron decays are generally unavoidable for LFV and branching fraction measurements, any hard requirement on \(p_{T}\) in spectroscopy analyses must be applied with care. Such cuts can induce large efficiency variations across the kinematic phase space, making them difficult to model accurately in simulation. In this study, we apply a relatively soft lower bound on \(p_{T}\), which introduces no significant efficiency shape variations over the invariant mass spectrum of the excited \(b\)-hadron candidates (e.g. \(m(B^+ K^-)\) for \(B_{sJ}^{\ast}\) studies).

\begin{figure}[t]
    \centering
    \includegraphics[width=0.325\linewidth]{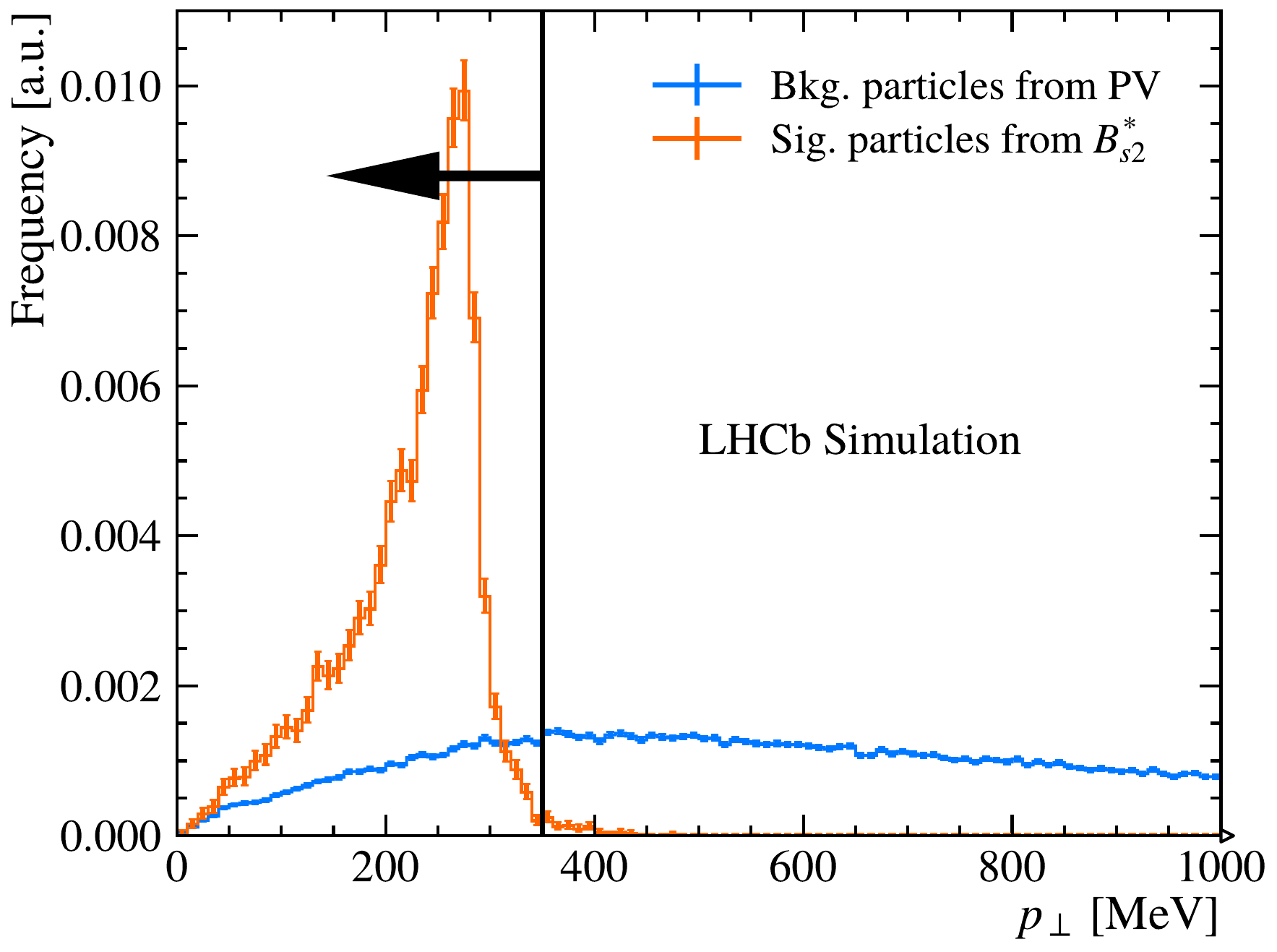}
    \includegraphics[width=0.325\linewidth]{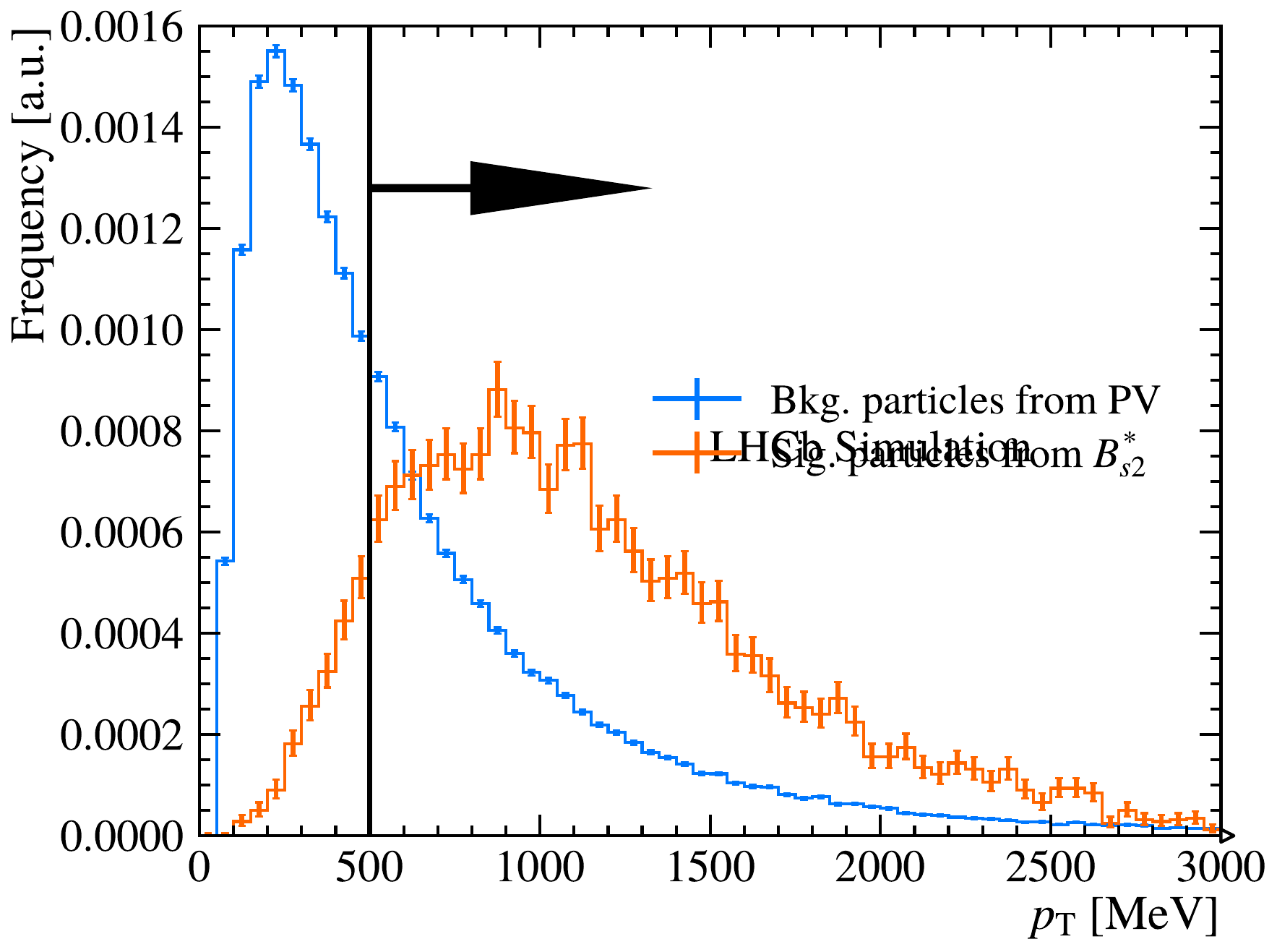}
    \includegraphics[width=0.325\linewidth]{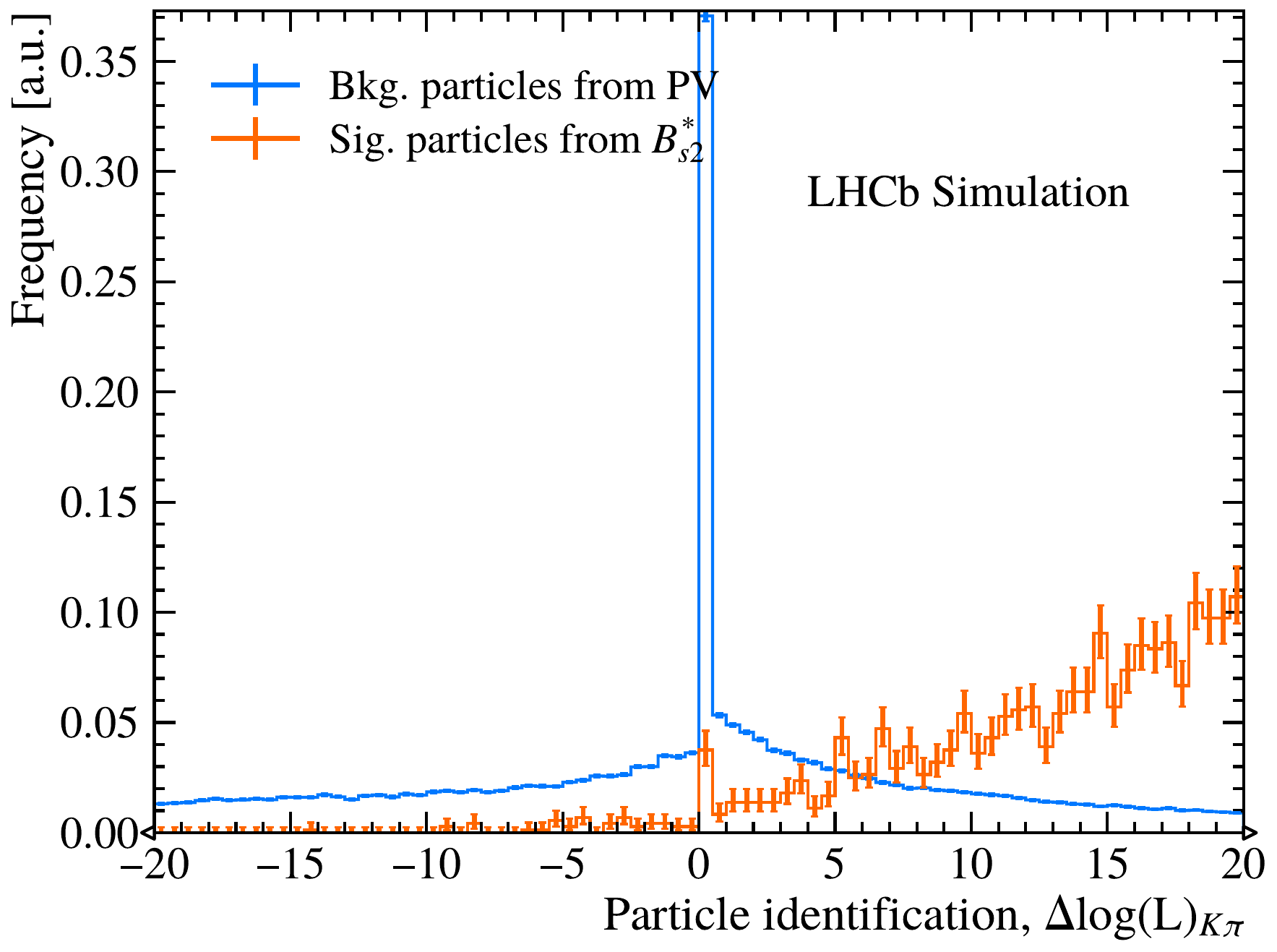}
    \includegraphics[width=0.4\linewidth]{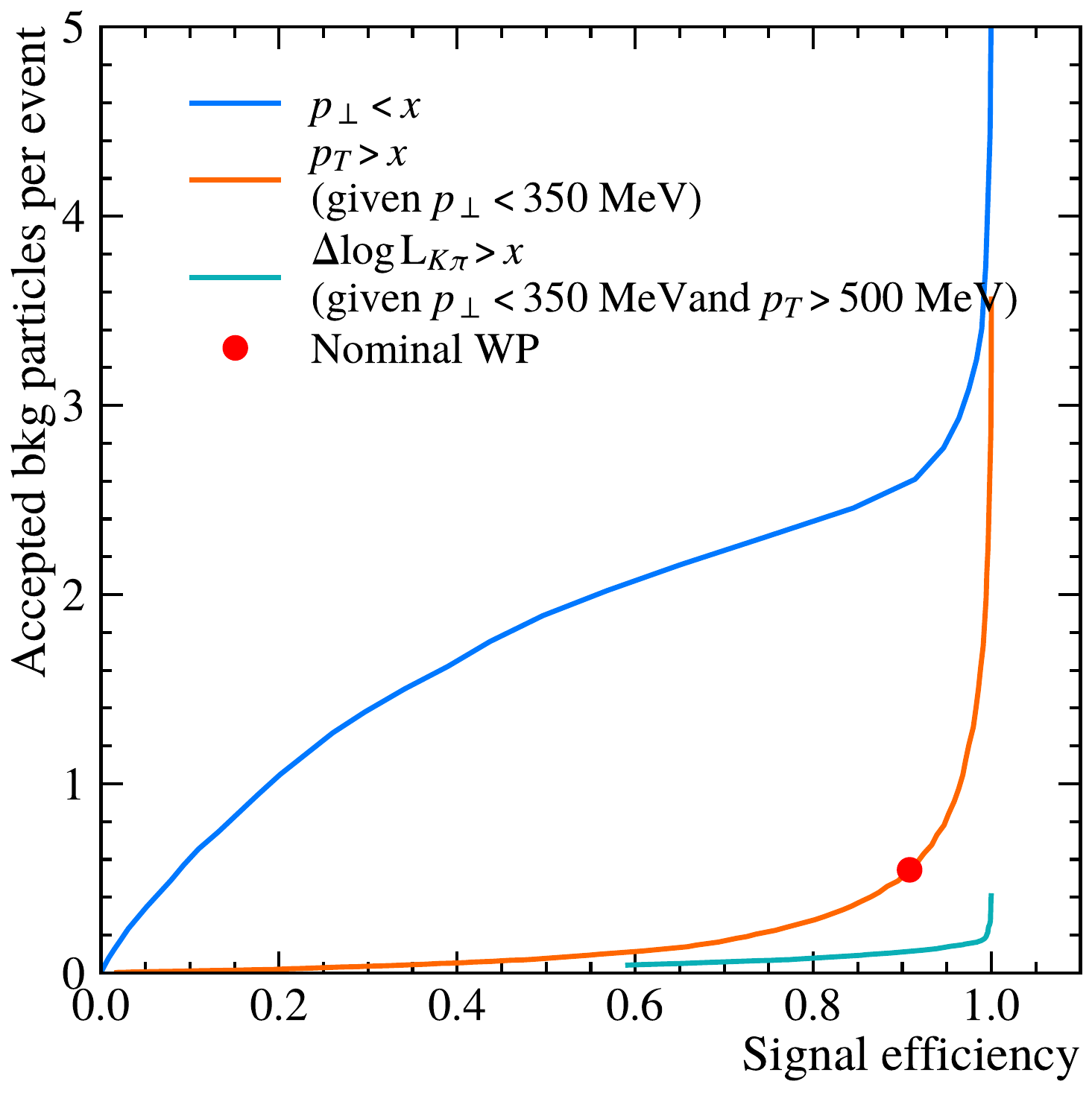}
    \caption{
        Top row (left to right):
        Distributions of transverse momentum (\(p_{T}\)), perpendicular momentum (\(p_{\perp}\)), and PID variable \(\Delta \log(\mathcal{L}_{K/\pi})\) for particles from prompt decays of excited \(b\)-hadrons (orange) and from primary-vertex background (teal) in a simulated Run~3 sample of \(B\!\to\! X_{u}\,\mu\nu_{\mu}\) decays.
        Bottom row: Average number of prompt background particles per event as a function of signal efficiency for three selection scans:
        (i) \(p_{\perp} < x\);
        (ii) \(p_{T} > x\) with \(p_{\perp} < 350~\text{MeV}\);
        (iii) \(\Delta \log(\mathcal{L}_{K/\pi}) > x\) with \(p_{T} > 500~\text{MeV}\) and \(p_{\perp} < 350~\text{MeV}\).
        The nominal selection is indicated by the solid black line in the top row and by the red marker in the bottom row.
    }
    \label{fig:excited_b_hadrons_features}
\end{figure}

Figure~\ref{fig:excited_b_hadrons_features} shows, from left to right, the distributions of \(p_{T}\), \(p_{\perp}\), and \(\Delta \log(\mathcal{L}_{K/\pi})\) for signal and background tracks in a simulated \(B \to X_{u}\,\mu\nu_{\mu}\) sample where the \(B\) meson originates from \(B_{s2}^{\ast}(5840) \to B^{+} K^{-}\) decays.  
The bottom panel presents the average number of background particles per event as a function of signal efficiency for three different selection strategies.
A cut on \(p_{\perp}\) alone does not sufficiently suppress the background, as it leaves too many background particles per event even at high signal efficiency.  
Introducing an additional requirement on \(p_{T}\) substantially improves background rejection.  
While the PID observable offers further suppression, it is omitted from the nominal selection to preserve flexibility in defining background-enriched control samples for later analyses.
The final two-dimensional selection is:
\[
p_{T} > 400~\text{MeV}, 
\qquad
p_{\perp} < 350~\text{MeV}.
\]
This choice retains over 90\% of prompt signal particles while accepting fewer than 0.5 background particles per event.

\section{Correlation matrix of input features for \IMI}
\label{app:correlation}

\begin{figure}[!htp]
    \centering
    \includegraphics[width=\textwidth]{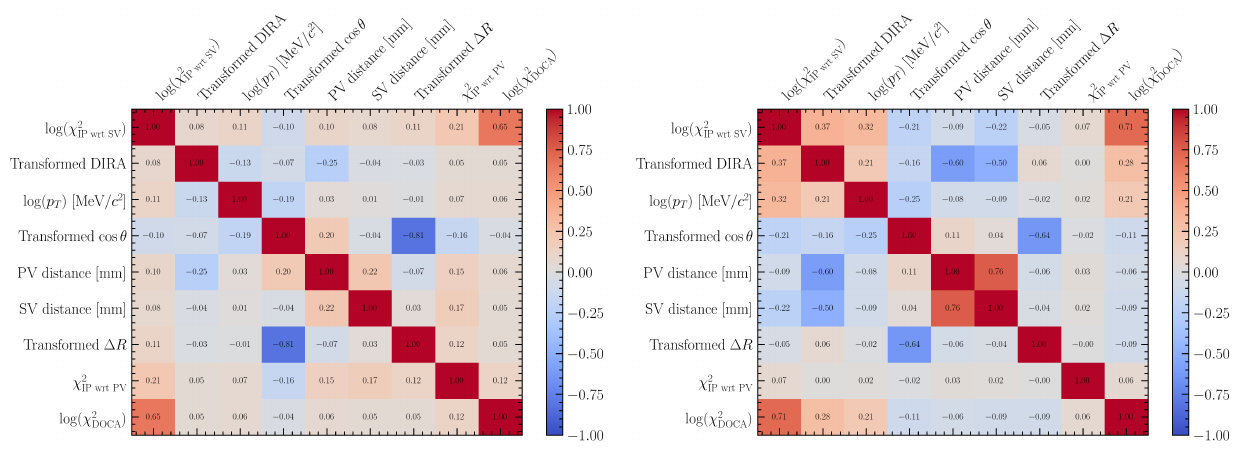}
    \caption{
    Correlation matrices of the input variables used in the \IMI algorithm, shown for signal (left) and background (right) particles.
    }
    \label{fig:correlation_matrix}
\end{figure}

Figure~\ref{fig:correlation_matrix} shows the correlation matrices of the input variables used in the \IMI algorithm, separately for signal and background particles. 
When a strong correlation appears in one class but not the other, both variables are retained in the training, as each can still contribute complementary information to the classifier.
Here we discuss only the most significant correlations (above 50\%) observed in both classes. 

There is an anti-correlation between the \textit{Transformed $\Delta R$} and \textit{Transformed $\cos\theta$} variables
(defined in Section~\ref{subsec:training_feature})
, which is more prominent for the signal class than for the background. This is expected since both describe the angular separation between the base and extra particle momenta -- small $\Delta R$ in detector $(\eta,\phi)$ space corresponds to a small opening angle and thus large $\cos\theta$ in three-dimensional momentum space, and vice versa. In signal class, extra particles are often kinematically related to the base particles through a common decay, enhancing this relationship, whereas in background events they are largely uncorrelated, diluting the pattern. We include both variables in the training because, despite their correlation, they probe angular separation in complementary coordinate systems, allowing the classifier to exploit subtle differences in signal and background topologies.

Additionally, a moderate correlation is observed between $\log(\chi^2_{\mathrm{IP\,wrt.\,SV}})$ and $\log(\chi^2_{\mathrm{DOCA}})$. This is expected since both quantify spatial compatibility of the extra particle with the base candidate: a small DOCA between the extra and base particle trajectories typically corresponds to a small impact parameter with respect to the secondary vertex, and vice versa. 
We include both variables in the training because they capture different geometric aspects: \(\chi^2_{\mathrm{DOCA}}\) measures the distance between the two particles, while \(\chi^2_{\mathrm{IP\,wrt.\,SV}}\) quantifies how well the extra particle is consistent with the secondary vertex reconstructed from the base particles.

Importantly, when we evaluated the signal efficiency across different exclusive channels, we observed a small but consistent improvement from retaining these two sets of correlated variables, compared to removing one from each pair.

\section{Signal and background efficiency as a function of isolation variables}
\label{app:isolation_score}

\begin{figure}[!htp]
    \centering
    \includegraphics[width=0.7\textwidth]{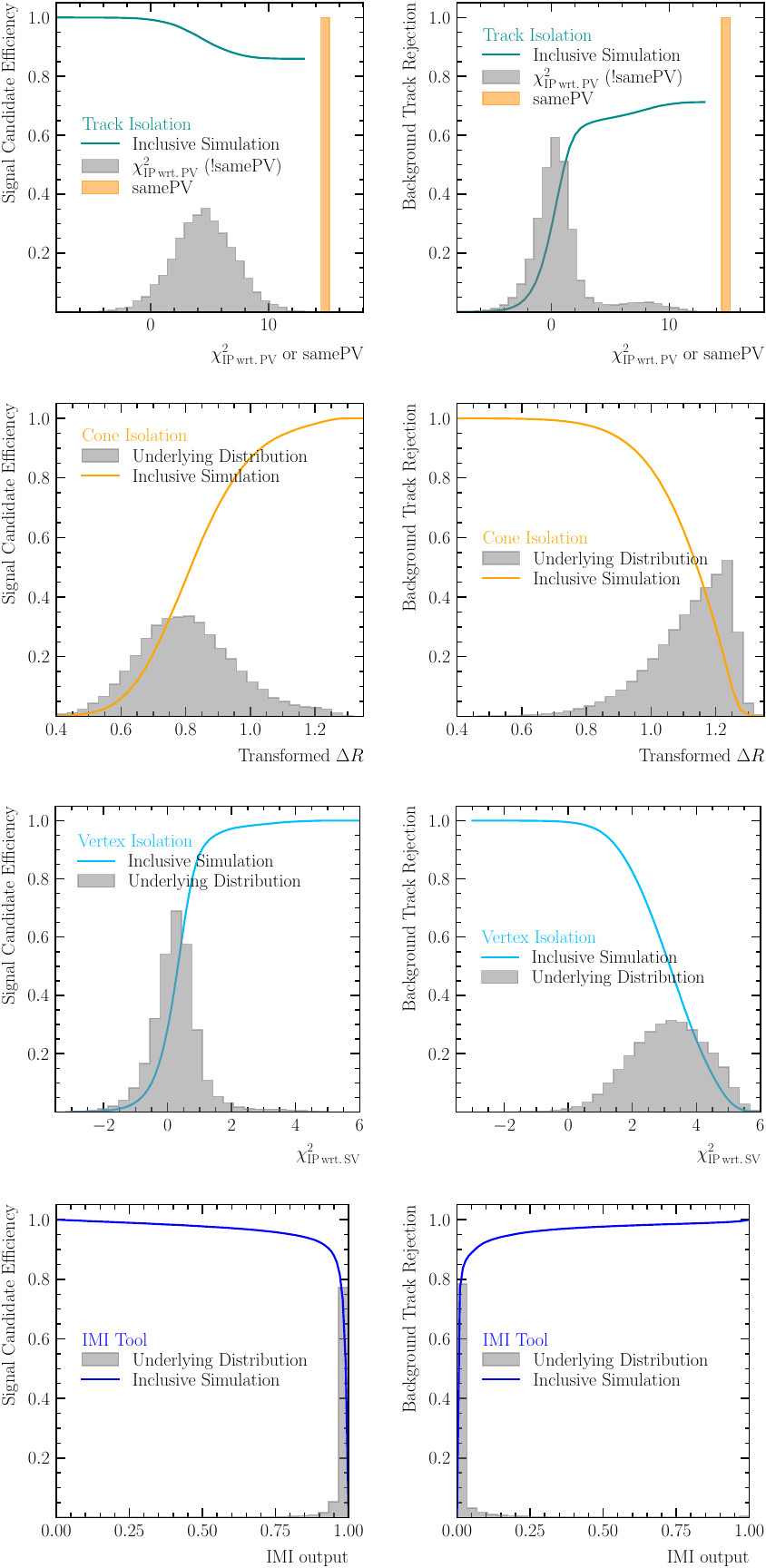}
    \caption{
        Signal efficiency (left column) and background rejection (right column) as a function of the isolation variable for the track (top row), cone (second row), vertex (third row), and \IMI\ (bottom row) algorithms, evaluated on the inclusive simulation.
    }
    \label{fig:efficiency_vs_isolation_score}
\end{figure}

Figure~\ref{fig:efficiency_vs_isolation_score} shows the signal efficiency and background rejection as a function of the isolation variable for the four algorithms: track, cone, vertex, and \IMI.
These have been evaluated using the inclusive simulation samples shown in Table~\ref{tab:samples_summary}, which include a variety of \(b\)-hadron decay modes and kinematic configurations.

For track isolation, the discriminating variable is $\chi^2_{\mathrm{IP\,wrt.\,PV}} \geq x$ or the \texttt{samePV} flag. 
The binary nature of the \texttt{samePV} flag sets the baseline signal efficiency and background rejection, 
while threshold $x$ determines further discrimination. 
When the threshold $x$ is very small, most tracks are retained, leading to a high signal efficiency but low background rejection. 
Increasing $x$ tightens the selection, reducing the signal efficiency while improving the background rejection.

For cone isolation, the relevant variable is $\Delta R < y$. 
An extremely small $y$ accepts only tracks very close in $(\eta,\phi)$ to the base particle, yielding low signal efficiency but strong background rejection. 
As $y$ increases, more tracks are accepted, resulting in higher signal efficiency but reduced background rejection.

For vertex isolation, the discriminating variable is $\chi^2_{\mathrm{IP\,wrt.\,SV}} < z$. 
Tight requirements (small $w$) retain only tracks highly compatible with the reconstructed secondary vertex, giving lower signal efficiency but stronger background rejection. 
Loosening the requirement increases the signal efficiency as more signal-side tracks pass, at the cost of admitting additional background tracks that are accidentally compatible with the SV.

For \IMI\ isolation, the discriminating variable is $\IMI > a$. 
Small values of $a$ accept most candidates, giving high signal efficiency but low background rejection. 
Raising $a$ increases the strictness of the selection, thereby lowering the signal efficiency and increasing the background rejection.

\section{Simulation vs data validation of input features}
\label{app:input_feature_validation}

\begin{figure}[!t]
  \centering
  \includegraphics[width=0.7\linewidth]{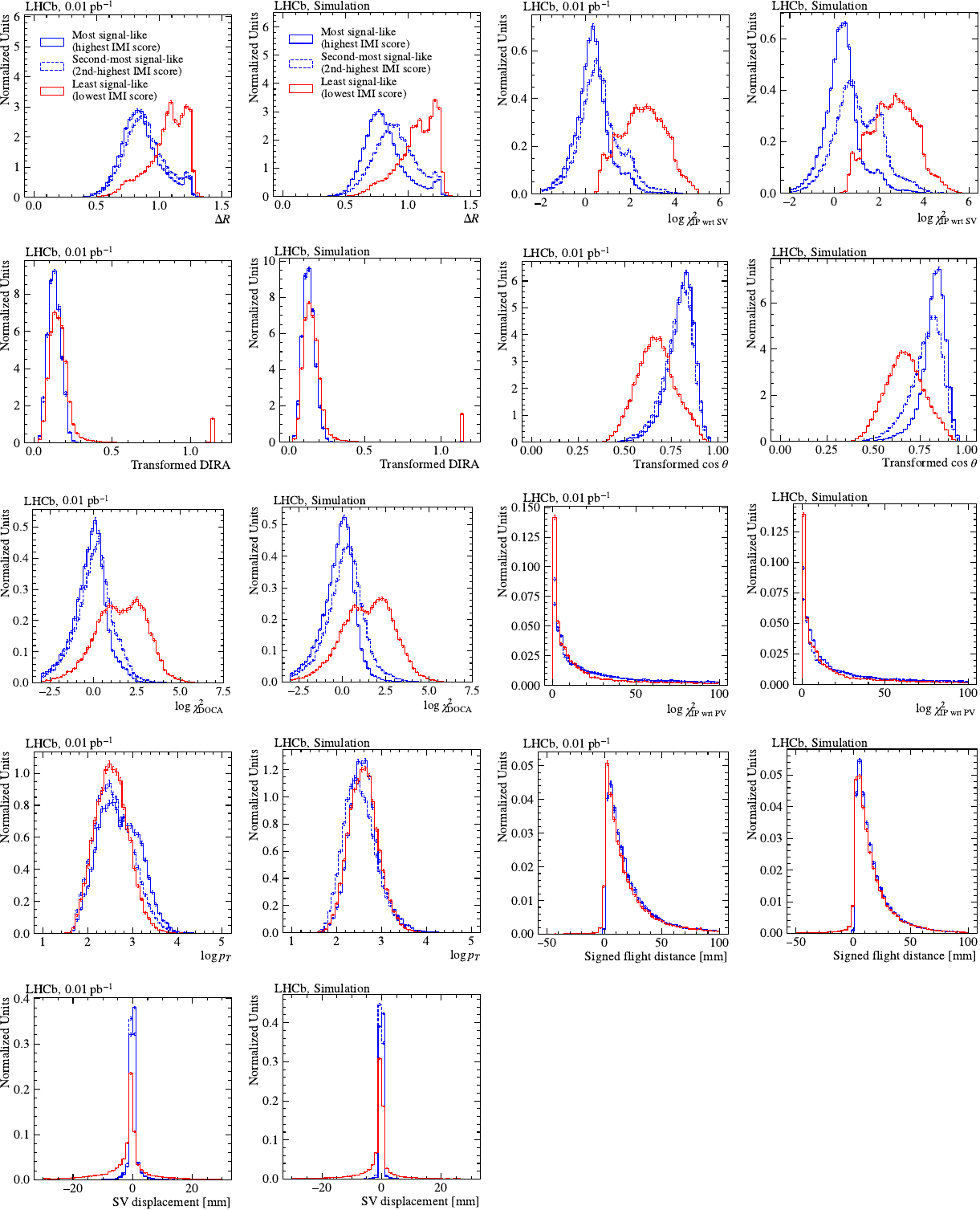}

  \caption{\small
    Distributions in partial Run~3 data (columns 1 and 3) and simulation (columns 2 and 4)
    of $B^{0} \to D^{*-}\ell^{+}\nu_{\ell}$ candidates, split by the \IMI\ ranking of the extra particle.
    Inputs shown (in order): cone angular separation (transformed $\Delta R$);
    impact parameter wrt\ SV ($\log(\chi^{2}_{\mathrm{IP\ wrt\ SV}})$);
    direction angle wrt\ flight direction (transformed DIRA);
    momentum angular separation (transformed $\cos\theta$);
    distance of closest approach significance ($\log\chi^{2}_{\mathrm{DOCA}}$);
    impact parameter wrt\ PV ($\chi^{2}_{\mathrm{IP\ wrt\ PV}}$);
    transverse momentum ($\log(p_T)$);
    signed flight distance ($d_{\rm PV}^{\rm signed}$);
    secondary vertex displacement ($d_{\rm SV}^{\rm signed}$).
  }
  \label{fig:ranking_all_inputs_compact}
\end{figure}

To further validate the \IMI\ input features (see Section~\ref{subsec:training_feature} for definitions), Figure~\ref{fig:ranking_all_inputs_compact} shows the distributions of all input features in partial Run~3 data and simulation of $B^{0} \to D^{*-}\ell^{+}\nu_{\ell}$ candidates, split by the \IMI\ ranking of the extra particle.
Overall, good agreement is observed between data and simulation for both high- and low-ranking extra particles, and the expected rank ordering is reproduced across all inputs.

Small differences are visible mainly in the tails and are more pronounced for the least signal-like particles.
In particular, the data distributions for the angular separation and vertex-related quantities (transformed $\Delta R$ and $\log\!\bigl(\chi^{2}_{\mathrm{IP\,wrt\,SV}}\bigr)$) are slightly broader, leading to increased overlap between ranks compared to simulation.
A modest broadening is also seen in displacement-sensitive observables (signed flight distance and SV displacement), consistent with additional background and resolution effects in data.
In contrast, variables such as transformed DIRA, transformed $\cos\theta$, $\log\!\bigl(\chi^{2}_{\mathrm{IP\,wrt\,PV}}\bigr)$ and $\log(p_{\mathrm{T}})$ show very similar shapes in data and simulation with only minor deviations.
These residual differences are due to the additional contributions present in data but not fully modelled in the signal simulation (various combinatorial and mis-identified background and feed-down from $D^{**}$ states), as well as imperfect modelling of underlying-event activity from soft QCD.

\newpage

\addcontentsline{toc}{section}{References}
\bibliographystyle{incl/LHCb}
\bibliography{main}
 
\end{document}